\documentclass[fleqn,usenatbib]{mnras}

\usepackage[T1]{fontenc}

\DeclareRobustCommand{\VAN}[3]{#2}
\let\VANthebibliography\thebibliography
\def\thebibliography{\DeclareRobustCommand{\VAN}[3]{##3}\VANthebibliography}

\usepackage{graphicx}
\usepackage{newtxtext,newtxmath}
\usepackage{lscape}
\usepackage{threeparttablex}
\usepackage{wasysym}
\usepackage{float}
\usepackage{diagbox}
\usepackage{capt-of}
\usepackage{natbib}
\usepackage{longtable}
\usepackage{booktabs}
\newcommand*{\distance}{\ensuremath{\mathrm{D}}}
\newcommand*{\sfr}{\rm \ensuremath{\mathrm{SFR}}}
\newcommand*{\stellarmass}{\ensuremath{\mathrm{{M}_{\star}}}}

\newcommand*{\distanceunits}{\ensuremath{\mathrm{Mpc}}}

\title[HECATEv2]{HECATEv2: An all-sky galaxy catalogue for multimessenger astrophysics}

\author[E. Kyritsis et al.]{
E. Kyritsis,$^{1,2}$\thanks{E-mail: ekyritsis@physics.uoc.gr}
A. Zezas,$^{1,2}$
K. Kovlakas,$^{4,5}$
C. Daoutis,$^{1,2}$
K. Kouroumpatzakis,$^{6}$
\newauthor
A. Hornschemeier,$^{7,8}$
A. Basu-Zych$^{9,10,11}$
\\
$^{1}$Physics Department, \& Institute of Theoretical and Computational Physics,  University of Crete, GR 71003, Heraklion, Greece\\
$^{2}$Institute of Astrophysics, Foundation for Research and Technology-Hellas, GR 71110 Heraklion, Greece
\\
$^{3}$Institute of Space Sciences (ICE, CSIC), Campus UAB, Carrer de Magrans, 08193 Barcelona, Spain\\
$^{4}$Institute of Space Sciences (ICE, CSIC), Campus UAB, Carrer de Magrans, 08193 Barcelona, Spain\\
$^{5}$Institut d’Estudis Espacials de Catalunya (IEEC), Edifici RDIT, Campus UPC, 08860 Castelldefels (Barcelona), Spain\\
$^{6}$Astronomical Institute, Academy of Sciences, Bocní II 1401, CZ-14131 Prague, Czech Republic\\
$^{7}$NASA Goddard Space Flight Center, Code 663, Greenbelt, MD 20771, USA\\
$^{8}$Department of Physics and Astronomy, Johns Hopkins University, 3400 N. Charles Street, Baltimore, MD 21218, USA\\
$^{9}$Center for Space Sciences and Technology, University of Maryland, Baltimore County, Baltimore, MD 21250\\
$^{10}$X-ray Astrophysics Laboratory, NASA/GSFC, Greenbelt, MD 20771\\
$^{11}$Center for Research and Exploration in Space Science and Technology, NASA/GSFC, Greenbelt, MD 20771
}

\date{Accepted XXX. Received YYY; in original form ZZZ}

\pubyear{\the\year{}}

\begin{document}
\label{firstpage}
\pagerange{\pageref{firstpage}--\pageref{lastpage}}
\maketitle

\begin{abstract}
We present HECATEv2, the second release of the Heraklion Extragalactic Catalogue (HECATE), an all-sky, value-added galaxy catalogue comprising 204\,733 galaxies from the HyperLEDA database with recession velocity $< 14\,000 \, \rm{km\,s^{-1}}$ (D$\lesssim$200 Mpc). This release focuses on qualitative upgrades of the provided information while maintaining the same parent galaxy sample as HECATEv1. Improvements include a new cosmology-based distance framework, expanded and homogenised optical and mid-infrared photometry from SDSS-DR17/NSA, PS1-DR2, and AllWISE, and new quality-control flags for stellar contamination, incorrect photometry, and coordinate inconsistencies. We also extend the galaxy-size coverage and derive stellar population parameters for a substantially larger fraction of the sample. Star-formation rates (\sfr{}) and stellar masses (\stellarmass{}) are now available for $>70$\% of galaxies using updated mid-IR/optical calibrations that account for stellar population age and dust attenuation, while gas-phase metallicities are derived for $\sim 90\%$. Activity classifications are provided for  $>50$\% of galaxies based on spectroscopic and/or photometric diagnostics, and supermassive black hole masses for $\sim 86\%$. In terms of $\rm L_{B}$, $\rm L_{Ks}$, \sfr{}, and \stellarmass{}, HECATEv2 is among the most complete local-Universe catalogues with spectroscopic redshifts. We also provide spatial completeness maps as a function of distance and luminosity, highlighting variations across the sky. Compared to other catalogues (e.g. GLADE+, NED-LVS), HECATEv2 offers broader (optical, near- and far-IR photometry, metallicity, activity classifications) or comparable (mid-IR photometry, \sfr{}, \stellarmass{}) coverage, making it a robust reference for studies of SMBH–host galaxy connections, gravitational-wave and high-energy transient hosts, population analyses, and rare galaxy subpopulations.

% This is a simple template for authors to write new MNRAS papers.
% The abstract should briefly describe the aims, methods, and main results of the paper.
% It should be a single paragraph not more than 250 words (200 words for Letters).
% No references should appear in the abstract.
\end{abstract}


%%%%%%%%%%%%%%%%%%%%%%%%%%%%%%%%%%%%%%%%%%%%%%%%%%

%%%%%%%%%%%%%%%%% BODY OF PAPER %%%%%%%%%%%%%%%%%%

\section{Introduction}
Large, modern galaxy catalogues are critical for advancing our understanding of galaxy demographics, their spatial distribution in the nearby Universe, and the physical processes that shape their evolution across cosmic time. Their scientific importance is increasing rapidly in the era of wide‑area and all‑sky, multi‑wavelength surveys (e.g. DESI, LSST, ZTF, Euclid, eROSITA) and with the advent of multi‑messenger observatories detecting gravitational waves (GWs), high‑energy transients, and neutrinos. In this context, value‑added galaxy catalogues that provide homogenised multi‑band photometry, robust distance estimates, and derived physical properties have become key tools for multimessenger astrophysics, time‑domain astronomy and statistical population studies.

Motivated by these needs, several major astronomical databases, such as the NASA/IPAC Extragalactic Database \citep[NED;][]{helou91}, SIMBAD \citep{wenger00}, and HyperLEDA \citep{makarov14}, have long supported extragalactic research by aggregating and organising basic galaxy information such as positions, redshifts/distances, photometric measurements, and morphological classifications. However, because these databases collect data from heterogeneous literature sources with varying apertures, depths, calibrations, and selection criteria, they are not optimised for deriving uniform galaxy‑wide physical parameters, such as star‑formation rate (\sfr{}), stellar mass (\stellarmass{}), gas‑phase metallicity, or nuclear activity class, across large, statistically representative samples.

Reliable stellar population information on large galaxy samples is especially important for the rapid identification of electromagnetic (EM) counterparts to GW events. The poor sky localisation of current GW detector networks, which frequently spans hundreds to thousands of square degrees, makes direct searches challenging \citep{petrov22}. Follow‑up strategies therefore prioritise galaxies within the GW 3D localisation volume not only by position and distance, but also by physically motivated weights that incorporate \sfr{}, \stellarmass{}, metallicity, or related parameters (i.e. specific-\sfr{}; sSFR = \sfr{}/\stellarmass{}) that trace the underlying stellar populations from which compact object binaries form \citep[e.g.][]{artale19,artale20}. The same value‑added information is also essential for broader astrophysical applications, including studies of supermassive black hole (SMBH) growth and the active galactic nucleus (AGN)–host connection \citep{kormendy13}, searches for intermediate mass black holes \citep[IMBH;][]{green20,sacchi24}, and the discovery of rare or extreme galaxy populations in the local Universe \citep{kouroumpatzakis24,kyritsis25}.

Over the past decade, several large galaxy catalogues have been assembled to support GW and time‑domain follow‑up by providing sky positions, redshift information, basic photometric data across wide-sky areas and galaxy physical properties (i.e. \sfr{}, \stellarmass{}). Examples include the Gravitational Wave Galaxy Catalog \citep[GWGC;][]{white11}, the Galaxy List for the Advanced Detector Era and its expanded release \citep[GLADE, GLADE+;][]{dalya18,dalya22}, the Census of the Local Universe survey \citep[CLU;][]{cook19}, and the NED‑Local Volume Sample \citep[NED‑LVS;][]{cook23}. While these catalogues are valuable for rapid galaxy prioritisation within GW localisation volumes they all lack systematically derived gas‑phase metallicities, despite the role of metallicity in compact binary formation efficiencies. In addition, they also do not provide nuclear classifications which are important for demographic studies of galaxies, or characterization of sources in multi-wavelength surveys. Furthermore, some of them (i.e. GLADE+) extensively use photometric redshifts even for galaxies with D$\leq$200 Mpc resulting in large uncertainties in their distance estimates. Finally, their reliance on monochromatic \sfr{} and \stellarmass{} calibrators which do not account for stellar population age and dust extinction effects, potentially biases the EM follow-up ranking schemes.

The first release of the Heraklion Extragalactic Catalogue \citep[HECATEv1;][]{kovlakas21} was designed to address many of these limitations by assembling a local‑Universe ($z\lesssim0.047$ , \distance$\lesssim 200$ \distanceunits{}) galaxy sample with cross‑matched multi‑wavelength photometry, redshift‑independent distances where available, and initial estimates of key stellar population quantities. Since its original release, HECATEv1 has been extensively used in a diverse range of astrophysical studies, including searches for hosts of GRBs and GWs \citep{giudice25}, fast radio bursts \citep{liu25,bhardwaj25}, and kilonovae \citep{franz25}. In addition, it has been used for the characterisation of the host galaxies of ultra-luminous X-ray sources \citep{kovlakas20} and the exclusion of interlopers in searches for isolated neutron star candidates \citep{kurpas26}, as well as in studies of normal galaxies \citep{vulic22,kyritsis25,kouroumpatzakis25}, galaxy clusters \citep{balzer25}, and AGN populations \citep{sacchi24,salvato25}. Owing to its all-sky nature and extensive multi-wavelength coverage, HECATEv1 has also been employed for the discovery of rare galaxy populations \citep{kouroumpatzakis24} and as a robust training set for the development of machine-learning-based diagnostics of galactic activity \citep{daoutis23,daoutis25}. Despite the wealth of value-added information included in the HECATEv1, its parameter coverage is limited. In particular, HECATEv1 provides \sfr{}, \stellarmass{}, gas-phase metallicity values, and activity classifications for $46{\%}$, $65{\%}$, $31{\%}$, and $32{\%}$, of the full sample, respectively. Improving the completeness and uniformity of these value‑added properties is essential for more reliable GW host localisation, for statistically robust studies of galaxy and SMBH co‑evolution, and for identifying rare populations in the nearby Universe. 

These requirements motivated a major upgrade of the HECATE catalogue. In this work, we present HECATEv2, an updated and substantially enhanced version of the HECATEv1. This new realease incorporates additional homogenised optical and mid‑IR photometry (explicitly accounting for the extended emission of nearby galaxies), and significantly improves the activity classifications through a combination of state-of-the art spectroscopic and photometric diagnostics. We also adopt updated \sfr{} and \stellarmass{} calibrations that account for stellar population age and dust extinction, expand and homogenise the gas-phase metallicity coverage, and implement refined distance estimates, and quality checks across the entire catalogue.

This paper is organised as follows. In Sect.~\ref{sec:sample_construction} we summarise the parent galaxy sample on which HECATE is built and in Sect.~\ref{sec:distances} we present the updated distance estimates. In Sect.~\ref{sec:multiwave_data} we present the compilation and homogenisation of multi-wavelength data and the quality checks for identifying and flagging photometric mis-associations. Identification of star contamination and treatment of galaxy sizes are discussed in Sect.~\ref{sec:star_contamination} and Sect.~\ref{sec:sizes}, respectively.  The derivation of value‑added physical parameters and the activity classifications are presented in Section~\ref{sec:physical_properties}. In Sect.~\ref{sec:Discussion} we evaluate the catalogue completeness, compare HECATEv2 with other galaxy catalogues, present potential science applications, and discuss the catalogue's limitations. Finally, Section~\ref{sec:conclusion} summarises our main results.

%--------------------------------------------------------------------
\section{Sample construction}\label{sec:sample_construction}

As the primary goal of this work is the qualitative upgrade of the HECATE catalogue, we used exactly the same galaxy sample as in the HECATEv1. While all the details of the sample selection are provided in \citet{kovlakas21}, we present here a brief description of the main steps followed to construct the galaxy sample. 

The basis of the HECATE (v1 and v2) is the HyperLEDA database \citep{makarov14} which includes, combines, and homogenises extragalactic data from the literature. By avoiding explicit flux or distance limits and addressing common issues such as misprints, duplications, poor astrometry etc., HyperLEDA serves as an ideal foundation for our catalogue. Since redshift-independent distance measurements were unavailable for the vast majority of the galaxies in the HyperLEDA, the selection of local Universe galaxies was made based on their radial heliocentric velocity, corrected for the peculiar motions of the Sun, and Milky Way including the Local Group infall to the Virgo Cluster ($\upsilon_{vir}$). For galaxies lacking radial velocity measurements in HyperLEDA, data were supplemented using NED \cite{helou91}. As a result, the local Universe galaxy sample in our catalogue includes all the objects characterised as individual galaxies ('objtype=G') in HyperLEDA with $\upsilon_{vir} < 14\,000 \, \rm{km\,s^{-1}}$, corresponding to $z{\lesssim}0.047$ and \distance$\lesssim 200$ \distanceunits{}. After excluding misclassified objects and/or duplicates through various manual inspections, the final sample of our catalogue consists of $204\,733$ objects. We note that a repetition of the HECATE catalogue construction, using the same criteria as in HECATEv1, showed that the underlying HyperLEDA database remained unchanged between HECATEv1 and HECATEv2.

\section{Updated distance estimates}\label{sec:distances}
Robust distance estimates are critical for the scientific goals of the HECATE catalogue, since they are used for the derivation of host galaxy properties. This is particularly important for local objects where peculiar velocities may lead to strong deviations from the Hubble flow. For that reason, distances in HECATEv1 were based on a combination of redshift-independent measurements from the NED distance database \citep[NED-D; i.e. Cepheids, RR Lyrae stars, Type Ia Supernovae etc;][]{Steer17}, and redshift-dependent distances derived through a regression approach trained on galaxies with known distances. Specifically, redshift-independent distances and their uncertainties were adopted directly from NED-D, with careful treatment of cases involving multiple distance measurements and their uncertainties, as well as missing values. For galaxies lacking redshift-independent distance measurements, distances and their corresponding uncertainties were estimated using a kernel regression method trained on the NED-D sample, with separate treatments for galaxies inside and outside the Virgo Cluster. For more details on the methods used for the derivation of the distances in the HECATEv1, see the Sect. 3 in \citet{kovlakas21}. 

However, in the prospect of the future expansion of the HECATE catalogue to higher redshifts, we update our distance estimation method in HECATEv2 by adopting the most recent cosmological model for galaxies lacking redshift-independent distances. In particular, for galaxies with available redshift-independent distances, we retained the measurements and associated uncertainties from HECATEv1. For the remaining galaxies, distances were assigned according to the following prioritisation scheme.

For galaxies outside the Virgo Cluster (hereafter VC), we computed distances using the Virgo-infall corrected radial velocity (v$_{\rm vir}$; provided in the HECATE, see Sect. 2 in \citealt{kovlakas21}) and the flat cold dark matter model adopting the cosmological parameters from \citet[][$H_0=67.66\rm\,km\,s^{-1}\,Mpc^{-1}$, $\Omega_m=0.30966$]{PlanckCollaboration18VI} implemented in the \texttt{astropy.cosmology.Planck18} python package. To estimate the associated uncertainties we modelled the scatter (i.e standard deviation ; $\sigma$) between the cosmological and redshift-independent distances (for galaxies with both measurements) as a function of v$_{\rm vir}$. This approach is motivated by the fact that the difference D$_{\rm cosmo}$-D$_{\rm z-ind}$ reflects deviations arising from peculiar motions, as well as from the Hubble-Lema\^itre flow in the local Universe. In particular, we performed a linear fit of the form:
\begin{equation}
    \rm \sigma(D_{\rm cosmo} - D_{\rm z-ind}) = \alpha \times (v_{\rm vir}/km\ s^{-1}) + \beta
    \label{eq:D_uncertainty}
\end{equation}

where D$_{\rm cosmo}$ is the distance derived from the cosmological model, D$_{\rm z-ind}$ is the redshift-independent distance, and v$_{\rm vir}$ is given in units of $\rm km\,s^{-1}$. The fit was performed in the range v$_{\rm vir}$ $\in [5,13\,999] \, \rm km\,s^{-1}$ using the \texttt{polyfit} function from the \texttt{numpy} python library \citep{harris20}. Using the best-fit parameters ($\alpha = 0.00285 $ and $\beta = 2.52$) we calculated the uncertainty for the redshift-dependent distance of these galaxies.

For galaxies located within the VC the cosmological model described above is not applicabl e due to the large peculiar velocities of Virgo members. To estimate the distances of galaxies within the VC that lack redshift-independent measurements, we performed a second-order polynomial fit using the subset of VC galaxies for which redshift-independent distances are available. The adopted polynomial has the following form:
\begin{equation}
    \rm D_{\rm VC\,,z-ind} = \alpha \times (v_{\rm vir}/km\ s^{-1})^{2} + \beta \times (v_{\rm vir}/ km\ s^{-1}) + \gamma
    \label{eq:D_VC_uncertainty}
\end{equation}

where D$_{\rm VC\,,z-ind}$ is the redshift-independent distance of the VC galaxies. By using the best-fit coefficients ($\alpha =1.632\times10^{-6} $, $\beta = 1.087\times 10^{-3}$, and $\gamma = 16.26$) we calculated distances for all the VC galaxies without redshift-independent distances (D$_{\rm VC}$). To quantify the uncertainties on these distances, we adopted a similar approach to that used for galaxies outside the VC by fitting a linear relation to the scatter of the difference between the calculated distances (D$_{\rm VC}$) and the redshift-independent distances (D$_{\rm VC\,,z-ind}$) as a function of the v$_{\rm vir}$. The resulting best-fit parameters ($\alpha = 2.362$ and $\beta = 2.54$) were then used to assign uncertainties to the distances of VC galaxies without NED-D measurements. These uncertainties reflect the impact of the large peculiar velocities that characterise VC members.

Finally, there is a small population (717 objects) with blue-shifted velocities (i.e. v$_{\rm vir}$$<$ 0) that are not members of VC, for which the cosmological model is not applicable. Moreover, the small number of such galaxies with redshift-independent distances (48), does not allow any reliable determination of a velocity-dependent relation. Therefore, for completeness, we assign to these galaxies a fiducial distance and associated uncertainty, derived from the mean and standard deviation of the redshift-independent distances of these 48 galaxies. However, we note that these distances are only indicative and recommend the users to treat them with caution. In Table~\ref{tab:Dmeth_stats} we present a summary of the different methods used for the calculation of the distances in the HECATEv2.

\begin{table}
% \centering
\caption{Summary of the different methods used for the Distance calculation in the HECATEv2 catalogue.}
\label{tab:Dmeth_stats}
\begin{tabular}{cc}
\hline\hline
Distance Methods in HECATEv2& Number of galaxies (\%) \\\hline
z-independent distance  & 20\,855 (10.19\%)  \\
Cosmological distance model & 182\,544  (89.16\%)\\
Virgo cluster distance model & 617 (0.3\%)\\
Fiducial distance for the blueshifted galaxies & 717 (0.35\%)\\
\hline
\end{tabular}
\end{table}

\section{Multiwavelength data}\label{sec:multiwave_data}
One of the main goals for the compilation of HECATEv2 is to signigicantly enchance the value-added information by incorporating updated photometric and spectroscopic data.  
In the following sections we describe the procedures adopted to include additional optical, and infrared photometric data from various surveys as well as the homogenisation approach employed to minimize systematics between the different surveys. We also present all the additional quality checks we implemented to ensure the reliability of the photometric information included in our catalogue. 

\subsection{Optical photometric surveys}\label{sec:optical_phot_surveys}

\subsubsection{SDSS-DR17}\label{subec:optical_phot_surveys_SDSS}
The Sloan Digital Sky Survey \citep[SDSS;][]{york01} is one of the most extensive optical surveys to date, providing deep imaging and precise photometry in five optical bands (u, g, r, i, z) across approximately one-third of the sky. While HECATEv1 utilizes data from the SDSS-DR12 catalogue, in this work, we revisit and update the optical photometric information using a more recent SDSS data release. Specifically, we adopt the SDSS-DR17 survey \citep{abdurrouf22} which provides several photometry types, including PSF magnitudes which are especially relevant for point sources (e.g. stars), and model-dependent magnitudes which are appropriate for extended objects (e.g. galaxies). For our galaxy catalogue, which mostly consists of various types of extended objects, we adopted the cModelMag and the fiberMag photometries. The former is based on a radial profile which takes the linear combination of the exponential and de Vaucouleurs fits in each band that best fits the image, providing us with the most reliable measurement of the galaxy's total flux. The latter is the flux within a 3\arcsec{}-aperture appropriate to the SDSS spectrograph, which is a good approximation of the flux in a galaxy’s nucleus (especially for the nearby galaxies) and crucial for the characterization of the nuclear activity of the galaxy \citep{daoutis23}. 

We cross-matched the HECATEv1 with the SDSS-DR17 photometric catalogue  using a search radius of 15\arcsec{} around the HECATEv1 coordinates, selecting only primary\footnote{Glossary of SDSS terminology}: \url{https://www.sdss.org/dr12/help/glossary/\#surveyprimary} objects. This initial circular search yielded in 125\,587 HECATE galaxies within the SDSS footprint (20\,582 galaxies with a single match and 105\,005 galaxies with multiple SDSS matches). We note here that although we could have used the elliptical area of each galaxy (R1,R2 see Table~\ref{tab:columns}), this more conservative circular search approach ensures that we include SDSS matches for very small and fuzzy galaxies. These objects represent the majority of the HECATE catalogue, and their size parameters are often highly uncertain (see Sect. \ref{subsec:wrong_phot_wrong_coord} for a more detailed discussion).

For galaxies with multiple matches, we selected either the brightest or the closest match depending on their size. Specifically, for galaxies with large semi-major axis (R1$> 25\arcsec{}$) we chose the brightest match within a circle of 20\arcsec{} radius. These are typically nearby spiral galaxies, and this search radius ensures that the brightest match corresponds to the galaxy's centre rather than bright knots associated with star-forming regions in the outer parts. For very small galaxies (R1$< 5\arcsec{}$) we selected the closest match within a 5\arcsec{} radius. For all other galaxies we opted for the brightest cModelMag (in the g band) match within a circular radius equal to R1.

\subsubsection{NASA-Sloan Atlas-NSA}\label{subec:optical_phot_surveys_NSA}
While SDSS-DR17 provides high-quality model-dependent photometry (cModelMag) for most of the galaxies in our sample, the fitted radial profile fails in some cases. Visual inspection of the these cases reveals that they mostly are nearby extended galaxies with very large sizes, where the model-dependent photometric fit either does not converge or produces incorrect values. 

To assign accurate optical photometry to these galaxies, we utilised the NASA-Sloan Atlas \footnote{NSA catalogue}: \url{http://nsatlas.org/} (NSA) which includes all known galaxies within the SDSS-DR8 footprint up to redshift $\rm{z} \lesssim 0.05$  ($\sim 140\,000$ galaxies). The NSA catalogue applies more advanced detection, deblending, and background subtraction techniques \citep{blanton11} than the standard SDSS pipeline, yielding improved image mosaics and photometric data in the SDSS optical bands as well as, GALEX far- and near- UV bands. It is currently the most accurate photometric catalogue for extended sources that provides photometric data in a self-consistent manner. Unlike SDSS, the NSA catalogue provides fluxes instead of the magnitudes in each band, along with inverse variance as an uncertainty measure for various photometry types (e.g. PETRO\_FLUX, SERSIC\_FLUX, FIBER\_FLUX etc.) Using these fluxes we calculated the Pogson magnitudes\footnote{SDSS flux and magnitude measurements}: \url{https://www.sdss4.org/dr17/algorithms/magnitudes/} and their corresponding uncertainties for each band\footnote{For the calculation of the Pogson AB magnitudes and their corresponding uncertainties we used the following equations:
\begin{equation*}
    m = 22.5 - 2.5log_{10}(f_{b}) \quad \rm{and} \quad \delta m = \frac{2.5}{\ln 10} \times \frac{1} {\sqrt{1/\sigma_{b}^{2}}}
\end{equation*}
where f$_{b}$ and 1/$\sigma_{b}^{2}$ correspond to the flux and the inverse variance per u, g, r, i, z band in units of nanomaggies. 
}.

To incorporate NSA photometry, we cross-matched the HECATEv1 with the NSA catalogue (\texttt{v1\_0\_1}) using a search radius of 10\arcsec{}, resulting in 108\,495 galaxies with NSA matches. Among these, 1\,117 had multiple NSA matches. We selected the brightest match apart from the cases we have blended or duplicated galaxies (see Sect.~\ref{subsec:wrong_phot_wrong_coord}). For those we selected the closest match to each on of the individual galaxies. From all the available photometry types in the NSA catalogue, we adopted the Petrosian flux and the corresponding 3\arcsec{}-aperture flux based on which we calculated the corresponding magnitudes and their uncertainties (hereafter PETRO\_MAG, FIBER\_MAG). As a result, for 32\,474 HECATE galaxies for which the cModelMag is either not available due to an unsuccessful fit or it is flagged as unreliable (i.e. clean = 0), we adopted NSA Petrosian and fiber magnitudes (where available).

\subsubsection{PanSTARRS DR2}\label{subec:optical_phot_surveys_PSDR2}
One of the main goals of HECATEv2 is to increase the coverage of the available photometric data by incorporating information from additional optical catalogues that cover larger sky areas than SDSS. To this end, we utilize the latest release of the Panoramic Survey Telescope \& Rapid Response System \citep[PanSTARRS1 or PS1-DR2 ;][]{chambers16,flewelling20}, a wide-area optical survey covering three-fourths of the sky. Designed for multi-epoch imaging, PS1-DR2 provides valuable photometric data in five optical bands (g, r, i, z, y) that complement SDSS measurements in the four common bands (g, r, i, z). In our analysis we adopted stack photometry (StackObjectThin table\footnote{Columns description of the PS1-DR2 StackObjectThin table:} \url{https://outerspace.stsci.edu/display/PANSTARRS/PS1+StackObjectThin+table+fields}), since it provides the best signal to noise ratio (S/N) according to \cite{magnier20}.

Similarly to SDSS-DR17, PS1-DR2 contains various photometric measurements for point-like sources such as PSF magnitudes and for extended sources such as model-dependent magnitudes (i.e. exponential, de Vaucouleurs, Sérsic etc.), curve of growth - derived magnitudes (i.e. Kron magnitude), and fixed-aperture magnitudes measured within predefined aperture radii (i.e. 1.03\arcsec{}, 1.76\arcsec{}, 3.00\arcsec{}, 4.63\arcsec{} etc.). Since we are interested in extended objects like galaxies, in our analysis we adopt the model-independent Kron magnitudes (KronMag) which is inferred from the curve of growth, after determining the Kron radius of the object \citep{kron80}. This choice is motivated by the fact that PS1-DR2 photometric pipeline is optimised for point sources, and its model-dependent surface brightness fits are unstable espcially for fainter extended objects. Furthermore, comparisons between different PS1-DR2 photometries  with those from SDSS shows better consistnecy between PS1-DR2 KronMag and and SDSS cModeMag magnitudes. Additionally, given that we need the corresponding photometry for the nuclear region for the activity classification of the galaxies in our catalogue, we also use the fixed-aperture flux (c6flxR5), measured within an aperture of r=3\arcsec{}, and we calculated the corresponding Pogson AB magnitudes (hereafter c6flxR5\_MAG) and their uncertainties for each band\footnote{For the calculation of the Pogson AB magnitudes and their corresponding uncertainties we used the following equations:
\begin{equation*}
    m = 8.9 - 2.5log_{10}(f_{b}) \quad \rm{and} \quad \delta m = \frac{2.5}{\ln 10} \times \frac{\delta f_{b}} {f_{b}}
\end{equation*}
where f$_{b}$ and $\delta f_{b}$ correspond to the flux and the flux uncertainty per g, r, i, z, y band in units of Jy. }. These 3\arcsec{}-aperture magnitudes are equivalent to the SDSS fiberMag photometry. 

We cross-matched the HECATEv1 with the PS1-DR2 photometric catalogue using a search radius of 15\arcsec{} around the HECATEv1 coordinates, selecting only primary objects with available Kron magnitudes at least in the g, and r bands. This cross-match yielded 102\,250 galaxies with a single match and 61\,137 galaxies with multiple sources. Following exactly the same methodology as in Sect. \ref{subec:optical_phot_surveys_SDSS} we selected either the closest or the brightest Kron magnitude (in the g band) resulting in 162\,932 galaxies with available PS1-DR2 photometric matches.

\subsubsection{Objects with wrong photometric matches and wrong coordinates}\label{subsec:wrong_phot_wrong_coord}
In order to select the unique optical SDSS-DR17 or PS1-DR2 photometric matches we used a circular search radius which ensures that we include matches even for the very small and fuzzy galaxies with uncertain sizes (see Sect. \ref{subec:optical_phot_surveys_SDSS}). However, in the case of galaxies with non-circular morphologies, this approach may still result in some incorrect photometric matches outside the galaxy’s ellipse. 

\begin{figure*}
        \includegraphics[width=\linewidth]{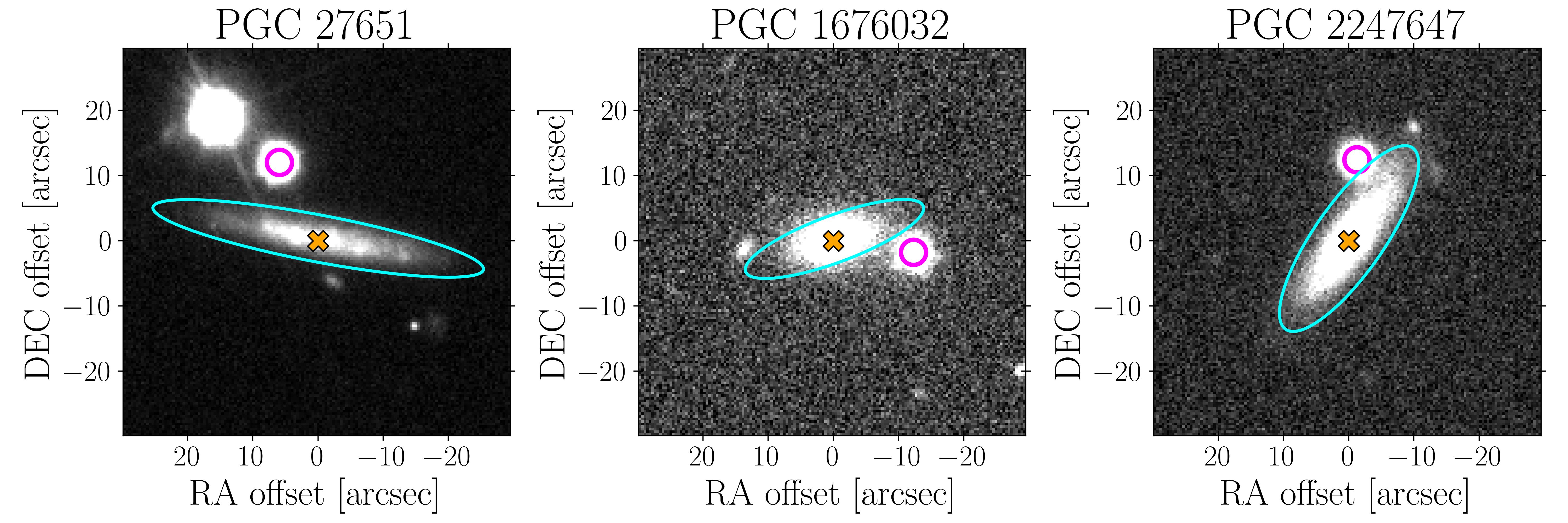}
        \includegraphics[width=\linewidth]{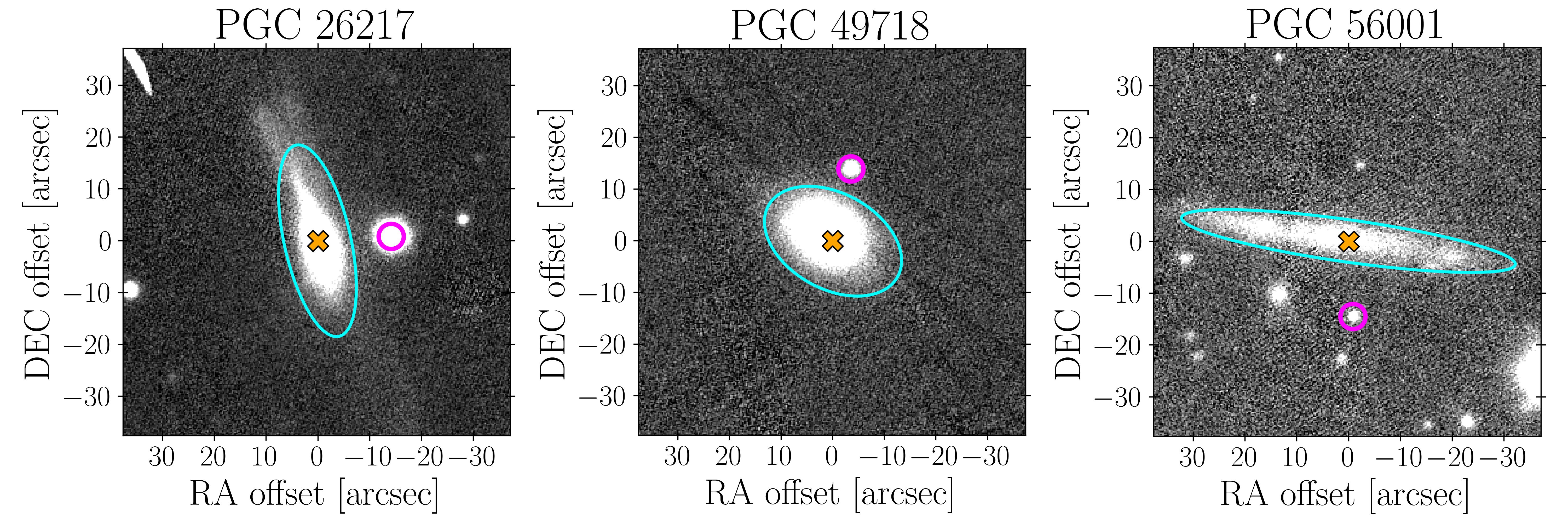}
    \caption{ A few examples of galaxies identified as having wrong photometric matches by our analysis desrcibed in Sect.\ref{subsec:wrong_phot_wrong_coord}. The top and bottom rows display galaxy images obtained from SDSS and the PS1-DR2 surveys, respectively, in the g band. The cyan ellipse represents the D$_{25}$ size of each galaxy, and
the magenta open circle marks the incorrect photometric match. The orange x symbol represents the HECATEv2's galaxy center (RA,DEC).}
    \label{fig:SDSS_PSDR2_wrong_matches}
\end{figure*}

Indeed, visual inspection of galaxies with elliptical shapes revealed that these mismatches usually involve elongated galaxies (small semi-minor axis; R2) for which the brightest match is either a foreground (i.e. star) or background source (e.g. QSO) outside the ellipse defined by the galaxy size parameters, but inside the circle of radius R1. To identify and flag such cases we checked whether the selected best SDSS or PS1-DR2 matches (from the circular search) fall within the galaxy's ellipse. In particular, 
we cross-match independently the HECATE-SDSS-DR17 and the HECATE-PS1-DR2 tables (containing multiple matches) with the corresponding tables from each catalogue of the selected best matches using the Sky Ellipses algorithm in the TOPCAT astronomical software\footnote{ TOPCAT software}: \url{https://www.star.bris.ac.uk/~mbt/topcat/}. For this, we used the elliptical radii (using the galaxy size parameters R1, R2, PA) for the HECATE galaxies, while assuming a fiducial positional error of 1.5\arcsec{} and 2.5\arcsec{} for the SDSS-DR17 and PS1-DR2 sources. Based on this analysis, we found 1\,017 and 1\,543 HECATE galaxies with SDSS-DR17 and PS1-DR2 matches, respectively, characterized as best photometric matches located outside the ellipse but within the circular search radius. We identified all of them as wrong photometric matches. We opted not to remove them and flag them instead, since screening of these matches would require visual inspection of each one in order to assess the reliability of the galaxy sizes for the more diffuse galaxies. In Fig. \ref{fig:SDSS_PSDR2_wrong_matches} we present a few examples of galaxies identified as having wrong photometric matches to demonstrate the effectiveness of our approach in identifying and flagging mismatched photometric associations in the HECATEv2 catalogue.

In addition, in the HECATEv1, there are a few cases where the assigned galaxy centroid coordinates exhibit an offset from the true galaxy's position. This is due to bad entries and/or mismatches in the parent databases. In order to systematically identify and flag these cases we used the results of the methodology for the detection of wrong photometric matches, by selecting galaxies that lack a match within the elliptical search region but have a match outside of it. In this way, we identified and flagged 166 HECATE galaxies as having wrong coordinates. 

\subsubsection{Adoption of the final homogenised optical photometries in the HECATEv2}\label{subsec:optical_phot_homogenization}
The optical photometric information in the HECATEv2 comes from multiple catalogues and surveys (i.e. SDSS-DR17, NSA, PS1-DR2), each introducing its own systematic offsets. To ensure consistent optical magnitudes for the largest possible set in the HECATEv2 it is crucial to homogenise the available optical photometry. For that reason, and following the analysis detailed in Appendix~\ref{append:opt_phot_rescal}, we rescaled all the optical photometries to match the reliable SDSS-DR17 photometry. 
Our results showed that no rescaling was needed between the SDSS and NSA integrated photometries, as they are in very good agreement (difference $<$0.1 mag). This reinforces the use of NSA photometry instead of the SDSS-DR1 when available. However, systematic differences were identified between SDSS/NSA and PS1-DR2 photometry (see Appendix.~\ref{append:opt_phot_rescal} for the method followed to correct for these offsets).

\begin{figure}
        \includegraphics[width=\columnwidth]{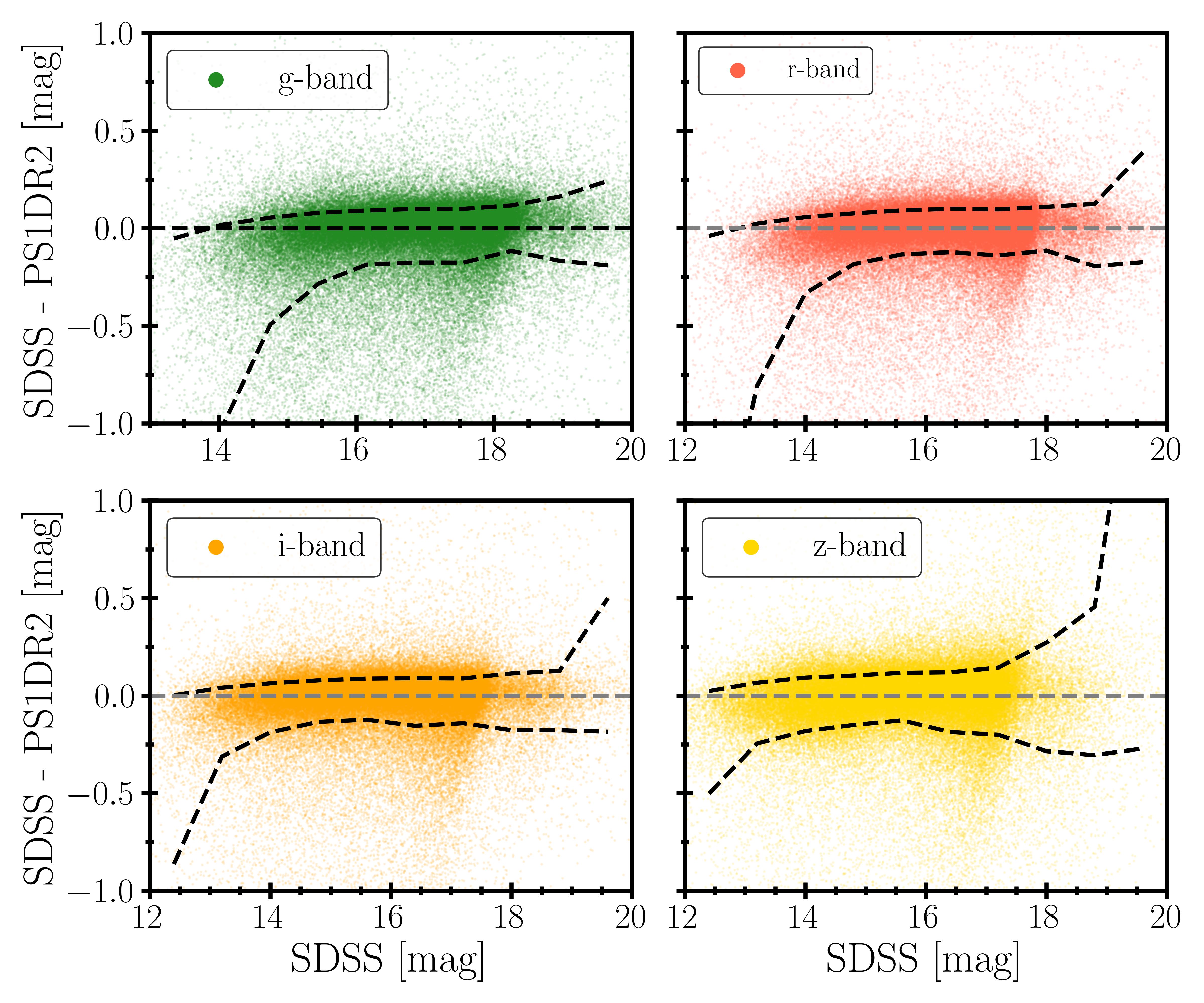}
    \caption{Comparison between the SDSS/NSA (cModelMag or PETRO\_MAG) and the rescaled PS1-DR2 Kron magnitudes as a function of the reference standard SDSS magnitudes, for the common bands (i.e. g,r,i,z) between the two optical surveys. The dashed dark gray line depicts the line of equality while the black dashed lines show the 16\% and 84\% percentiles. The two photometries are highly consistent, with a scatter of $\sim 0.25$ mag around the equality line.}
    \label{fig:SDSS_vs_PS1-DR2}
\end{figure}

After applying all the rescaling factors listed in Table~\ref{tab:PS1DR2-rescaling_factors} to bring the photomerties to the SDSS-DR17 photometry, we present in Fig. \ref{fig:SDSS_vs_PS1-DR2} the difference between the SDSS (cModelMag or PETRO\_MAG) and the rescaled PS1-DR2 Kron magnitudes as a function of the reference standard SDSS magnitudes. The 16\% and 84\% percentiles (black dashed lines) indicate that, after rescaling, the two photometries are highly consistent, with scatter around the equality line within $\sim 0.25$ mag. In summary, we adopted the SDSS-DR17 photometry as the reference optical photometry for HECATEv2. Where available, we used NSA photometry instead, to better account for the extended emission of nearby galaxies. Finally, in cases where neither SDSS-DR17 nor NSA data were available, we supplemented them  with PS1-DR2 photometry on a per-band basis, after correcting for the relevant offsets as needed.

Since the photometric data come from multiple surveys, the associated quality flags differ. Specifically, in SDSS-DR17, photometry is considered reliable when the 'clean' column is set to 1 and unreliable when it is 0\footnote{{Tutorial of SDSS flags}: \url{https://www.sdss4.org/dr14/tutorials/flags/}. In NSA, the quality is indicated by the 'DFLAGS' column, where a value of 0 means a reliable measurement, while 1 indicates an unreliable one \footnote{Description of the NSA catalogue columns}: \url{https://nsatlas.org/data}}. Finally, in PS1-DR2, photometry is reliable when the value of the 'QualityFlag' column is $\le$ 63\footnote{Tutorial of PS1-DR2 flags}: \url{https://outerspace.stsci.edu/display/PANSTARRS/PS1+Object+Flags}. In order to have a consistent photometric quality flag across all surveys, we adopted the SDSS-DR17 convention. Specifically, we assigned a value of 1 to all reliable data from different surveys for each band (i.e. u, g, r, i, z, and y) and 0 to the unreliable ones. The value 2 is used to indicate missing data for each band. 

\subsection{Infrared photometric surveys}
\subsubsection{Mid-infrared: AllWISE}\label{subsec:midir_phot_surveys_AllWISE}

The Wide-field Infrared Survey Explorer mission \cite[WISE ;][]{wright10} is the most sensitive mid-infrared survey up to now, covering 99.86\% of the entire sky in four broad bands: W1 (3.4 $\rm{\mu}$m), W2 (4.6 $\rm{\mu}$m), W3 (12 $\rm{\mu}$m), and W4 (22 $\rm{\mu}$m). By combining data from the WISE cryogenic and NEOWISE \citep{mainzer11} post-cryogenic survey phases the AllWISE Source Catalogue\footnote{Description of the AllWISE catalogue columns}: \url{https://wise2.ipac.caltech.edu/docs/release/allwise/expsup/sec2_1a.html} provides new products that have enhanced photometric sensitivity and accuracy, and improved astrometric precision compared to the earlier All-Sky Data Release Catalogue.

The HECATEv1 provides mid-IR photometric data obtained from the forced photometry catalog by \cite{lang16} who extracted fluxes from unWISE coadds \citep{lang14}. The unWISE coadds are full-depth combinations of WISE and NEOWISE imaging that preserve the native angular resolution of the original WISE observations, enabling accurate forced photometry at the positions of SDSS-DR10 photometric objects using the same apertures as SDSS. While this catalogue provides reliable photometric data for both point-like and extended sources, it is limited to the SDSS footprint, resulting in an incomplete census of the mid-IR emission for HECATE galaxies. In order to increase the completeness of the mid-IR photometric information, in the HECATEv2 we utilise the 
AllWISE Source Catalogue.

To that end, we cross-matched the HECATEv1 with the AllWISE Source Catalogue using an adaptive search radius that accounts for the size of each galaxy, selecting the closest match. More specifically, for galaxies smaller than the AllWISE angular resolution (R1 $<$ 6.5\arcsec{}; at W3) we used a fixed search radius of 10\arcsec{}. For larger galaxies with R1$>$20\arcsec{}, we increased the search radius to 25\arcsec{}. Finally, for galaxies with 6.5\arcsec{}$\leq$R1$\leq$20\arcsec{}, we performed an elliptical search based on their size parameters (i.e. R1, R2, PA) to ensure accurate matches. Using this approach, we identified a total of 200\,783 matches within the AllWISE Source Catalogue.

The AllWISE survey provides a large variety of fixed-aperture and profile-fit photometries. Given the broad range of galaxy sizes and distances in our catalogue, in the updated version of the HECATE we consider two types of photometry: the w?mag\_2, and w?gmag (where the question mark represents the four WISE bands: 1,2,3 and 4). The former is a reliable tracer of integrated photometry for more distant, unresolved galaxies, while the latter fully captures the emission from nearby, resolved galaxies. Specifically, the w?mag\_2 photometry is measured within a fixed circular aperture of 8.25\arcsec{} radius centered on the source position in each AllWISE band. Although these measurements do not include any curve-of-growth correction, we accounted for aperture effects by applying appropriate corrections of 0.261, 0.319, 0.825 and 0.576 mag, for W1, W2, W3 and W4, respectively\footnote{Apperture corrections for AllWISE magnitudes}: \url{https://wise2.ipac.caltech.edu/docs/release/allsky/expsup/sec4_4c.html\#resolved}. 
In contrast, the w?gmag\_2 is based on flux measurements within elliptical apertures with parameters derived by the 2MASS Extended Source Catalogue (XSC) \citep{jarrett00}. To ensure the broadest possible mid-IR coverage in HECATEv2 while accounting for all the different galaxy sizes, we adopted a hybrid photometric scheme using  w?gmag photometry whenever available (86\,159 galaxies), and for the remaining 114\,624 galaxies we adopted the w?mag\_2. However, before we adopt them we rescaled them using as a reference the reliable forced photometry (WF) \citep{lang16} from HECATEv1. This analysis is presented in the Appendix \ref{append:WISE_hyb_rescal}. In Fig. \ref{fig:WF_vs_Whyb} we show the difference between the forced photometry (WF) and the rescaled hybrid photometric scheme (W$_{\rm hyb}$) magnitudes as a function of the reference standard WF magnitudes. The 16\% and 84\% percentiles indicate again, after rescaling, the two photometries are in strong agreement, with scatter of less than 0.5 mag. 

The quality flags per band provided by the AllWISE Source Catalogue can be divided in two main categories: detector artifacts (i.e. w?flg\_2 or w?gflg columns) and image artifacts (i.e. cc\_flags). The former indicate whether one or more image pixels in the measurement aperture for a given band are affected by contamination of saturated or unusable pixels, confusion with nearby objects, or is an upper limit. A value greater than 1 in these columns marks potentially unreliable data. The latter is a contamination and confusion indicator flagging the sources whose photometry and/or position measurements may be biased due to proximity to image artifacts. A non-zero value in this column suggests possible contamination. To combine these different types of quality flags, we created a new eight-character flag where each character is either 0 or 1. The first four characters (corresponding to bands W1,W2,W3,W4) represent  detector artifact flags, while the last four correspond to image artifact flags. Following the SDSS-DR17 convention, a value of 1 indicates reliable photometry, while 0 denotes unreliable data. For instance, a quality flag of "10101010" means that the photometry is reliable in the W1 and W3 bands but unreliable in W2 and W4.

\begin{figure}
        \includegraphics[width=\columnwidth]{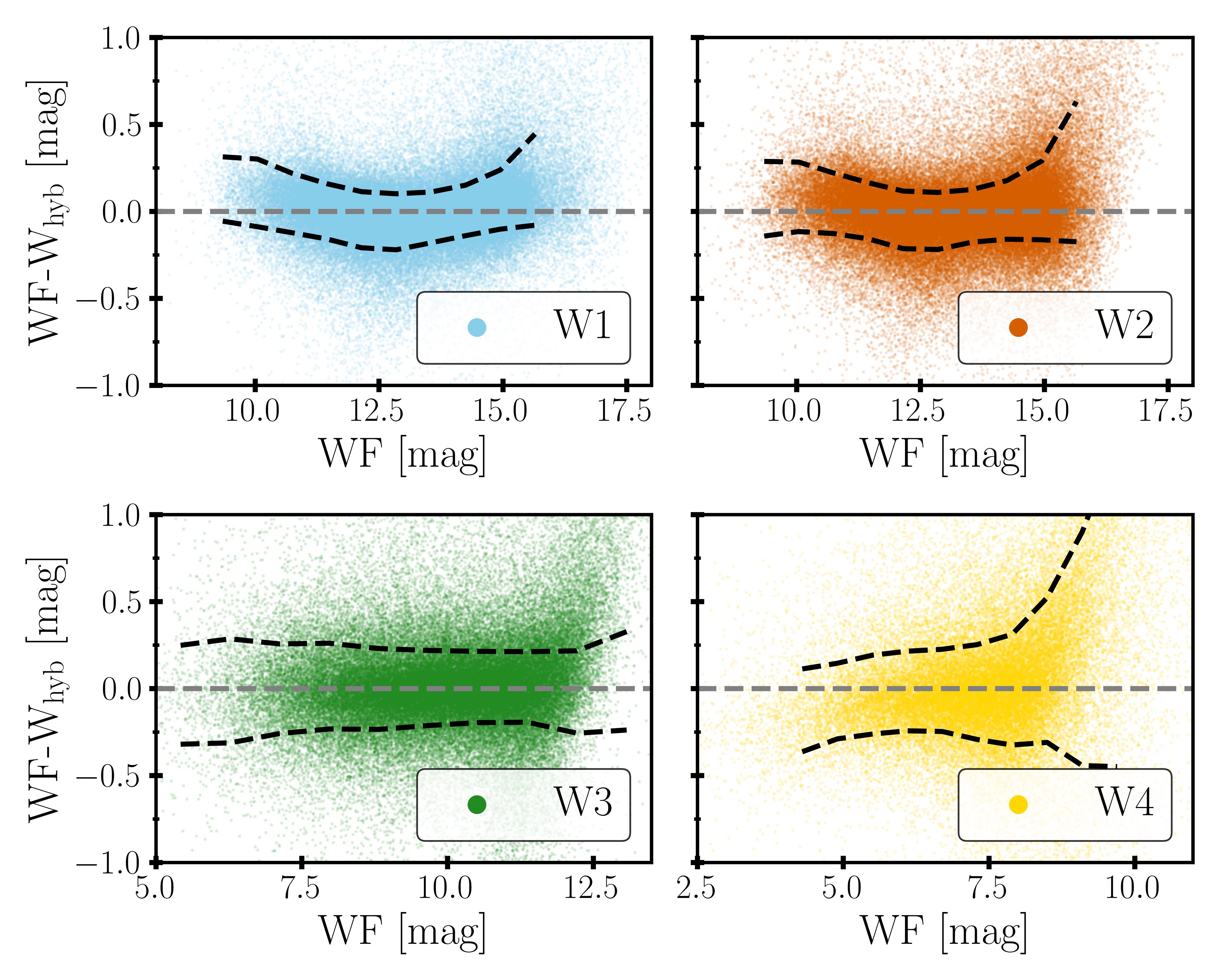}
    \caption{Comparison between the forced photometry magnitudes and the rescaled magnitudes derived from our hybrid photometric scheme (W$_{\rm hyb}$), as a function of the reference standard WF magnitudes. The dashed dark gray line indicates the line of equality while the black dashed lines show the 16\% and 84\% percentiles. The two photometries are in strong agreement with a scatter less than $\sim 0.5$ mag around the equality line.}
    \label{fig:WF_vs_Whyb}
\end{figure}

\subsubsection{Far-infrared \& Near-infrared: IRAS \& 2MASS}
In HECATEv2, the far-infrared and near-infrared photometry has been directly adopted from HECATEv1. Specifically, the far-infrared photometry is based on the cross-match of HECATEv1 with the Infrared Astronomical Satellite (IRAS) survey. We provide far-infrared fluxes at 12, 25, 60, and 100 $\mu$m for 19\,082 galaxies, using data from the Revised IRAS-FSC Redshift Catalogue \citep[RIFSCz;][]{wang14} and the IRAS Revised Bright Galaxy Sample \citep[IRAS-RBGS;][]{sanders03}, ensuring proper coverage of extended sources.

The near-infrared photometry is obtained from the cross-match of the HECATEv1 with three 2MASS catalogues, following a priority order to ensure the most reliable flux measurements for resolved galaxies: (i) Large Galaxy Atlas \citep[2MASS-LGA;][]{jarrett03}, (ii) Extended Source Catalogue \citep[2MASS-XSC;][]{skrutskie06}, and (iii) Point Source Catalogue \citep[2MASS-PSC;][]{cutri12}. As a result, HECATEv2 provides near-infrared 2MASS photometry in the JHK bands for 143\,546 galaxies. We note that when no uncertainty is provided, the listed magnitudes should be considered as upper limits. For further details on the cross-matching process, see the HECATEv1 paper by \cite{kovlakas21}.

\subsection{Optical spectroscopic data}\label{subsec:optical_spec_data}
The spectroscopic data provided in HECATEv2 originate from three different sources and they are used to derive metallicities and classify galaxies based on their nuclear activity. The vast majority of the spectroscopic measurements come from the MPA-JHU DR8 catalogue \citep{kauffmann03,brinchmann04,tremonti04}, which provides emission line fluxes for all galaxies in the SDSS-DR8 sample after subtracting the underlying stellar component. To incorporate these data into HECATEv2, we cross-matched (based on the SDSS objID identified from the cross-match of HECATE with the SDSS photometric catalogues; Sect.\ \ref{subec:optical_phot_surveys_SDSS}) our catalogue with the MPA-JHU DR8 catalogue, selecting only reliable measurements (RELIABLE=1) and excluding galaxies for which the continuum subtraction failed, leading to negative fluxes for the measured spectral lines. This resulted in a final set of 70\,898 galaxies with available emission line measurements. 

We also include 61 additional galaxies with spectra obtained and analyzed for other projects \cite[e.g. the study of the X-ray emission from star-forming galaxies with eROSITA;][]{kyritsis25}. In particular, we retrieved 56 spectra from the SDSS-DR17 spectroscopic database, and we measured the fluxes by fitting their spectral lines after subtracting the contributions from stellar populations. Furthermore, we also obtained five spectra from observing runs at Skinakas Observatory\footnote{Skinakas Observatory}: \url{https://skinakas.physics.uoc.gr/en/home/} in Crete. These spectra were processed following standard spectroscopic reduction procedures, including bias and flat-field corrections, wavelength calibration, and spectral extraction, using the IRAF astronomical software. The spectral line fluxes were then measured following the same methodology as for the 56 SDSS-DR17 galaxies. A more detailed description of the spectral analysis will be provided in a forthcoming work (Kyritsis et al. in prep.). In total, HECATEv2 provides spectroscopic information for 70\,959 galaxies.  

\section{Star contamination}\label{sec:star_contamination}
One of the biggest challenges in constructing large galaxy catalogues like HECATE is ensuring that all included objects are correctly classified as galaxies. While the vast majority of HECATEv1 objects are identified as galaxies there are a few cases involving stars misclassified as galaxies or galaxies assigned incorrect radial velocities due to contamination from nearby stars. Identifying and characterising such cases systematically is crucial, since wrong velocity measurements can lead to incorrect distance estimations, which in turn affect derived luminosities and other physical properties of the galaxies.

For this reason, in HECATEv2 we utilized the morphological and spectroscopic classification information provided by SDSS-DR17. Specifically, the morphological classification is based on the difference between the cModel and the PSF magnitudes, classifying an object as extended source (i.e. galaxy) if psfMag - cmodelMag $>$ 0.145, otherwise it is classified as a point-like source (i.e. star). The spectroscopic classification, on the other hand, relies on spectral template matching to categorize objects as galaxies, stars, QSOs etc. To systematically identify misclassified objects, we cross-matched HECATEv2 (using the objID from the photometric cross-match, Sect. \ref{subec:optical_phot_surveys_SDSS}) with the SpecPhoto Table\footnote{Columns description of the SpecPhoto table}: \url{https://skyserver.sdss.org/dr7/en/help/browser/browser.asp?n=SpecPhoto\&t=U} from SDSS-DR17, which contains all SDSS objects with clean photometry and spectra. 

By selecting only objects with spectra that have no known issues (zWarning $\leq$ 1), we identified 136 HECATEv2 objects that were classified as stars by both methods, morphologically (type = 6) and spectroscopically (class = 'STAR'). To ensure the stellar nature of these objects, we visually inspected all of them and confirmed that 89 were indeed stars. We flagged these objects as stars. The remaining cases were either HECATE galaxies wrongly associated with a nearby star (i.e wrong photometric match) or objects with wrong coordinates. We flagged these cases and we included them in the relevant HECATEv2 columns described in Sect. \ref{subsec:wrong_phot_wrong_coord}. Additionally, we identified 1\,243 objects which exhibit contradictory classifications, namely objects that were morphologically classified as galaxies (type = 3) but spectroscopically identified as stars (class = 'STAR'). These cases were flagged as candidate stars. 

To identify as many misclassified stars as possible in HECATEv2, we also searched for problematic galaxies that are not covered by SDSS-DR17 footprint but have coverage from the PS1-DR2 survey. To achieve this, we cross-matched the HECATEv2 with the PS1 Point Source Catalogue (PS1-PSC) from \cite{tachibana18} using the objid from the photometric PS1-DR2 matches (see Sect.\ref{subec:optical_phot_surveys_PSDR2}). This catalogue provides morphological classifications of unresolved point sources in the PanSTARRS1 survey (PS1). In their work, \cite{tachibana18} combined flux and morphological properties measured in the PS1 stack images across five filters to develop a machine-learning model that estimates point-source probabilities for 1.5 billion objects in the PS1 survey. Their model assigns a probabilistic ranking to all sources (ps\_score), where a value of 0 corresponds to extended objects and a value of 1 corresponds to point sources. Based on this table, we found 545 additional HECATE objects with ps\_score=1 and we flagged them as candidate stars.

We note here, that visual inspection of the objects characterized as candidate stars revealed that the majority are indeed extended objects that were spectroscopically classified as stars either due to the projection of star on the galaxy or because of poor-quality spectra that confused the template matching. As a result, the cases flagged as candidate stars should, in principle, be considered as galaxies, but they should be used with caution as they may be contaminated by nearby or overlapping stars. In total, HECATEv2 contains 91 objects identified as stars and 1\,788 objects flagged as candidate stars.

\section{Galaxy sizes}\label{sec:sizes}
\subsection{Sizes in HECATEv2}\label{subsec:sizes}

 The size of a galaxy is typically measured using the 25 mag\,arcsec$^{-2}$ surface-brightness isophote in the optical B band \cite[D$_{25}$;][]{paturel91,paturel97}, and defines an ellipse characterised by a semi-major axis (R1), a semi-minor axis (R2), and a position angle (PA). Knowing that size is crucial, especially in large galaxy catalogues suchs as HECATE, as it has important implications for morphological classification, as well as for time-domain and multimessenger astrophysics. For instance, accurate size information can be used for identifying the host galaxy of multi-wavelength sources and more importantly the EM counterparts of gravitational wave events \citep{ducoin20,kovlakas21}.

To maintain uniformity over the full range of galaxy sizes, in HECATEv2 we retained the size measurements from HECATEv1. However, in Sect. \ref{subsect:size_comparison} we compare them with sizes from other, more detailed surveys, assesing their reliability and robustness. The vast majority of these size measurements ($\sim$80\%) were adopted from the HyperLEDA database \citep{makarov14} and correspond to the D$_{25}$ isophote. For the remaining galaxies, size information was drawn from various catalogues, with the most significant contributions coming from SDSS \citep{aguado19} and 2MASS \citep{jarrett00} and they have been rescaled to match the D$_{25}$ size in HyperLEDA (see Appendix B.1 from \cite{kovlakas21}). 

Althought the completness of the size infromation in the HECATEv1 is $>98$\%, size measurements are missing for 4\,837 galaxies. These galaxies lie outside the main survey footprints used in the compilation of the HECATEv1. To recover size information for these cases, we utilised photometric information from the PS1-DR2 survey using data from the StackObjectThin and StackModelFitSer tables. Specifically, we used the Kron radius (gKronRad) and the best-fitted Sérsic axis ratio (gSerAb; defined as minor-to-major axis) in the g band to derive new elliptical sizes. We adopted the gKronRad as a proxy for the semi-major axis of the ellipse and for the derivation of the semi-minor axis we multiplied the gKronRad by the gSerAb. The choice of the gKronRad over the best-fitted Sérsic radius was motivated by the fact that the latter requires the best-fitted Sérsic index, which was unavailable due to a corrupted column in the PS1-DR2 catalogues. The position angle of the ellipse was assigned using the best-fitted Sérsic major-axis position angle (gSerPhi). Before adopting the Kron radius in our analysis we compared it with the R1 which is the reference size in our catalogue. By computing the gKronRad/R1 ratio for all the HECATE galaxies with available R1, R2 and gKronRad we found that there is a systematic offset ($\sim$0.2'), with the gKronRad being systematically smaller than the R1. To correct for this offset, we calculated the median value of the gKronRad/R1 ratio, and we rescaled the gKronRad accordingly. For galaxies without available gSerAb measurements we assumed circular sizes using the gKronRad as the radius. Finally, for objects lacking both gKronRad and gSerAb, we assigned circular sizes with a radius equal to the median gKronRad of the galaxies with available measurements. In this way, 1\,313 galaxies without size information in HECATEv2 received new elliptical apertures, while 3\,524 galaxies were assigned new circular apertures. Of these 1\,996 received circular apertures based on their individual gKronRad values, while the remaining 1\,528 galaxies were assigned circular apertures using median gKronRad value. 

Visual inspection of a representative set of these galaxies revealed that the adopted elliptical apertures provide a good proxy for the galaxy size in most cases. Galaxies assigned circular apertures are generally very faint and fuzzy, making the precise size measurement very challenging. In these cases, visual inspection showed that the gKron photometric measurements where often based on bright knots within the galaxy body, resulting in incorrect gKronRad measurements. Therefore, we suggest users to interpret these sizes with caution. 

\subsection{Comparison of galaxy sizes with other surveys}\label{subsect:size_comparison}
Given the various sources of size information incorporated in HECATEv2, it is important to compare its size measurements with other studies providing morphological information for galaxies. For that reason we cross-matched the HECATEv2 with the Siena Galaxy Atlas SGA-2020\footnote{SGA-2020 catalogue}: \url{https://sga.legacysurvey.org/} compiled by \cite{moustakas23} using a search radius of 10\arcsec{}. This optical and infrared imaging atlas includes $\sim$400\,000 galaxies with D$_{25}>$20\arcsec{} (in the B band) within the 20\,000 deg$^{2}$ footprint of the Dark Energy Spectroscopic Instrument (DESI) Legacy Imaging Surveys Data Release 9 \citep[LS/DR9;][]{dey19}. Through a detailed and homogenized analysis of the grz optical bands from DESI and the 3.4-22 $\mu$m infrared bands from unWISE coadds, SGA-2020 provides precise coordinates, multiwavelength mosaics, model images, photometry, optical surface-brightness profiles and galaxy sizes for the full sample, making it an ideal comparison sample for assessing the accuracy of the HECATEv2 size measurements. 

For our comparison we opted to use the semi-major axis length at the 25 mag\,arcsec$^{-2}$ isophote (i.e. SMB\_SB25) provided in the SGA-2020 catalogue, as it is the closest to the R1 semi-major axis measurement in HECATEv2. However, since the SMB\_SB25 is derived from the r-band, while the R1 is based on the B-band, a direct comparison may introduce biases due to differences in bandpasses. To mitigate this, we calculated the corresponding SMA\_SB25 radius in the g-band of DESI (i.e. SMA\_SB25\_G), which is closer to the B-band than the r-band. To achieve this, we utilised the curve of growth for the g-band (measured at the 22, 22.5, 23, 23.5, 24, 24.5, 25, 25.5, 26 mag isophotal radii) provided by SGA-2020. From these, we computed the surface-brightness profile which was then interpolated to determine the radius corresponding to 25 mag\,arcsec$^{-2}$ in the g-band. 

\begin{figure}
        \includegraphics[width=\columnwidth]{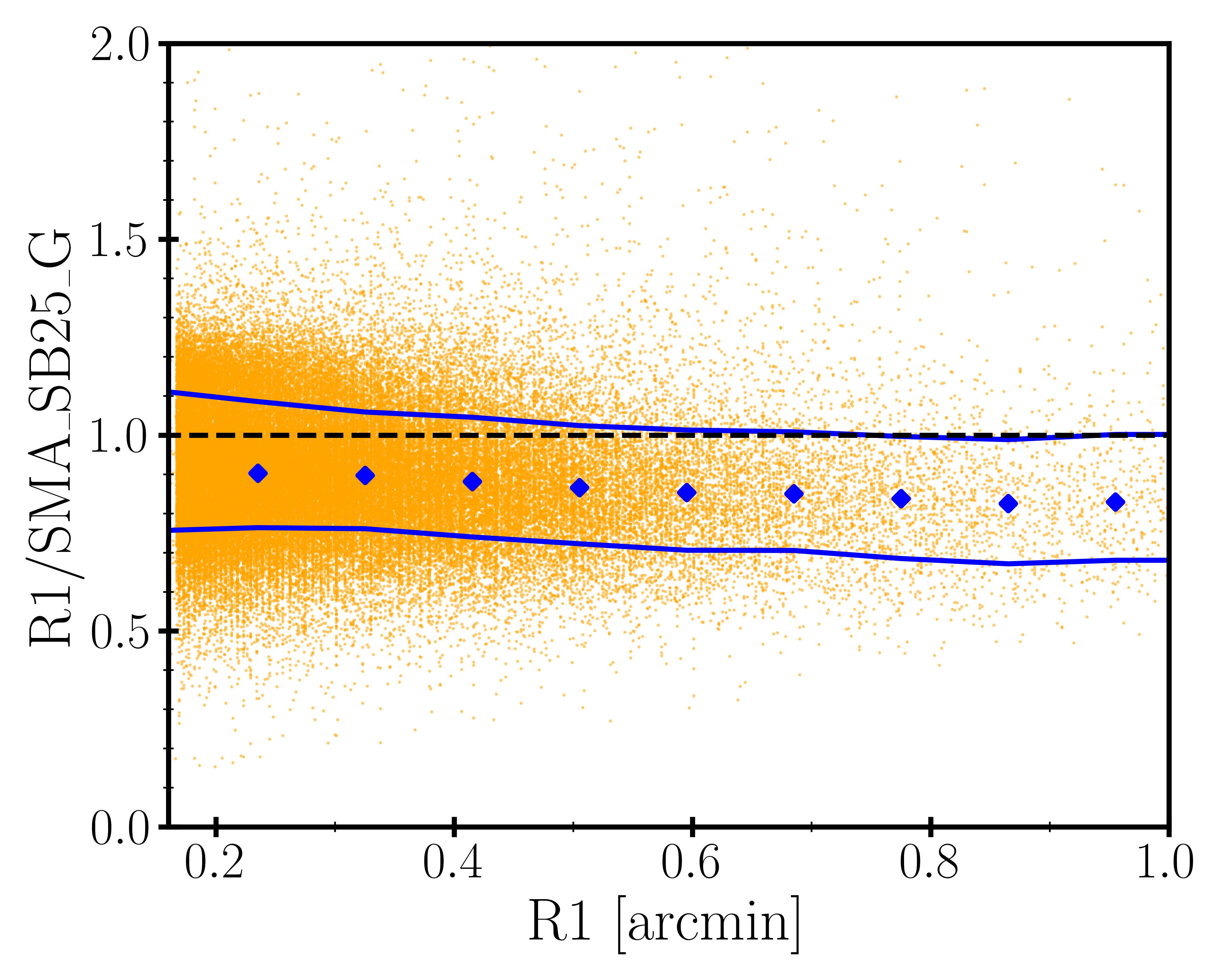}
    \caption{ The ratio of the R1 radius (as provided by HECATE) to the SMA\_SB25\_G radius (from the SGA-2020 catalogue) is shown as a function of R1. Blue diamonds represent the median ratio in 9 equally spaced R1 bins, while the blue solid lines indicate the 16th and 84th percentiles. The black dashed line denotes the one-to-one relation. Overall, the two size measurements are in good agreement, with a small systematic offset (less than $\approx$20\%) likely due to the higher signal-to-noise of the deeper optical imaging used in SGA-2020.}
    \label{fig:R1_vs_SMA_SB25}
\end{figure}

In Fig. \ref{fig:R1_vs_SMA_SB25} we compare the R1 semi-major axis available in HECATEv2 with the g-band 25 mag arcsec$^{-2}$ photometric radius calculated for the galaxies available in the SGA-2020 catalogue. As shown, the HECATE sizes (R1) are in a reasonable agreement with the SGA-2020 sizes (SMA\_SB25\_G). The small systematic difference (less than $\approx$20\%) between the two measurements can be attributed to the higher S/N of the deeper optical images used in SGA-2020. This finding is also consistent with a similar comparison presented in \cite{moustakas23}. Although we could adopt the size measurements from SGA-2020  (for the region of the sky covered by the LS/DR9), given that they are based on better-quality optical images, the imposed size cut (D$_{25}>$20\arcsec{}) excludes a $\sim 65\%$ of HECATE galaxies. For that reason, and also to maintain uniformity over the full range of galaxy sizes, we retained the size measurements from HECATEv1 in HECATEv2, as their good agreement with the SGA-2020 results confirms their robustness. 

\subsection{Sizes of interacting galaxies and galaxy mergers}\label{subsec:sizes_of_interacting}
Measuring the size of galaxies that are members of interacting systems or galaxy mergers is an extremely challenging task. In these systems the regular structure of galaxies is distrupted, producing tidal tails, bridges, and asymmetries. This results in non-elliptical or fragmented isophotes, making it difficult to define a consistent diameter. In addition, in closely interacting systems the blending of light profiles complicates the separation of individual galaxies. Finally, projection effects can also affect the surface-brightness profiles when one galaxy partially overlaps another. Since there is no reliable systematic methodology to define the galaxy sizes in such systems, in HECATEv2, we identified and flagged these galaxies.

To identify as many problematic cases as possible, we performed a self-match of HECATEv2 using the Sky Ellipses algorithm in TOPCAT. For the first instance of HECATEv2 we used the galaxy size ellipse parameters (R1, R2, and PA), while for the second instance, we used a 3\arcsec{} search radius in order to obtain a list of galaxies with centers located within the D$_{25}$ eliipses of neighboring galaxies. This resulted in 1\,635 groups of potentialy problematic galaxy sizes in the HECATEv2 catalogue. Approximately 90\% of these groups contain two galaxy members, while only 1 group contain more than 50 galaxies which are within the Sagittarius Dwarf Spherodial. For all these groups, we visually inspected their optical images in the g band from the PS1-DR2 syrvey. Since most of these problematic cases are expected to have unreliable  galaxy sizes in HECATEv2, we overplotted their assigned D$_{25}$ ellipses on their optical images. For comparison we also plotted their corresponding sizes from the Siena catalogue (SMA\_SB25\_G). Based on our visual inspection we identified and flagged several cases where strong blending effects were present, galaxy sizes were overestimated, or even the D$_{25}$ was totally wrong. Furthermore, for a few cases we adopted the SMA\_SB25\_G from SIENA catalogue as it better represents the galaxy shape. Finally, our visual inspection revealed a few cases where one or more HECATE galaxies were duplicates of another galaxy (i.e. the same galaxy assigned with two PGC names). Although we did not remove these duplicate entries, we flagged them accordingly.

\section{Derived Physical Properties}\label{sec:physical_properties}
The following paragraphs describe the methods we used to derive the physical properties of the galaxies in the HECATEv2. By combining optical, mid-IR photometric and spectroscopic data we provide reliable estimations of the star-formation rate (\sfr{}), stellar mass (\stellarmass{}), gas-phase metallicity and nuclear activity classification for a much larger set of galaxies in comparison to HECATEv1. 

\subsection{Stellar Masses and star-formation rates in HECATEv2}\label{subsec:SFR_Mstar_S18_K23}
The \stellarmass{} and \sfr{} values provided in HECATEv1 were derived using far-infrared photometry from 2MASS and mid-infrared forced photometry (WF; see Sect.~\ref{subsec:midir_phot_surveys_AllWISE}) combined with the infrared emission calibrations of \citet{bell03} and \citet{kennicutt12}, respectively. However, the low signal-to-noise ratio of the 2MASS data, along with the limited availability of the WF photometry, which is restricted to galaxies within the SDSS footprint, resulted in \sfr{} and \stellarmass{} coverage for only 46\% and 65\% of the HECATEv1 galaxy sample, respectively. Moreover, a significant limitation of these IR-based monochromatic calibrations is that they do not fully account for biases introduced by stellar populations of different age or the effects of dust extinction within the galaxies. 

To increase the coverage and provide more reliable estimates of \sfr{} and \stellarmass{} in HECATEv2, we employ multiple methods for deriving the physical properties of the galaxies. In the following paragraphs, we describe these methods in detail, along with the homogenisation procedure adopted to bring the various \sfr{} and \stellarmass{} calibrations onto a consistent reference scale.

\subsubsection{SED-based \sfr{} and \stellarmass{}}\label{subsec:SFR_Mstar_S18}
Spectral energy distribution (SED) fitting is the most robust approach in estimating \sfr{} and \stellarmass{}, since it simultaneously models the galaxy emission across multiple wavelengths from UV to far-infrared, accounting for both the stellar-population age and extinction. Based on this method \citet[][hereafter S18]{salim16,salim18} compiled the GALEX-SDSS-WISE Legacy Catalogue 2 (GWSLC-2), which provides robust \sfr{} and \stellarmass{} estimates for $\sim 230\,000$ galaxies within the SDSS footprint using UV, optical, and IR data. To include the most-reliable \sfr{} and \stellarmass{} measurements available, we cross-matched our catalogue with the GSWLC-2 using the SDSS objID, obtaining SED-based measuremns of \sfr{} and \stellarmass{} for 75\,672 galaxies. 

However, the SED-based \sfr{} and \stellarmass{} require extended photometry across multiple bands from UV to IR, which is not available for the entirety of large galaxy catalogues such as HECATE. In practice, most SED-based estimates are limited to galaxies within the SDSS footprint, with only small subsets covered elsewhere on the sky. To increase the coverage of the \sfr{} and \stellarmass{} in HECATEv2 we leveraged the sensitivity and the all-sky coverage of the AllWISE survey.

\subsubsection{Mid-infrared-based \sfr{} \& \stellarmass{} calibrations}\label{subsec:SFR_Mstar_K23}
A limitation of most monochromatic or multi-band \sfr{} calibrations is that they do not account for the contribution of older stellar populations in the dust heating, or for the unobscured  star-forming activity when UV data are not available.  Similarly, most \stellarmass{} calibrations using WISE bands do not account for the effect of stellar population ages. 

To mitigate this, \cite{kouroumpatzakis23} (hereafter K23) developed updated mid-IR calibrators based on a combination of WISE W1 and W3 bands for \sfr{} and W1 for \stellarmass{} along with optical (u-r) or (g-r) colors. Using SED fitting, K23 constructed a large grid of galaxy SEDs spanning a wide range of star formation conditions, dust content, and interstellar medium properties. By fitting the relationships between \stellarmass{} or \sfr{} with mid-IR and optical photometry from these model SEDs, K23 derived calibrations that account for the contribution of old stellar populations to dust emission in galaxies with low \sfr{} while also incorporating the effects of dust extinction. Using the color excess E(B-V) as an extinction metric they provide more reliable \stellarmass{} and \sfr{} estimates across a broad range of star-forming environments. In order to also account for cases where extinction measurements are unavailable, they also provide extinction-independent calibrations which implicitly account for the dust effect on average since they are calibrated based on the SED-based galaxy models including a wide range of extinction representing that present in the galaxy population. 

To calculate the \stellarmass{} and \sfr{} using the K23 calibrations, the first step is to determine the color excess $E(B-V)$. To that end, we computed $E(B-V)$ values for all the HECATEv2 galaxies with available positive \ion{H}{$\alpha$} and \ion{H}{$\beta$} emission line fluxes, using the conversion of \cite{dominguez13}: 
\begin{equation}
    E(B-V) = 1.97 \rm{log}\Big[\frac{(f_{H\alpha}/f_{H_{\beta}})}{2.86}\Big]
\end{equation}
and the extinction law of \cite{calzetti00}. To ensure that we only included galaxy spectra without problematic starlight subtraction we only use in this calculation spectra with $\frac{f_{H\alpha}}{f_{H_{\beta}}} > 2.86$. 

To caclulate the stellar mass, we used the mass-to-light calibration from K23, as presented in their Equation 6. This calibration uses the W1 mid-IR band, the optical (u-r) or (g-r) colours, and the $E(B-V)$ extinction term. In cases where extinction measurements are unavailable, we used the calibration involving only the W1 and (u-r) or (g-r) colours. The best-fit parameters for both the extinction-dependent and extinction-independent calibrations are provided in the Table 3 of K23.

In our analysis we included only galaxies with good-quality optical and mid-IR photometric data (i.e. photometric quality flags = 1; see Sect.\, \ref{subsec:optical_phot_homogenization}, \ref{subsec:midir_phot_surveys_AllWISE}) in the  u, g, r, and W1 bands that are required for the appplication of the K23 mass-to-light calibration.  More specifically, we only considered objects with S/N $>$5 in the optical bands (u, g, and r). In addition, we excluded objects that: i) have wrong photometric matches or wrong coordinates (Sect.\ref{subsec:wrong_phot_wrong_coord}), and ii) are characterised as secure stars (Sect.\ref{sec:star_contamination}). To prevent extrapolation beyond the parameter space covered by the galaxy-SED grid of K23, we included only galaxies with optical colors satisfying u-r $>$ -0.6 and g-r $>$ -0.25. 

After applying these quality and S/N cuts, we used the extinction-dependent calibration of K23 for all galaxies with E(B-V) measurements with S/N$\geq5$. Wherever available, we adopted the (u-r) colour as the color term. When u-band photometry was unavailable we used the (g-r) colour instead. For objects lacking SDSS photometry, we used the (g-r) colour from PS1-DR2. Finally for galaxies without available colours from either SDSS-DR17 or PS1-DR2, we adopted the median (u-r) SDSS-DR17 colour. For galaxies with  E(B-V) S/N $<5$ or without extinction measurements, we applied the extinction-independent mass-to-light calibration of K23 following the same prioritization scheme for selecting the colour term.

For the calculation of the SFR, we utilise the calibrations from K23 as presented in their Equation 4. This relation uses the W1 and W3 mid-IR bands, as well as the E(B-V) extinction measurement. The best-fit parameters for both the extinction-dependent and extinction-independent calibrations are listed in Table 1 of K23. Applying the same quality and selection criteria as those used for the \stellarmass{} calculation, we excluded from our analysis all objects with problematic photometric measurements and/or star contamination, and proceeded only with galaxies that have good-quality W1 and W3 data. The \sfr{}s were then derived using both the extinction-dependent and the extinction-independent forms of the calibration from K23, based on the same $E(B-V)$ S/N criteria adopted for the stellar mass estimates. 

\begin{figure*}
 \centering
        \includegraphics[width=0.80\textwidth]{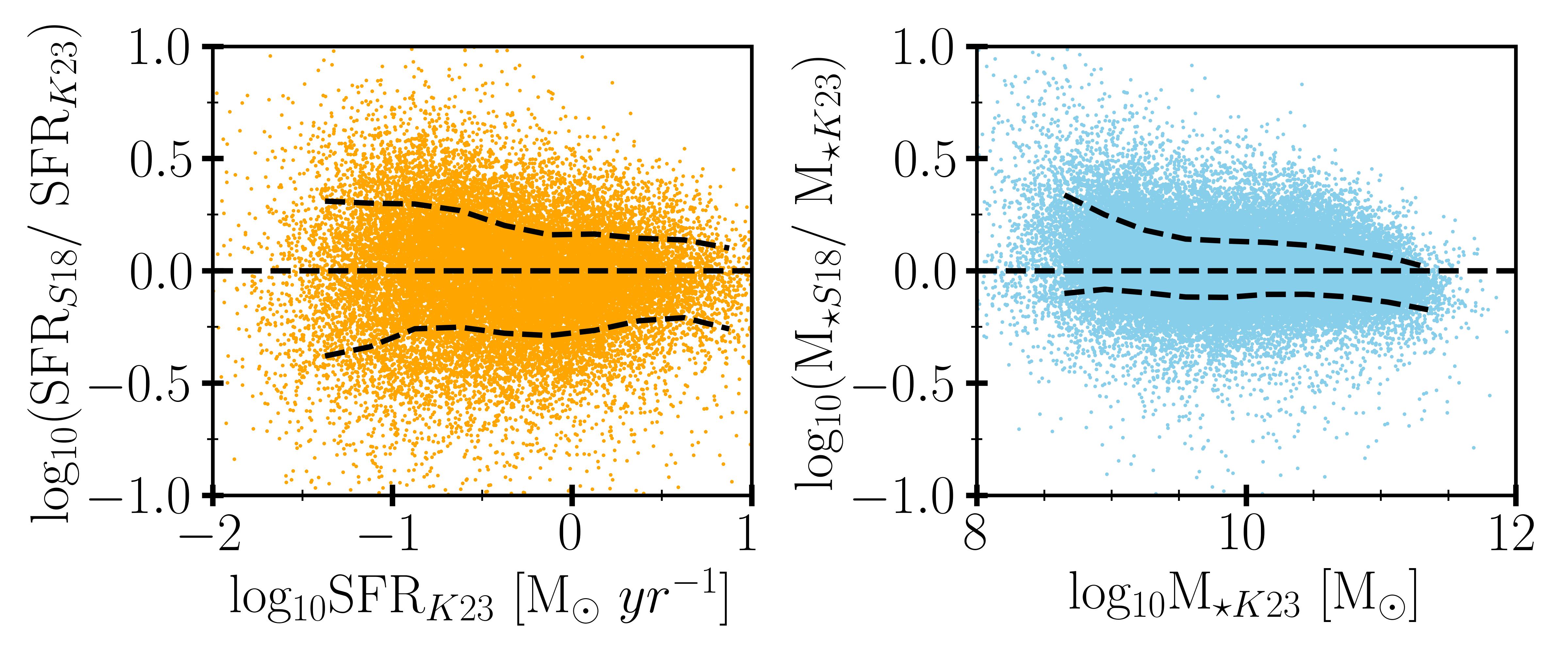}
        \includegraphics[width=0.80\textwidth]{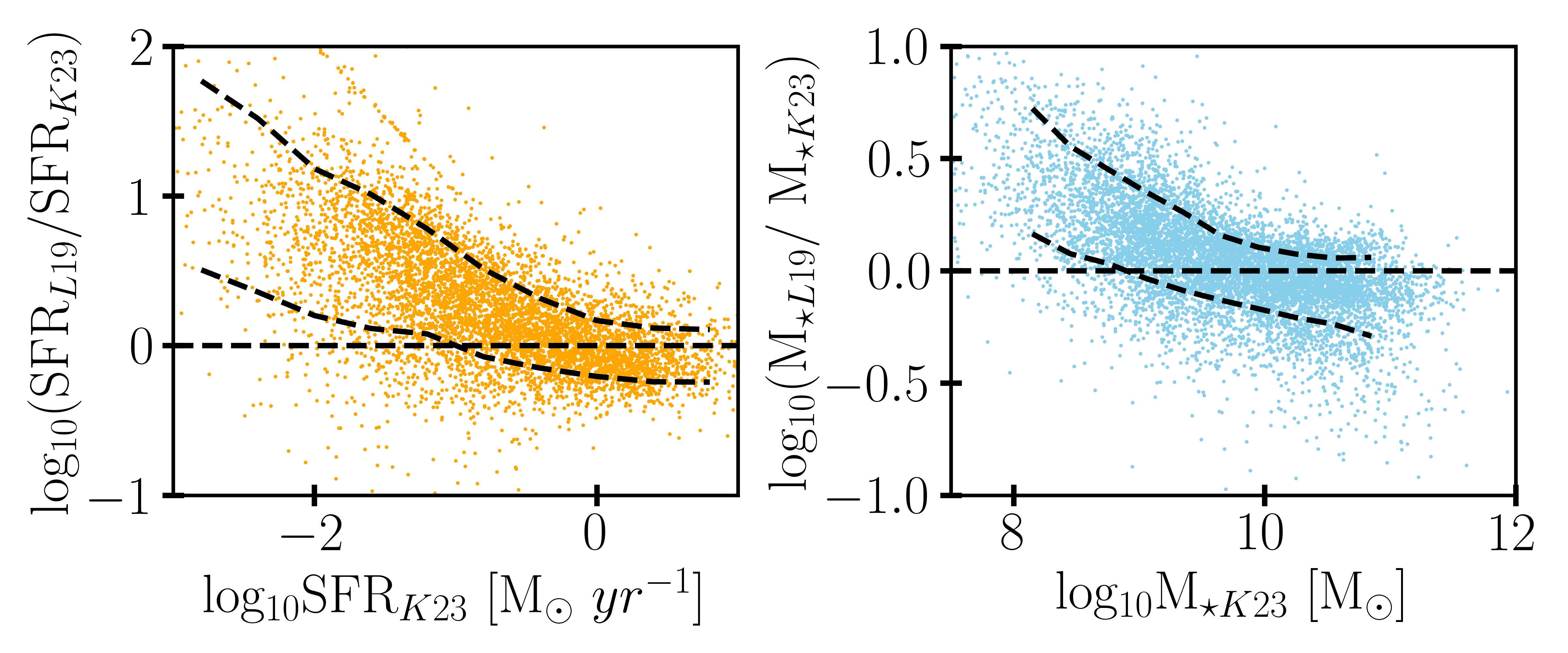}
    \caption{Top panel: Comparison between the rescaled \sfr{} (left panel) and \stellarmass{} (right panel) estimates from S18 with respect to the reference K23 estimates. The y-axis shows the logarithmic ratio of the values (i.e. $\textrm{log}_{10}\big(\frac{SFR_{S18}}{SFR_{K23}}\big)$ and $\textrm{log}_{10}\big(\frac{M_{\star S18}}{M_{\star K23}}\big)$ ), with the black dashed horizontal line at zero
indicating equality between the two methods.  The black dashed lines indicate the 16\% and 84\% percentiles of the distributions. As shown the homogenised \sfr{} and \stellarmass{} estimates are in a good agreement.\\
Bottom panel: Comparison between the original \sfr{} (left panel) and \stellarmass{} (right panel) estimates from L19 with respect to the reference K23 estimates. The difference in the lower \sfr{} and \stellarmass{} regimes are primarly driven by aperture effects. }
    \label{fig:SFR_M_K23_S18_L19_comp}
\end{figure*}

\subsubsection{\sfr{} and \stellarmass{} for nearby extended galaxies}\label{subsec:SFR_Mstar_L19}
Measuring the \sfr{} and \stellarmass{} for the most nearby galaxies becomes more challenging requiring special care. Due to their proximity, these galaxies appear very large on the sky, recuiring uniform multiband photometric measurements obtained with apertures that properly account for their extended emission. As a result, most of the published source catalogues (i.e. GALEX, AllWISE) do not provide reliable flux measurements for the closest more extended galaxies. 

For this reason, \cite{leroy19} (hereafter L19) developed the "z = 0 Multiwavelength Galaxy Synthesis" (z0MOGS) survey, an atlas of UV and IR images for $\sim 15\,750$ nearby galaxies (D$\le$50 \distanceunits{}) covered by GALEX and WISE surveys. Using these images, they defined customised appertures that account for the extended emission of these galaxies and measured integrated fluxes in the UV and IR bands to derive robust \sfr{} and \stellarmass{}. Following various calibration methods, they rescaled their \sfr{} and \stellarmass{} estimations using the GSWLC data as a reference sample, enabling a direct comparison between local galaxies and the full galaxy population observed by SDSS. 

To include accurate \sfr{} and \stellarmass{} measurements for the nearby galaxies in our catalogue, we cross-matched the HECATEv2 with the Table 4 of L19, using the PGC identifier. This resulted in 15\,685 HECATEv2 galaxies with available \sfr{}, and \stellarmass{} in the L19 catalogue. We note that a population of very local dwarf galaxies ($\sim 11\,000$) included in HECATEv2 is missing from the L19 catalogue due to the L19 sample selection criteria. For these we estimated \sfr{} and \stellarmass{} using one of other methods as described in the previous section.  

\subsubsection{Adoption of the final homogenised \sfr{} and \stellarmass{} in HECATEv2}\label{subsec:homo_SFR_Mstar}
One of the main goals for the compilation of HECATEv2 is to increase the catalogue's completeness by incorporating updated and robust estimates of the \sfr{} and \stellarmass{}. For that reason, in the HECATEv2 we included SED-based measurements from S18, as well as mid-IR/optical-based \sfr{}s and \stellarmass{} using the calibrations from K23 (see Sect.~\ref{subsec:SFR_Mstar_S18}, and Sect.\ref{subsec:SFR_Mstar_K23}). In addition, we also used \sfr{} and \stellarmass{} from L19 to account for the extended emission from the most nearby galaxies (see Sect.~\ref{subsec:SFR_Mstar_L19}). 

However, different calibrations introduce systematic offsets \citep[e.g.;][]{kennicutt12,kouroumpatzakis23}. Therefore, homogenising all available \sfr{} and \stellarmass{} measurements both for differences in the assumed distances and the used calibrations is essential before adopting the final values. To achieve this, we rescaled all the estimations from S18 using the K23 method as the reference standard, following the methodology described in the Appendix \ref{append:SFR_Mstar_rescales}. As we discussed in Sect.\ref{subsec:SFR_Mstar_S18} although SED-based methods offer the most robust estimates of \sfr{} and \stellarmass{}, they require extensive multiwavelength photometric coverage, which is not available for large samples of galaxies, especially in larger distances. In contrast, K23 demonstrated that their mid-IR/optical calibrations are consistent with the SED-based estimates from S18. For this reason, by anchoring all the other methods used in HECATEv2 to the K23 calibrations facilitates future extensions of the catalogue to higher redshifts, where optical and mid-IR sky surveys are easily available (e.g. in comparison to UV photometry), while maintaining an accuracy comparable to that of SED-based methods.

After rescaling the \sfr{} and \stellarmass{} from S18 using the scaling factors calculated in Appendix \ref{append:SFR_Mstar_rescales}, we compared them with the reference estimates of K23. The results of this comparison are shown in the top panel of Fig. \ref{fig:SFR_M_K23_S18_L19_comp}. As shown, the rescaled \sfr{}s are in very good agreement, with a scatter less than 0.25 dex, in the higher \sfr{} regime (i.e $\sim$ \sfr{}$>$ 1 M$_{\odot}\,yr^{-1}$), and less than 0.3 dex at lower \sfr{}s. This slight increase of the scatter at low \sfr{}s is expected, as galaxies with lower star-formation activity tend to be photometrically fainter, making it more challenging to obtain reliable measurements. For the \stellarmass{} we also find very good agreement between the two calibrations, with a constant scatter of $\sim$ 0.25 dex across the entire \stellarmass{} range. 

In the bottom pannel of Fig. \ref{fig:SFR_M_K23_S18_L19_comp} we present the comparison between the \sfr{} and \stellarmass{} estimates from L19 with the corresponding estimates from K23. As shown, for \sfr{}$>$ 0.5 M$_{\odot}\,yr^{-1}$, the two calibrations are broadly consistent. At lower \sfr{}s, however, the scatter increases and a systematic offset between the two methods emerges, reaching $\sim$1 dex at the lowest \sfr{} values. This difference is primarily driven by aperture effects, as these objects are mostly very nearby galaxies for which the customised photometric analysis of L19 provides more accurate measurements. The larger apparent sizes captured by the tailored approach of L19 account for the extended emission of these galaxies, leading to higher \sfr{} estimates and, consequently, systematically higher SFR and \stellarmass{} values compared to those from K23. As a result, since this difference arises from variations in the different apertures used in the two methods rather than discrepancies in the underlying calibration approaches, we opted to retain the more reliable L19 estimates without applying any rescaling apart from distance differences. 

\begin{figure}
        \includegraphics[width=0.47\textwidth]{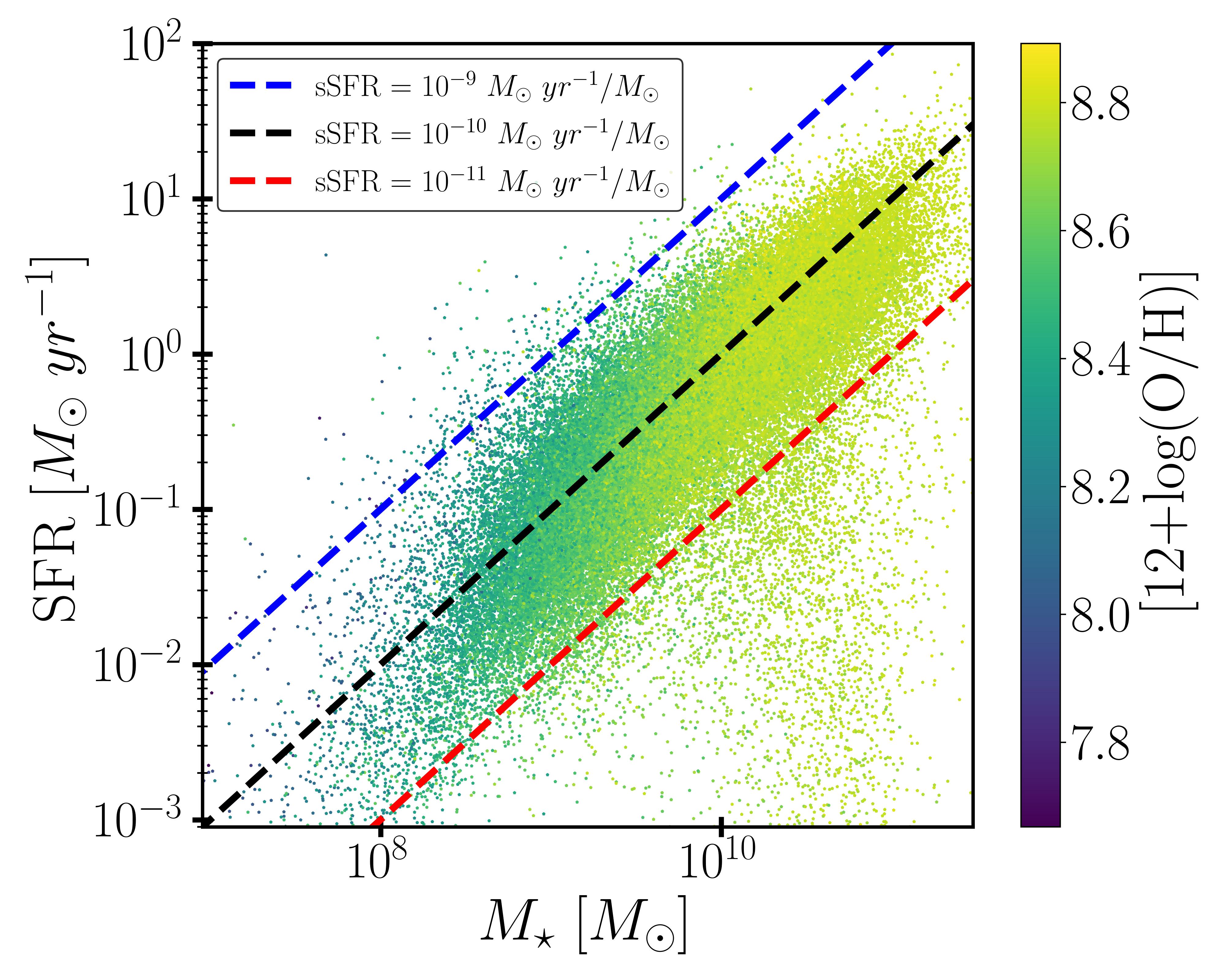}
    \caption{ The distribution of HECATEv2 galaxies in the \sfr{}-\stellarmass{} plane, color-coded with their gas-phase metallicity. The diagonal dashed lines indicate three different sSFR values (i.e. sSFR: $10^{-9}, 10^{-10},$ and $10^{-11}$ $\rm{M_{\odot}}\,\rm{yr^{-1}}/\rm{M_{\odot}}$). We exlude from this galaxy sample the AGNs since their \sfr{} and metallicity estimates are not robust. }
    \label{fig:HECv2_MS}
\end{figure}

After following the homogenisation procedure described above, we adopted the final \sfr{} and \stellarmass{} estimates in HECATEv2 based on the following prioritisation scheme. First, we selected the estimates from L19 where available, due to their more accurate, customised photometric analysis. In these cases, we also retained the original \sfr{} and \stellarmass{} quality flags as defined in L19. For galaxies not included in L19, or where the measurements were flagged as unreliable, we adopted the SED-based \sfr{} and \stellarmass{} values from GSWLC-2 when available. Finally, for the remaining galaxies, we used the estimates derived from the extinction-dependent and extinction-independent calibrations of K23, following the methodology described in Sect.~\ref{subsec:SFR_Mstar_K23}. In this way, HECATEv2 provides \sfr{} and \stellarmass{} estimates for 143\,439 and 183\,917 galaxies, corresponding to coverage of 70$\%$ for \sfr{} and 89$\%$ for \stellarmass{}, respectively. In Table \ref{tab:SFR_Mstar_statistics} we present the number statistics of the \sfr{} and \stellarmass{} in the HECATEv2, grouped by the method used. In Fig. \ref{fig:HECv2_MS}, we also present the distribution of the HECATEv2 galaxies in the \sfr{}-\stellarmass{} plane, color-coded with their gas-phase metallicity. The diagonal dashed lines indicate three different sSFR values (i.e. sSFR: $10^{-9}, 10^{-10},$ and $10^{-11}$ $\rm{M_{\odot}}\,\rm{yr^{-1}}/\rm{M_{\odot}}$). As shown, the majority of the HECATEv2 galaxies follow well the Main-Sequence relation.

A small fraction of HECATEv2 galaxies exhibit very low values of \sfr{} and \stellarmass{}. This arises because \sfr{} estimates are provided for all galaxies regardless of their activity type based on the current data\footnote{Some of these classifications may change in future releases of our catalogue}. For example, most galaxies with extremely low \sfr{} (i.e. \sfr{}$<10^{-5}$ $\rm M_{\odot} \ yr^{-1}$) are passive systems with negligible star-forming activity. Additionaly, we suggest to the users of our catalogue to treat with caution the \sfr{} for AGN-host galaxies, as AGN emmision can significantly contribute in the photometric bands used for \sfr{} calibration. Similarly, 1\,254 galaxies show unusually low \stellarmass{} estimates (i. e. \stellarmass{}$<10^{6}$ M$_{\odot}$). For approximately 50\% of these, the optical photometry used in the stellar mass-to-light calculation either relies on a fiducial average value (see Sect. \ref{subsec:SFR_Mstar_K23}) or is flagged as potentially contaminated by stars (see Sect. \ref{sec:star_contamination}) leading to unreliable \stellarmass{}. Visual inspection of the remaining objects revealed that they are very low-surface-brightness dwarf galaxies with weak infrared emission which makes unreliable the mid-IR calibrators applied for their \stellarmass{} estimation. For these reasons, we flagged all the 1\,254 HECATEv2 galaxies with \stellarmass{}$< 10^{6}$ M$_{\odot}$ and we suggest the users to treat these values with caution. 

\begin{table}
\centering
\caption{Summary of the physical quantities provided in the HECATEv2 catalogue.}
\label{tab:SFR_Mstar_statistics}
\begin{tabular}{ccc}
\hline\hline
\multicolumn{3}{c}{\sfr{}} \\
Method & N$_{\textrm{tot}}$ & Reference\\
(1) & (2) & (3) \\
      & & \\
\hline      
L19 & 15\,420 & \cite{leroy19}\\
S18 & 72\,175 & \cite{salim16,salim18}\\
K23 & 55\,844 & \cite{kouroumpatzakis23}\\
\hline
Total & 143\,439 & Coverage = 70\%\\
\hline
\multicolumn{3}{c}{\stellarmass{}} \\
% \hline      
L19 & 15\,420 & \cite{leroy19}\\
S18 & 72\,175 & \cite{salim16,salim18}\\
K23 & 96\,322 & \cite{kouroumpatzakis23}\\
\hline
Total & 183\,917 & Coverage = 90\% \\
\hline
\multicolumn{3}{c}{Metallicity} \\
C20-O3N2 & 61\,155 & \cite{curti20} \\
C20-MZ & 104\,616 & \cite{curti20} \\
\hline
Total & 165\,771 & Coverage = 81\%\\

\hline

\end{tabular}
\medskip
\parbox{\columnwidth}{\footnotesize
Note: Columns (1)–(2) indicate the method used for deriving each physical quantity and the corresponding number of galaxies. Column (3) provides the reference for each methodology.
}
\end{table}

\subsection{Metallicity estimates}\label{subsec:metal_estimates}

The gas-phase metallicity of a galaxy, provided in terms of oxygen abundance $[\log_{10}(\textrm{O/H}) + 12]$, can be measured using various calibrations, including theoretical methods based on photoionisation models \citep[e.g.][]{kewley02}, and empirical methods that rely on electron temperature measurements derived from metallicity-sensitive optical emission lines \citet[henceforth PP04]{pettini04}. Although comparison among different metallicity calibrations revealed large discrepancies, \cite{kewley_elison08} showed that the empirical calibration of PP04 is among the most robust metallicity calibrators. Based on the $\log_{10}$([\ion{O}{III} $\rm{\lambda}5007$]/H$\beta$ / [\ion{N}{II} $\rm{\lambda}6584$]/H$\alpha$) ratio, the PP04-O3N2 calibrator has the lowest scatter compared to other methods, and it is less sensitive to extinction effects than other metallicity calibrators. However, since the PPO4-O3N2 method was derived from a relatively small and unrepresentative galaxy sample, it is only applicable to sources with O3N2 $>2$, thus excluding a significant fraction of low-metallicity galaxies. 

\begin{figure*}
 \centering
        \includegraphics[width=0.47\textwidth]{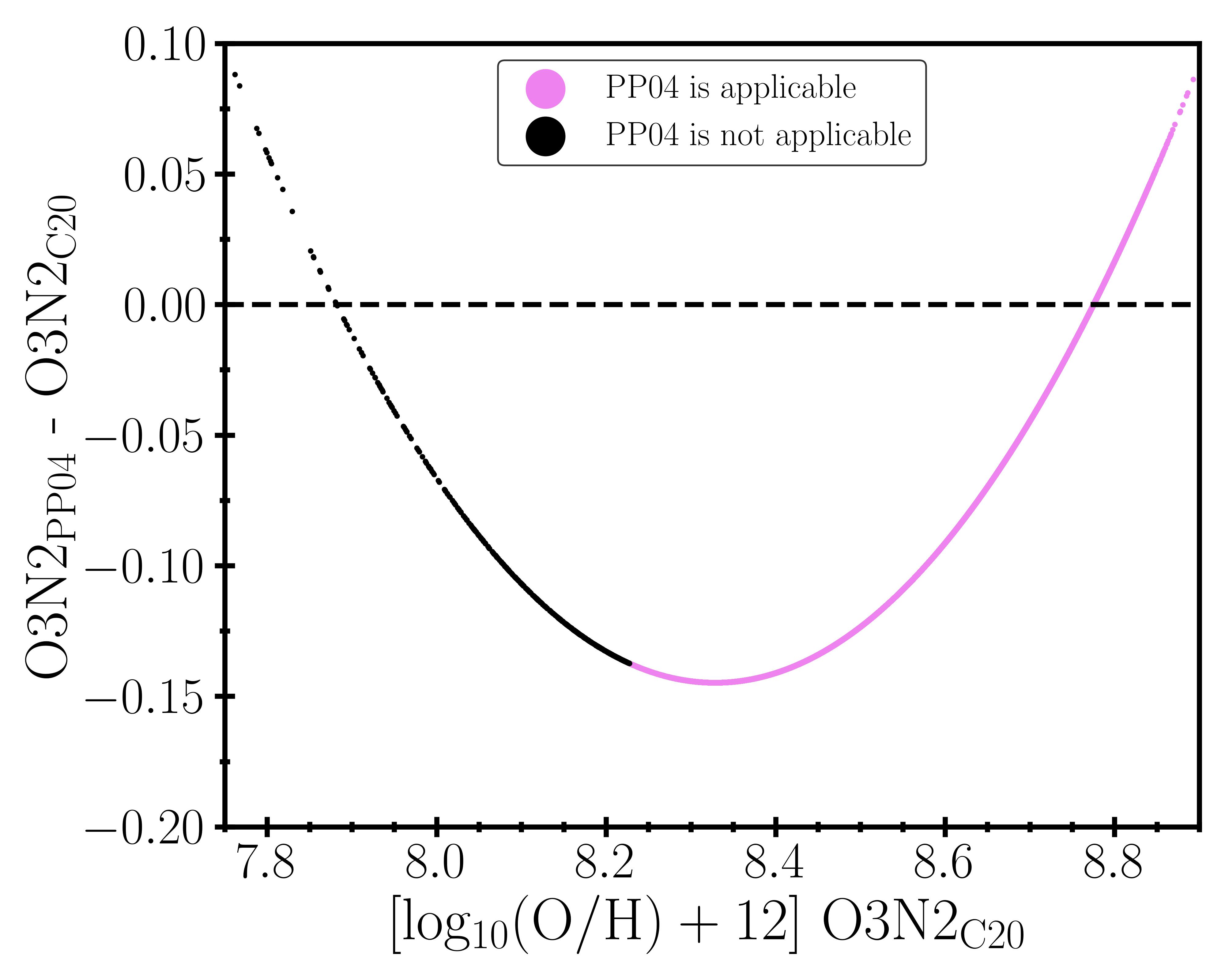}
        \includegraphics[width=0.47\textwidth]{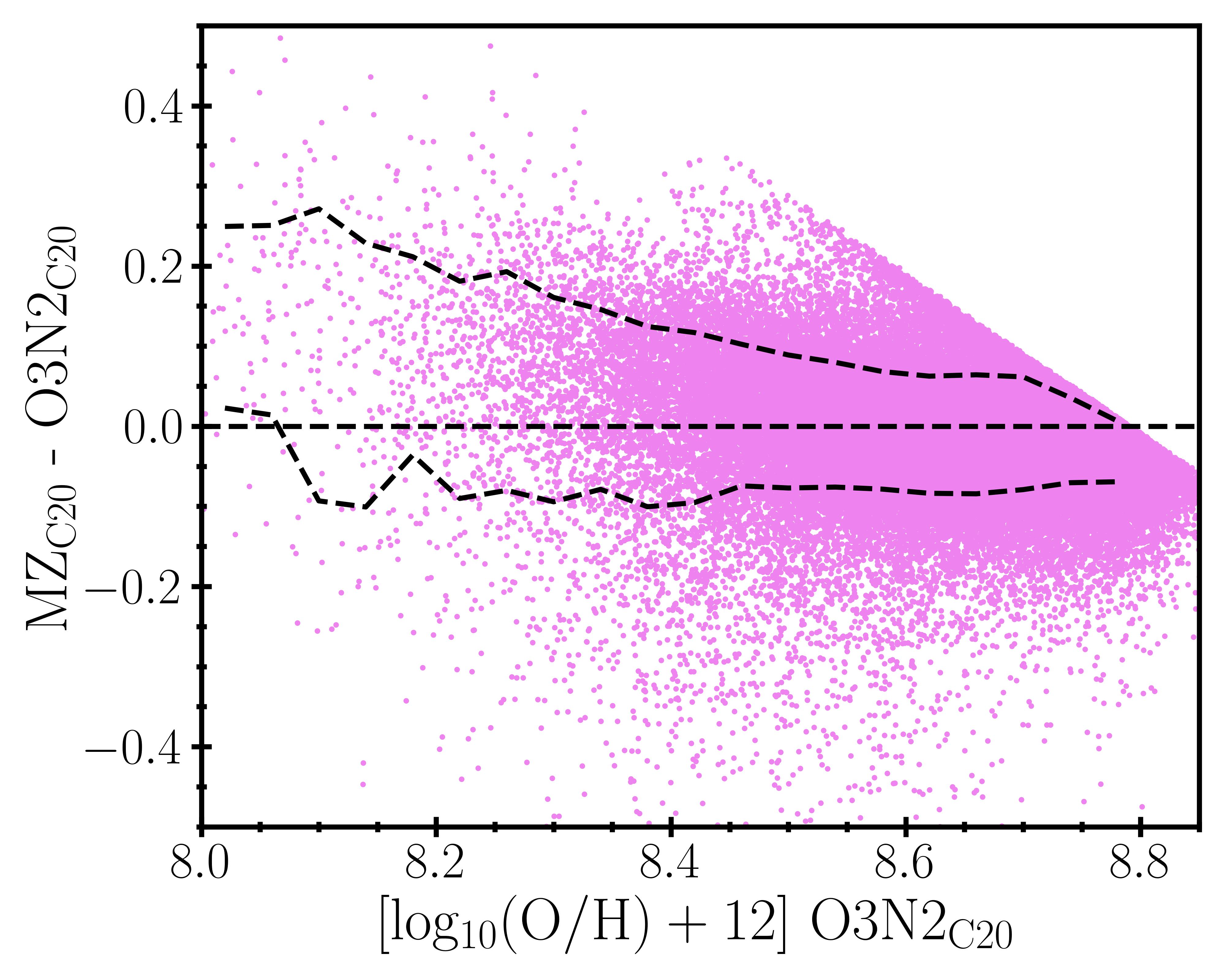} 
    \caption{Left panel: Difference between the O3N2 metallicity calibrator proposed by \citet[][PP04-O3N2]{kewley_elison08} and \citet[][C20-O3N2]{curti20} plotted as a function of the metallicity derived using the latter. Purple dots indicate galaxies with O3N2$>2$, for which the PP04-O3N2 calibrator is valid. The dashed black line represents the equality between the two methods. As shown, the two calibrations differ by up to $\sim$ 0.15 dex. Unlike PP04-O3N2, the C20-O3N2 calibration is applicable across the full metallicity range, including the low-metallicity regime. Therefore, C20-O3N2 was adopted for HECATEv2. 
    Right panel: Difference between the metallicity derived from the mass-metallicity relation from \citet[][C20-MZ]{curti20} and O3N2 spectroscopic metallicity calibrator proposed by \citet[][C20-O3N2]{curti20} as a function of the metallicity derived using the latter. The horizontal dashed black line represents the equality between the two methods while the remaining lines correspond to the 16th and 84th percentiles, respectively. The two methods are consistent across the entire metallicity range.}
    \label{fig:Metal_comparison}
\end{figure*}

To overcome this limitation, in the HECATEv2 we adopt the prescription from \cite{curti20} (C20) who recalibrated the O3N2 metallicity diagnostic using electron temperature-based metallicity measurements from both individual galaxies and stacked spectra, drawn from the much larger SDSS-DR7 sample \citep{Abazajian09}. This updated calibration (hereafter C20-O3N2) removes the O3N2 $>2$ constraint, allowing its application to galaxies with lower metallicities. In the left panel of Fig.\ref{fig:Metal_comparison} we present the comparison between the PPO4-O3N2 and the C20-O3N2 metallicity calibrations. Galaxies with O3N2 $>2$, for which the PP04-O3N2 method is not applicable, are indicated with black dots. As shown, the two calibrations differ by up to $\sim0.15$~dex, with PP04-O3N2 systematically yielding lower metallicities. Furthermore, while the C20-O3N2 calibration remains valid across the entire metallicity range, the PP04-O3N2 cannot be used to the lower-metallicity end. As a result, for all HECATE galaxies with available emission-line measurements and S/N$\geq 3$ we calculate the gas-phase metallicity using the best-fit parameters of the C20-O3N2 calibrator (as listed in Table 2 of \citealt{curti20}).

Althought the C20-O3N2 is one of the most robust metallicity calibrators, its application is limited only to galaxies with available spectroscopic data. To calculate metallicities for the HECATE galaxies lacking spectroscopic information, we utilised the revisited mass-metallicity relation from C20 (hereafter C20-MZ). Using their Equation 2 and the corresponding best-fit parameters, we calculated the metallicity for all HECATE galaxies with available \stellarmass{} but lacking spectra or with emission-line measurements having S/N$< 3$. In the right panel of Fig. \ref{fig:Metal_comparison} we compare the C20-O3N2 and the C20-MZ metallicity calibrations that we adopted in the HECATEv2. As shown, the two methods are consistent across the entire metallicity range, and no homogenisation is needed. The diagonal sharp feature visible in Fig. \ref{fig:Metal_comparison} is artificial and arises because the C20-MZ relation is calibrated within the range $[\log_{10}(\textrm{O/H}) + 12] \in [7.6, 8.9]$. By combining the spectroscopic metallicity calibrator C20-O3N2 and the mass-metallicity relation from C20 (Table \ref{tab:SFR_Mstar_statistics}), HECATEv2 provides gas-phase metallicity measurements for $\sim 80\%$ of the catalogue, increasing its coverage by a factor of $\sim$2.5. 

\subsection{Activity classification}\label{subsec:activity_class}
One of the main objectives for the compilation of the HECATE catalogue was to establish a reference catalogue for studying large samples of galaxies hosting AGN as well as normal galaxies (i.e. those without AGN activity). These samples combined with the value-added information provided by the HECATEv2, will enable studies into the connection between different galaxy classes and the stellar populations they host. For that reason, is essential to provide reliable nuclear activity classifications for the largest possible fraction of the HECATE galaxies. In HECATEv1, the activity classification was based on spectroscopic data, which limited its coverage to the SDSS footprint, providing classifications for $\sim 30\%$ of the entire catalogue. To expand this coverage, HECATEv2 leverages the larger sky coverage of PS1-DR2 and the all-sky coverage of AllWISE surveys, incorporating new mid-infrared/optical photometric diagnostics along with the spectroscopic data from the SDSS survey. The following sections outline the steps taken to obtain activity classifications in HECATEv2. 

\subsubsection{Spectroscopic activity classification}\label{subsec:spectroscopic_activity_class}
Using the emission-line data described in Sect. \ref{subsec:optical_spec_data}, we classify the HECATEv2 galaxies based on the emission-line ratio diagnostic of \cite{stampoulis19}. This is an extension of the generally used Baldwin-Phillips-Terlevich \citep[BPT;][]{baldwin81} diagnostics of \cite{kewley01}, \cite{kauffmann03}, and \cite{schawinski07} which however, incorporates simultaneously all available line ratios. This way it maximises the use of spectral information while mitigating contradictory classifications that arise from the traditional two-diamensional line ratio BPT diagrams. The classification approach employs a soft clustering scheme (soft data driven allocation method; SoDDA) by fitting multivariate Gaussian distributions to the four-dimensional emission-line ratio space: $\log_{10}$([\ion{N}{II} $\rm{\lambda}6584$]/H$\alpha$), $\log_{10}$([\ion{S}{II} $\rm{\lambda\lambda}6717,6731$]/H$\alpha$), $\log_{10}$([\ion{O}{I} $\rm{\lambda}6300$]/H$\alpha$), and $\log_{10}$([\ion{O}{III} $\rm{\lambda}5007$]/H$\beta$). This method classifies the galaxies into four distinct activity classes: star-forming, Seyfert, LINER, and composite. Moreover, due to the probabilistic nature of this diagnostic, we not only obtain the most probable activity classification for each galaxy but also derive its probability of belonging to each of the other activity classes, providing a more comprehensive characterization of nuclear activity. In addition, for the cases where the generally weak [\ion{O}{I} $\rm{\lambda}6300$] line is not available, \cite{stampoulis19} provide an alternative diagnostic based on the Support Vector Machine (SVM) algorithm \citep{cortes95}. This method uses multidimensional decision boundaries to classify the galaxies based on their distribution within a three-dimensional emission-line ratio space: $\log_{10}$([\ion{N}{II} $\rm{\lambda}6584$]/H$\alpha$), $\log_{10}$([\ion{S}{II} $\rm{\lambda\lambda}6717,6731$]/H$\alpha$), and $\log_{10}$([\ion{O}{III} $\rm{\lambda}5007$]/H$\beta$). 

While HECATEv2 adopts the same diagnostic framework as HECATEv1 for spectroscopic activity classification, we refine the analysis by incorporating corrections to the emission-line measurements and imposing more stringent S/N criteria on the emission-line fluxes, ensuring a more robust and reliable classification. To that end, before applying the activity diagnostic of \cite{stampoulis19} we applied the corrections of \cite{juneau14} to the emission line uncertainties reported in the MPE-JHU DR8 catalogue. Additionally, we also corrected for the stellar continuum the flux of the H$\beta$ emission line following the recommendations of \cite{groves12}.  After applying these corrections, we selected only objects with H$\alpha$/H$\beta > 2.86$ to exclude cases affected by problematic starlight subtraction and errors in the emmision-line measurements. Finally for galaxies with S/N $>3$ in all characteristic emission lines, we apply the four-diamensional SoDDA diagnostic. In cases where all lines met the S/N $>3$ threshold except for [\ion{O}{I} $\rm{\lambda}6300$], we applied the three-dimensional SVM diagnostic. In total, we spectroscopically classified 53\,352 galaxies.Among them, 39\,247 were classified using the SoDDA method, while the remaining 14\,105 were classified using the SVM3d method

\subsubsection{Photometric activity classification}\label{subsec:phot_activity_class}

To obtain activity classifications for a larger fraction of HECATE galaxies, HECATE v2.0 provides activity classifications based on the mid-IR/optical photometric diagnostics developed by \cite{daoutis23}. This method is based on a Random Forest (RF) machine-learning classifier \citep{louppe14} that utilizes mid-IR colors (W1-W2 and W2-W3 obtained from the AllWISE survey), and the optical g-r color (obtained from the SDSS-DR17 survey) to distinguish galaxies into five activity classes: star-forming, AGN, LINER, composite, and passive. The classifier is trained on a sample of galaxies with high-quality spectra from the MPA–JHU DR8 catalogue, where emission-line objects are characterised using the \cite{stampoulis19} diagnostics, while passive galaxies are identified from MPA–JHU DR8 spectra which lack detectable emission-lines. To account for aperture effects, the classifier is trained on a hybrid mid-IR photometric scheme identical to the one provided in HECATEv2, based on AllWISE photometry (Sect. \ref{subsec:midir_phot_surveys_AllWISE}). This new photometric diagnostic offers a comprehensive and robust approach to increase the completeness of activity classifications in HECATEv2. Its applicability on a much larger volume of galaxies, its sensitivity in identifying passive galaxies, and its overall accuracy of $\sim 80\%$, make it the most suitable option for this task. Furthermore, the probabilistic nature of the RF algorithm provides the full probability distribution for the classification of each galaxy, similar to the SoDDA algorithm used for the activity classification (Sect. \ref{subsec:spectroscopic_activity_class}). 

Using the fiber optical photometries and the hybrid mid-IR photometric scheme described in Sect. \ref{sec:optical_phot_surveys} and Sect. \ref{subsec:midir_phot_surveys_AllWISE}, respectively, we selected all HECATEv2 galaxies that simultaneously satisfy the S/N thresholds of S/N $>5$ in W1 and W2 bands, S/N $>3$ in the W3 band, and S/N $>5$ in the g and r oprical bands. The lower S/N limit for the W3 band was adopted due to the lower sensitivity of data in this band, without however, significantly reducing the reliability of the classifications. In addition, we excluded from our analysis objects with bad quality flags (see Sect. \ref{subsec:midir_phot_surveys_AllWISE}, \ref{subsec:optical_phot_homogenization}) in any of the bands used in the diagnostic as well as objects with incorrect photometric matches (Sect. \ref{subsec:wrong_phot_wrong_coord}). We note that while HECATEv2 provides rescaled, aperture-corrected magnitudes, the photometric diagnostic was applied using the unrescaled, non-aperture-corrected magnitudes, as the classifier was trained on these magnitudes. Following these selection criteria, we applied the photometric diagnostic to a total of 84\,825 galaxies in the HECATEv2.

\subsubsection{Homogenized activity classifications}
The activity classification in the HECATEv2 is derived from two different diagnostics: one based on spectroscopic data and another on photometric information provided by the catalogue. To ensure a consistent and homogenized classification for the largest possible sample in the updated HECATE, it is essential to combine the results from both methods into a single classification column. However, given the different nature of these diagnostics, it is crucial to first compare their performance on a common set of galaxies before adopting them. 

\begin{figure}
        \includegraphics[width=\columnwidth]{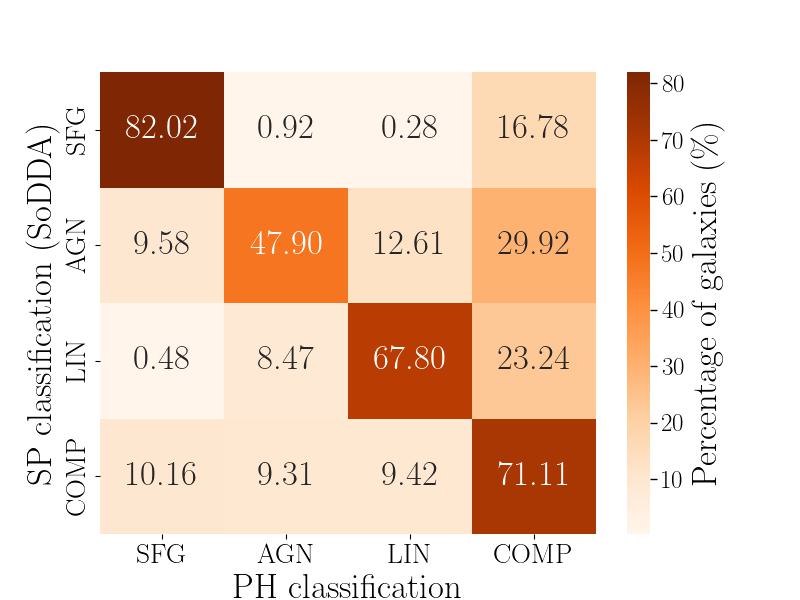
        }
    \caption{Confusion matrix comparing the two activity classification methods implemented in HECATEv2. The x-axis corresponds to classifications based on the photometric diagnostic of \citet{daoutis23} (PH classification), while the y-axis shows classifications derived from the spectroscopic diagnostic of \citet{stampoulis19} (SP classification). The two methods demonstrate good agreement, with an accuracy consistent with the level reported by \citet{daoutis23}.}
    \label{fig:SP_PH_comparison}
\end{figure}

To that end, in Fig. \ref{fig:SP_PH_comparison} we compare in the form of a confusion matrix the classifications obtained from the spectroscopic diagnostic (SoDDA) and the mid-IR/optical photometric method (PH). We selected only galaxies with S/N$>5$ in their emmision lines, consistent with those that were used in the training of the photometric classifier. As shown in Fig. \ref{fig:SP_PH_comparison}, the two methods agree very well, with an accuracy level similar to that reported by \cite{daoutis23}. Specifically, the accuracy in case of star-forming galaxies is $\sim 82\%$ indicating an excellent agreement between the two methods. Although the classifier identify only $\sim48\%$ of the spectroscopic AGN as such, considering the Composite and LINER galaxies that can be considered as AGN candidates, the photometric method identifies  $\sim90\%$ of the spectroscopic AGN as AGN or AGN candidates. 

To further assess the performance of the photometric classifier, we repeat the comparison using galaxies with S/N$<3$ for the emission line fluxes, which is the threshold adopted for the application of the \cite{stampoulis19} diagnostics in the HECATEv2 galaxies. In this case, we find a slightly lower performance dropping to $\sim75\%$ for star-forming galaxies, $\sim44\%$ for AGN, $\sim71\%$ for LINER, and $\sim69\%$ for composite galaxies. This behavoir can be attributed to the fact that the photometric classifier was trained on spectroscopic data with higher S/N emmision-line measurements, making it less reliable for galaxies with noisy spectra. 

Based on these findings, we adopted the spectroscopic classification where available and supplemented the remaining galaxies with the photometric classification. This resulted in 53\,291 galaxies classified spectroscopically and 53\,064 galaxies classified photometrically. In this way, HECATEv2 provides robust activity classifications for a total of 106\,355 galaxies, increasing the catalogue's activity classification completeness from  $31\%$ to $52\%$. 

\section{Discussion}\label{sec:Discussion}

So far, we have presented the major updates implemented in HECATEv2, the second release of the HECATE catalogue. These include a new distance estimation framework based on a cosmological model for galaxies without redshift-independent distances, with special treatment for Virgo Cluster members. This approach ensures consistency and enables future extensions of the catalogue to higher redshifts. The photometric coverage has been significantly improved by incorporating and homogenizing optical and mid-IR data from SDSS-DR17/NSA, PS1-DR2, and AllWISE, accounting for aperture effects in nearby galaxies. Additional quality checks, including flags for stellar contamination, incorrect matches, and coordinate inconsistencies, enhance the reliability of the photometry. The completion of the coverage for the galaxy size , as well as the validation of all the galaxy sizes against measurements derived from other more detailed analysis ensures consistency and completeness of our size information provided in our catalogue. 

Using this homogenized photometric framework, we derived stellar population parameters for a large fraction of the sample. Updated mid-IR calibrations accounting for stellar population age and dust attenuation were applied to estimate \sfr{} and \stellarmass{} for over 70\% of galaxies, while gas-phase metallicities were computed for $
\sim$90\% of the catalogue using spectroscopic data and the mass–metallicity relation. Furthermore, improved activity classifications, based on combined photometric and spectroscopic diagnostics, are provided for more than 50\% of the sample. Together, these enhancements establish HECATEv2 as a robust value-added reference catalogue for local Universe studies. In the following sections, we evaluate the completeness of HECATEv2, compare it with other major galaxy catalogues, and present representative science applications.

\subsection{Completeness}\label{sec:completeness}
A robust assessment of HECATE’s completeness is challenging, primarily due to the absence of a well-defined selection function for HyperLEDA and the additional selection biases introduced by the other catalogues used for the compilation of HECATEv2. However, we can still approximate it by comparing the observed galaxy luminosities in various bands and distance bins to the expectations from the corresponding Luminosity Functions (LFs). For the calculation of the completeness, we adopted the methodology proposed by \cite{cook23} who integrated the LF over all luminosities rather than to a luminosity down to the $\sim 60\%$ of the "knee" of the LF \citep{gehrels16,dalya18}. This approach offers a more accurate and representative measure of the true galaxy census in the local Universe. Specifically, by dividing our catalogue in 20 equally-spaced distance bins, we computed the completeness of HECATEv2 in each bin using the relation:

\begin{equation}\label{eq:completeness_per_Dbin}
    \text{Completeness} = \frac{\sum_{i=1}^{N}L_{\rm{band}}^{i}}{p_{L}\times V_{\rm{D,bin}}}
\end{equation}

where the numerator represents the total luminosity of galaxies in the HECATEv2 catalogue within a given distance bin, and the denominator corresponds to the expected luminosity in the bin, calculated as the total luminosity density $p_{L}$ multiplied by the volume of the bin, $V_{\rm{D,bin}} = \frac{4}{3}\pi(D_{\rm{max}}^{3} - D_{\rm{min}}^{3})$, where D is the luminosity distance. The total luminosity density of the galaxies in the local Universe can be derived from their LF, $\Phi(L)$. This function is usually described by a Schechter form, which is characterised by three parameters: $\phi^{\ast}$, the normalization factor, L$^{\ast}$ the characteristic luminosity after which the number of galaxies rapidly drops, and the the faint-end power law slope, $\alpha$. By integrating the  $\Phi(L)$ over all luminosities, the total luminosity density can be expressed analytically as:

\begin{equation}
p_{L} = \int_{0}^{\infty}L\Phi(L)dL  = \phi^{\ast}L^{
\ast} \Gamma(\alpha + 2) \,\,\, [L_{\odot} \, Mpc^{-3}]
\end{equation}

In our analysis, we estimated the completeness in two photometric bands, the optical B-band and the near-infrared Ks - band. The former was chosen because it is the most populated optical band in our catalogue, with B magnitudes available for 183,204 galaxies ($\sim 90 \%$ of the HECATEv2 catalogue). Although the Ks-band includes fewer galaxies with measured photometry, it enables direct comparison with other published galaxy catalogues where it has been used as a completeness indicator. For the B-band and Ks-band LFs we adopted the parametrizations from \cite{gehrels16} and \cite{bell03}, respectively.\footnote{Schechter parameters for the B-band and Ks-band LFs : $\phi^{\ast}{=}(1.6\pm0.3){\times}10^{-2}\,h^{3} \rm{Mpc}^{-3}$, $a{=}({-}1.07\pm0.07)$, $L_\star{=}(1.2\pm0.1){\times} 10^{10}\,{\rm L_{B,\odot}}$ \citep{gehrels16} and $\phi^{\ast}{=}(0.0143\pm0.0007)\,h^{3} \rm{Mpc}^{-3}$, $a{=}({-}0.77\pm0.04)$, $L_\star{=}(4.2\pm0.2){\times} 10^{10}\,{\rm L_{Ks,\odot}}$ \citep{bell03}.} For the conversion of the magnitudes to luminosities we adopted the following solar absolute magnitudes: B = 5.44 (Vega) mag and Ks = 3.27 (Vega) mag \citep{willmer18}. To account for the uncertainties associated with the adopted LFs into our completeness estimates, we performed 1\,000 Monte Carlo realizations. For each realization, we randomly sampled values of the three LF parameters, namely $\phi^{\ast}$, $\alpha$, and L$^{\ast}$, from normal distributions centered on their best-fit values with standard deviations equal to their corresponding uncertainties and assuming independence. Completeness was then computed for each realization, and the final completeness in each distance bin was taken as the mean of the resulting distribution. The corresponding standard deviation was adopted as the uncertainty. 

\begin{figure}
        \includegraphics[width=\columnwidth]{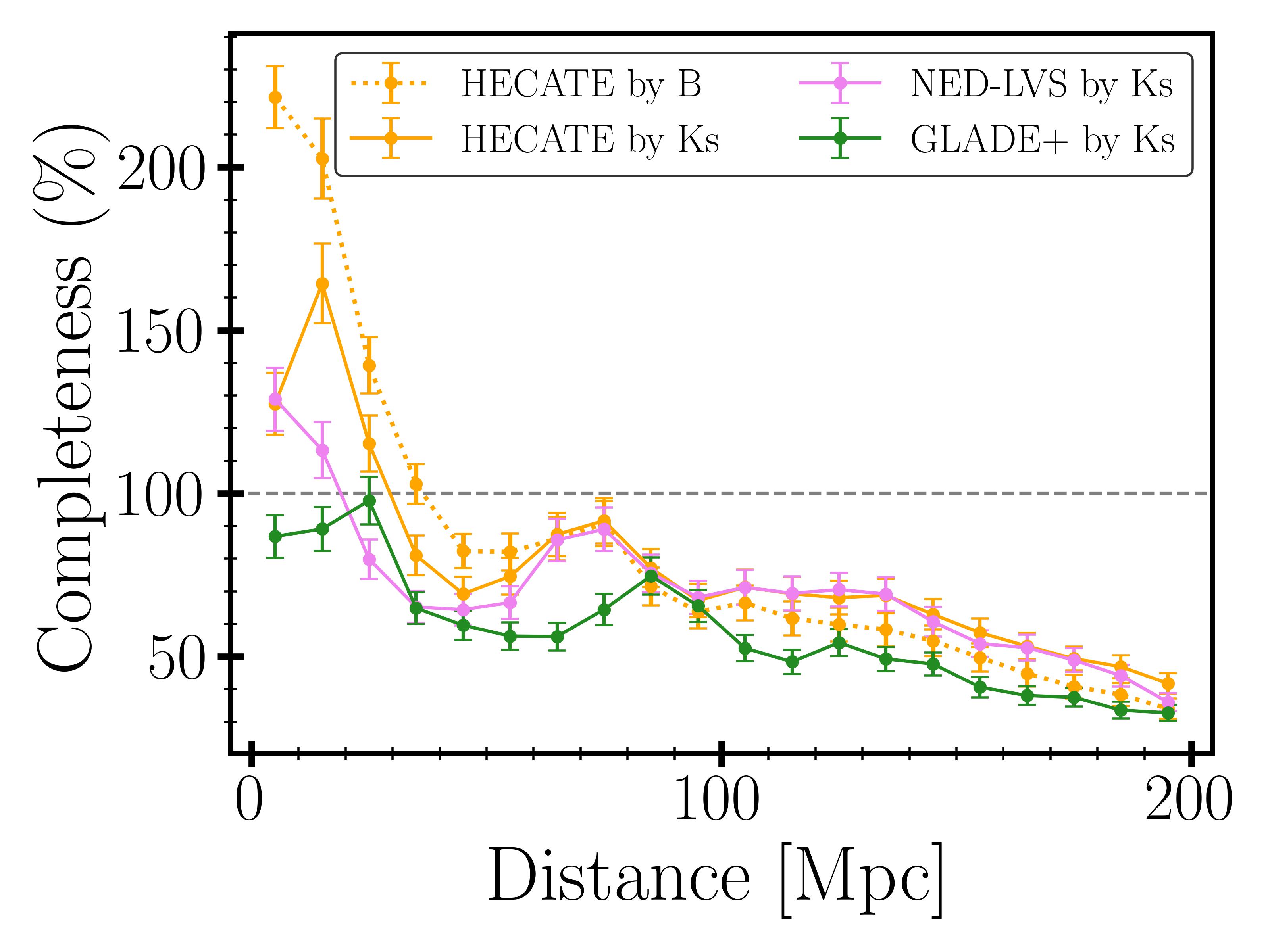}
        \includegraphics[width=\columnwidth]{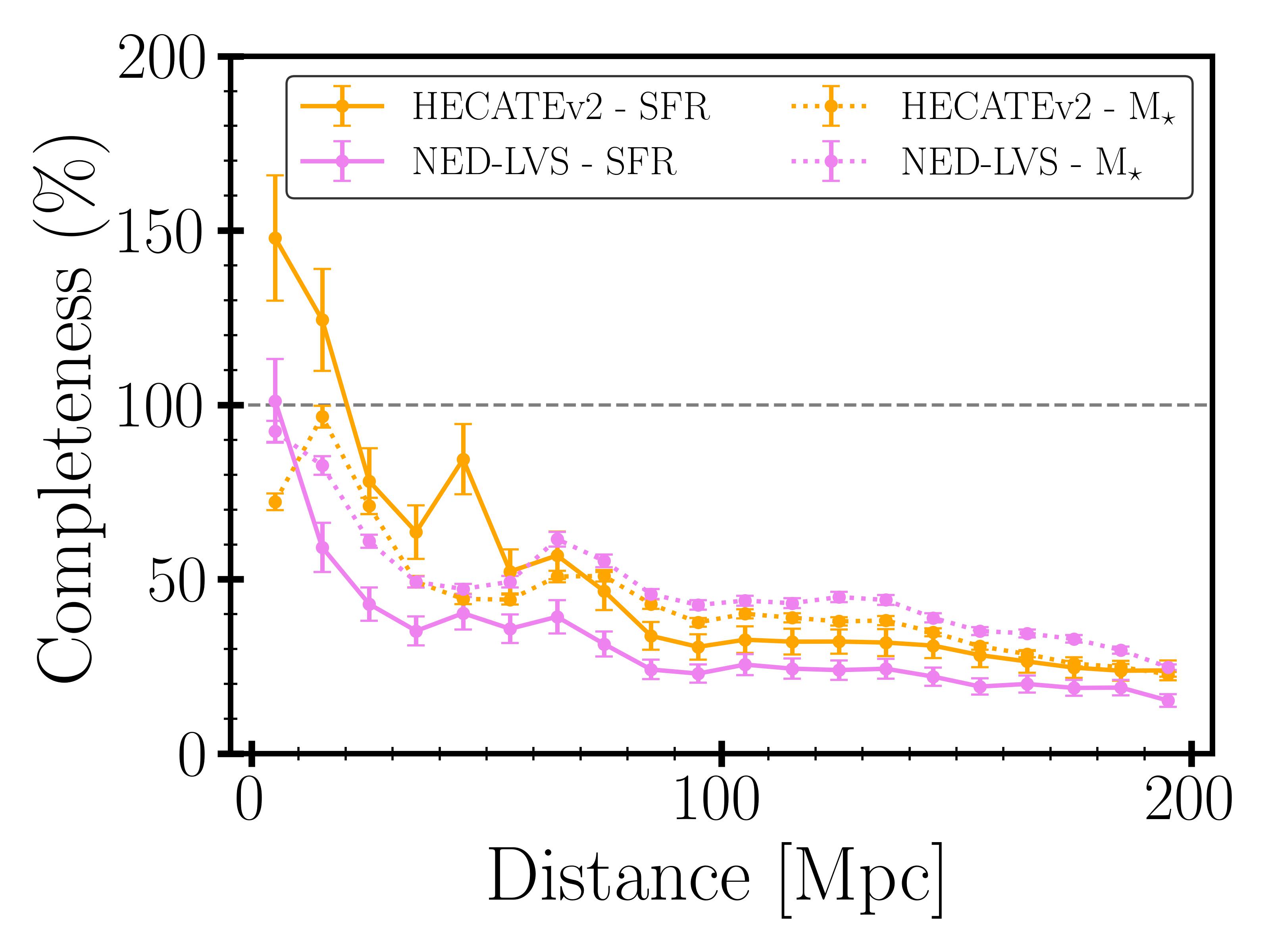}
        
    \caption{Top panel: The completeness of the HECATEv2 in terms of the B-band and Ks-band luminosity with respect to the expectation from observational estimates in different distance bins. For comparison we also show the completeness in the Ks-band of the GLADE+ (green line) and the NED-LVS (pink line) catalogues for the same distance bins.
    Bottom panel: Completeness of HECATEv2 as a function of distance in terms of \sfr{} (orange solid line) and \stellarmass{} (orange dotted line). For comparison, we also show the completeness of the NED-LVS catalogue for the same physical quantities. The completeness in \sfr{} for NED-LVS is shown by the solid pink line, while the dotted pink line represents the completeness in terms of \stellarmass{}.
}
    \label{fig:Completeness_per_Dist_B_Ks_SFR_Mstar}
\end{figure}

In the top panel of Figure \ref{fig:Completeness_per_Dist_B_Ks_SFR_Mstar} we present the B-band ($L_{\rm B}$) and the Ks-band ($L_{\rm Ks}$) completeness of HECATEv2. We find that the catalogue is 100\% complete in the B-band for distances less than 40 Mpc and $>$75\% complete for distances out to $\sim 75$ \distanceunits{}. Beyond this range, the completeness gradually declines, reaching $\sim$50\% at distances up to 170 Mpc. The Ks-band completeness follows a similar trend, being $>$70\% complete for distances within 120 Mpc. The apparent completeness exceeding 100\% at very small distances (D$<$30 Mpc) is the result of the local overdensity of galaxies in the vicinity of the Milky Way. Furthermore, the slight increase in completeness observed in the 70-80 Mpc distance bin is due to the presence of the Hydra galaxy cluster within that volume.

Since many of the applications of HECATEv2 are related to the stellar population properties of galaxies, we also calculated the completeness of the catalogue in terms of \sfr{} and \stellarmass{}. To do so, we applied Equation \ref{eq:completeness_per_Dbin} where instead of the luminosity we considered the \sfr{} or the \stellarmass{}. In the case of \sfr{} we adopted a local Universe star formation density of $\rho_{\rm SFR} = (15 \pm 1.7) \times 10^{-3} \,\, \rm M_{\odot}\,yr^{-1} \, \rm Mpc^{-3}$ \citep{madau14}. Similarly, for the stellar mass we used a stellar mass density of $\rho_{M_{\star}} = (5.26 \pm 0.12) \times 10^{8} \,\, \rm M_{\odot} \, \rm Mpc^{-3}$ \citep{bell03}. The numerator of Equation \ref{eq:completeness_per_Dbin} was computed by using the homogenised \sfr{} and \stellarmass{} from HECATEv2.  In the bottom panel of Figure \ref{fig:Completeness_per_Dist_B_Ks_SFR_Mstar} we present the result of this analysis. As shown, our catalogue is more than 70\% and 50\% complete in \sfr{} and \stellarmass{}, respectively, out to distance of 50 Mpc. Beyond this, the completeness gradually declines, with the \stellarmass{} completeness remaining at approximately 50\% up to 100 Mpc. Table~\ref{tab:HECv2_Lb_complet_per_D_bin} and Table~\ref{tab:HECv2_SFR_Mstar_complet_per_D_bin} present a subset of the L$_{B}$, \sfr{}, and \stellarmass{} completeness per distance bin. The full tables of the B-band luminosity, \sfr{}, and \stellarmass{} completeness per distance bin are available as supplementary online material (see Appendix~\ref{append:completeness_per_lum_bin_sky_patch}).

To further study the completeness of our catalogue, in Appendix \ref{append:completeness_per_lum_bin_sky_patch} we calculate the completeness of HECATEv2 as a function of different luminosity and distance bins in the B and Ks bands, as well as different sky areas. Our results showed that HECATEv2 is 100\% complete down to $L_{B} \sim 10^{7.1}\ L_{B,\odot}$ within 10 Mpc and down to $L_{\rm Ks} \sim 10^{8.3} \ L_{\rm Ks,\odot}$ within 20 Mpc. At greater distances (D $>$100 Mpc), completeness remains high ($\gtrsim$50\%) for galaxies brighter than $L_{\rm B} \sim 10^{11.1} \ L_{\rm B,\odot}$ and $L_{\rm Ks} \sim 10^{9} \ L_{\rm Ks,\odot}$. In addition, spatial completeness maps show significant variation across the sky due to the inhomogeneous depth and coverage of the contributing surveys. Full results of the completeness estimates per sky region in terms of B-band luminosity, \sfr{}, and \stellarmass{} are provided as supplementary online material (see Appendix~\ref{append:completeness_per_lum_bin_sky_patch}).

\subsection{Comparison with other catalogues}
\begin{figure}
        \includegraphics[width=\columnwidth]
        {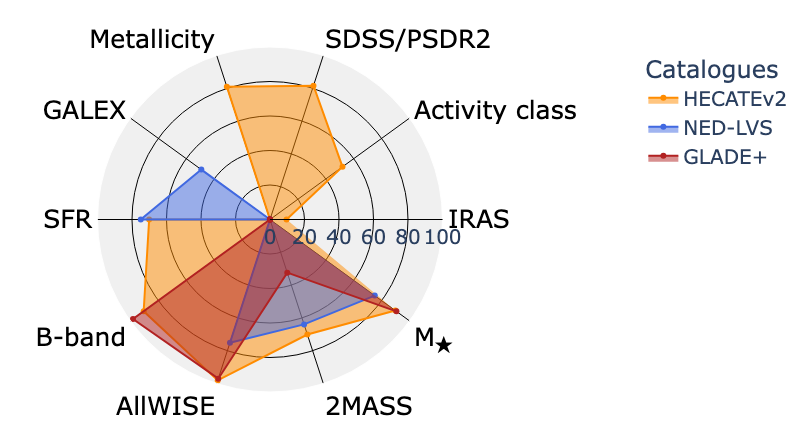}
    \caption{Comparison of the coverage fractions of photometric and derived physical properties for the three catalogues. HECATEv2 (orange shaded area), GLADE+ (red shaded area), and NED-LVS (blue shaded area) within their common distance range (0–200 Mpc). HECATEv2 exhibits the highest overall coverage across most properties, highlighting its broader multi-wavelength and physical parameter completeness.}
    \label{fig:radar_plot}
\end{figure}

In the past decade several galaxy catalogues have been developed aiming primarily in the characterization of multi-messenger sources. Each of these catalogues focuses on different aspects, such as redshift span, and/or available information. In this Section we evaluate the properties of our catalogue in comparison to two major such catalogues: GLADE+ and NED-LVS. In Fig. \ref{fig:radar_plot} we compare the key characteristics of each catalogue, restricted to the same distance range as HECATEv2. 

The GLADE+ catalogue \citep{dalya22}, is the extended version of the GLADE galaxy catalogue introduced by \cite{dalya18}. It compiles data from six astronomical catalogues including HyperLEDA, the Gravitational Wave Galaxy Catalog (GWGC; \citealt{white11}), the 2MASS-XSC, the 2MASS Photometric Redshift catalogue (2MPZ; \citealt{bilicki14}), the WISExSCOSPZ Photometric Redshift catalogue \citep{bilicki16}, and the SDSS DR16 Quasar catalogue \citep{lyke20}. The basic properties provided in GLADE+ include names, positions, distances, and fluxes in the optical B band, in near-IR from 2MASS, and in mid-infrared from WISExSCOS. Contrary to the previous versions of GLADE catalogue, GLADE+ also provides stellar masses and individual neutron star merger rates for galaxies with available W1 magnitudes. 

A direct comparison between the GLADE+ and HECATEv2 within the same distance range (0$<$z$<$0.048) reveals that GLADE+ contains approximately twice as many objects (449\,802) as HECATEv2 (204\,733). However, only 16\% (74\,968) of the GLADE+ entries have redshift-independent distances or spectroscopic redshift measurements. The remaining galaxies rely on photometric redshifts, which, while useful in the absence of more accurate alternatives, are highly uncertain in the Local Universe (D$<$200 Mpc) providing less reliable distance estimates. To ensure a fair comparison of the completeness between HECATE and GLADE+ catalogue, we restricted the analysis only to galaxies with redshift-independent distances or spectroscopic redshift measurements, as these are consistent with the methods used to derive distances in HECATEv2. 

In the top panel of Fig. \ref{fig:Completeness_per_Dist_B_Ks_SFR_Mstar} we show the completeness of GLADE+ in the Ks-band as a function of distance (green line). As shown the completeness of GLADE+ is systematically lower than that of HECATEv2 (orange line) across all distance bins, with an average offset of $\sim$20\%. This is mostly because of the limited number of GLADE+ galaxies with available spectroscopic redshifts, and consequently, reliable Ks-band luminosities. When we compute the completeness of GLADE+ using the entire sample (including objects with photometric redshifts) we find that at larger distances (D$>$160 Mpc), its completeness appears higher than of HECATEv2. However, this apparent excess is driven by the inclusion of many objects with unreliable photometric distances. 

By comparing the basic properties of the two catalogues, in Fig. \ref{fig:radar_plot} we find that both HECATEv2 (orange shaded region) and GLADE+ (red shaded region) have comparable coverage in B-band optical fluxes (90.53\% and 98.18\%, respectively). However, HECATEv2 also includes homogenised optical photometry from SDSS-DR17 and PS1-DR2 compared to the B-band magnitudes providing by GLADE+. As a result HECATEv2 offers a more uniform and extended coverage across the optical spectrum. In the mid-IR, both catalogues have similar completeness via AllWISE or WISExSCOS data (98.07\% and 97.03\%, respectively), however in the near-IR HECATEv2 provides 2MASS fluxes for 70.11\% of its galaxies compared to only 32.41\% in GLADE+. In terms of stellar population parameters, both catalogues offer stellar masses for $\sim90\%$ of their respective galaxy sample. However, the stellar masses in GLADE+ use monochromatic W1 mid-infrared calibrations, which do not fully account for biases introduced by varying stellar population ages or the effects of internal dust extinction. In contrast, HECATEv2 addresses these limitations by employing stellar mass estimates derived from either SED fitting or mid-IR/optical calibrations, which incorporate corrections for these effects (see Sect.~\ref{subsec:SFR_Mstar_S18_K23}). Moreover, HECATEv2 covers a larger range of the stellar population parameter space by providing \sfr{}s, metallicities and galaxy nuclear activity classifications for 70.06\%, 80.96\% and 51.95\% of the catalogue, respectively. 

The NASA/IPAC Extragalactic Database (NED) Local Volume Sample (NED-LVS; \citealt{cook23}) is a value-added galaxy catalogue which contains $\sim$1.9 million galaxies out to a distance of 1000 Mpc. The basic properties provided by the NED-LVS include coordinates, distances and object types. By incorporating data of various multiwavelength catalogues they provide: NUV and FUV fluxes from the two GALEX source catalogues that have been integrated into NED: the All-Sky Survey Catalogue (GASC) and the Medium Imaging Survey Catalogue (GMSC), near-IR fluxes from 2MASS (J,H,Ks bands from the LGA, XSC, and PSC catalogues), and mid-IR fluxes from the AllWISE Source Catalogue (W1,W2,W3, and W4 bands). Special care was taken for the highly extended galaxies by incorporating GALEX and WISE fluxes from the z0MOGS catalogue of \cite{leroy19}. NED-LVS also provides additional stellar parameters such as \sfr{} and \stellarmass{} using combined UV and mid-IR and calibrators. 

A direct comparison between NED-LVS and HECATEv2 shows that NED-LVS contains $\sim$30\% more galaxies (286\,270) than HECATEv2 (204\,733) within the same distance range (D $<$ 200 Mpc). However, as in the case of GLADE+, around 25\% of the NED-LVS (70\,036 objects) have distances derived from photometric redshifts. When accounting only for objects with redshift-independent distances or spectroscopic redshift measurements, the two catalogues contain a comparable number of galaxies in the same distance range: N(NED-LVS)= 216\,234 and N(HECATEv2) = 204\,733. The remaining difference arises from the selection criteria applied during the construction of each catalogue. Specifically, HECATEv2 includes only objects explicitly classified as galaxies in HypeLEDA (objtype='G'). On the other hand NED-LVS imposes no such classification restriction, resulting in the inclusion of more unclassified sources.

In the top panel of Fig. \ref{fig:Completeness_per_Dist_B_Ks_SFR_Mstar} we show the completeness of NED-LVS in the Ks-band as a function of distance (violet line). As with GLADE+, the analysis is restricted to the subset of NED-LVS galaxies with available redshift-independent distances or distances derived from spectroscopic redshifts. As shown, the NED-LVS exhibits slightly lower completeness than HECATEv2 for distances up to 60 Mpc, while at larger distances the completeness of the two catalogues is nearly identical. We note here that when the full NED-LVS sample is considered (including objects with photometric redshifts) the completeness exceeds that of HECATEv2 by a factor of 10\%-20\% for D$>$80 Mpc, while below that distance the two catalogues have similar completeness. This difference reflects once again the impact of including galaxies with uncertain photometric distances in the completeness estimates of a catalogue. 

In the bottom panel of Fig. \ref{fig:Completeness_per_Dist_B_Ks_SFR_Mstar} we also present the completness of NED-LVS catalogue in terms of \sfr{} (violet open stars) and \stellarmass{} (violet open circles). Comparing the two catalogues we find that HECATEv2 is more complete in \sfr{} (orange filled stars) by a factor of $\sim$15\% at distances beyond 40 Mpc, while at smaller distances, both catalogues are essentially complete in \sfr{}. The systematically higher completeness of HECATEv2 at larger distances is due to the broader availability of the photometric bands used by its \sfr{} calibrators. Specifically, HECATEv2 estimates \sfr{} using the W1 and W3 mid-IR bands (see Sect. \ref{subsec:SFR_Mstar_S18_K23}), which are available for a larger fraction of galaxies compared to the NUV/FUV and W4 bands used in NED-LVS. The limited sensitivity of the UV and W4 bands at greater distances results in fewer galaxies with measurable \sfr{} in NED-LVS. In terms of \stellarmass{} both catalogues have similar levels of completeness, with NED-LVS showing slightly higher completeness at distances beyond 120 Mpc. This small difference can be attributed again to the fact that the stellar mass calibrator used in HECATEv2 requires additional optical colour information alongside the W1 band, whereas the NED-LVS estimates rely solely on W1 photometry.

In Fig. \ref{fig:radar_plot} we also compare the multiwavelength and stellar population properties of NED-LVS (blue shaded area) with those of HECATEv2. As shown, NED-LVS includes UV fluxes for the 49.2\% of its sample while HECATEv2 does not have UV coverage. In contrast, NED-LVS does not offer optical fluxes, whereas HECATEv2 includes optical photometry in B-band and the u,g,r,i,z, and y bands from SDSS-DR17 and PS1-DR2. Both catalogues provide mid-IR and near-IR fluxes from the AllWISE and 2MASS, respectively. However, NED-LVS has a slightly lower coverage with 75.1\% for AllWISE and 63.9\% for 2MASS, compared to HECATEv2. In addition, HECATEv2 includes far-IR photometry for 9.6\% of its galaxies, while NED-LVS does not include any far-IR data. In terms of stellar population parameters, both catalogues provide estimates of \sfr{} and \stellarmass{}. HECATEv2 includes \sfr{} estimates for 70\% and \stellarmass{} estimates for 89.8\% of its galaxies, while NED-LVS offers \sfr{} and \stellarmass{} estimates for 74\% and 75\% of its sample, respectively.

\subsection{Science Use Cases}

\subsubsection{Study the connection between SMBH and the host for large representative galaxy samples}\label{sec:SMBH_host_properties}
Supermassive black holes, located at the centers of galaxies, are closely linked to several properties of their host galaxies, as well as their morphological type, suggesting a co-evolutionary relationship \citep[][and references therein]{kormendy13}. In recent years, the advent of large optical, mid-IR, and X-ray surveys has led to several studies of the properties of SMBHs (BH mass distribution, accretion rate, etc.) as a function of the host galaxy properties \citep[e.g. \sfr{}, \stellarmass{}; e.g.][]{mullaney12,rodighiero15,mountrichas24}. However, all these studies are based on limited galaxy samples lacking a complete census of the galaxy population in the local Universe. HECATEv2 addresses this limitation by offering morphological classifications, activity diagnostics, and homogenized measurements of key galaxy properties for the vast majority of the galaxies within a distance of 200 Mpc. As such, it enables robust and statistically representative studies of the connection between SMBH growth and galaxy evolution, making it the most suitable catalogue for these investigations in the local Universe.

To further facilitate such studies, we derived the black hole masses (M$_{\rm BH}$) of the galaxies in HECATEv2 and we explore their connection with the host galaxy characteristics. For the calculation of the M$_{\rm BH}$ we used the empirical \stellarmass{} - M$_{\rm BH}$ relation provided by \cite{green20}: 
\begin{equation}
    \rm log_{10}(\rm M_{BH}) = \alpha + \beta \rm log(M_{\star}/M_{0})+\epsilon
\end{equation}
where M$_{0} = 3 \times 10^{10} M_{\odot}$ and the parameters $\alpha$ and $\beta$ are the best-fit coefficients, and $\epsilon$ is the intrinsic scatter  for different morphological types, as provided in the Table 5 of \citet[][; online supplementary material]{green20}. In our analysis we used the relation derived also considering upper limits. To apply these relations, we divided the HECATEv2 galaxies into subsamples based on their activity classification, which serves as a proxy for their morphological type. Specifically, for the star-forming and composite galaxies we adopted the scaling relation corresponding to the late-type galaxies (tenth row in their Table 5), while for passive galaxies we used the prescription for early-type galaxies (eighth row in their Table 5). For AGN and LINERs we applied the scaling relation from the fit to all galaxies independent of the morphological type (i.e. global; seventh row of their Table 5) from \cite{green20}. Our analysis was restricted to galaxies with reliable stellar mass estimates (i.e. LOG\_MSTAR\_HEC\_QF = 0 or null), to ensure the robustness of the derived M$_{\rm BH}$ values. The derived M$_{\rm BH}$ along with the method used for their calculation are provided as additional columns in the HECATEv2 catalogue. The intrinsic scatter of the scaling relations used for their derivations is $\sim$0.65 dex for both early and late-type galaxies, and about $\sim$0.80 dex for the global relation that does not account for the morphological type. 

\begin{figure}
        \includegraphics[width=\columnwidth]{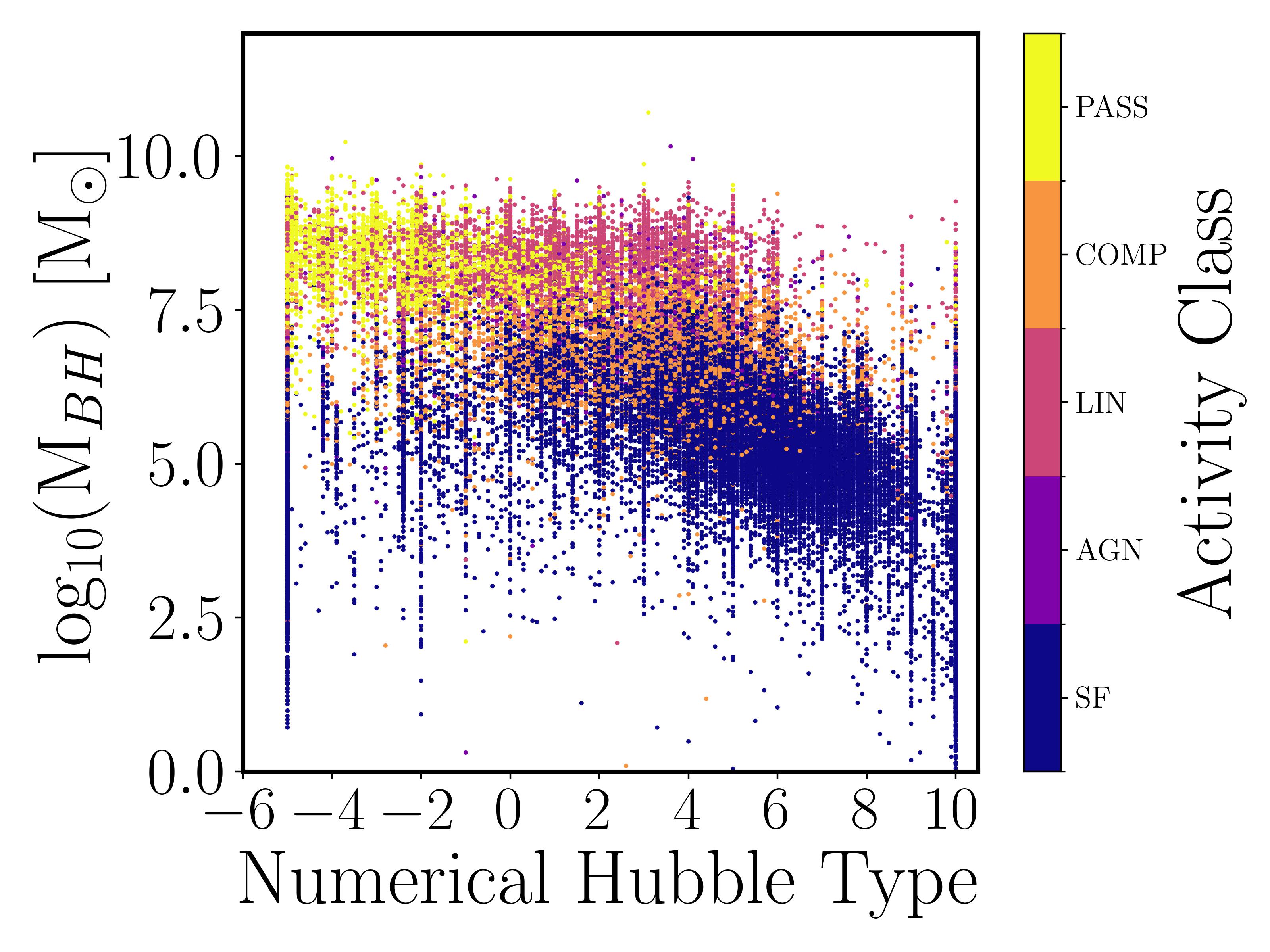}
    \caption{The decimal logarithm of the M$_{\rm BH}$ of HECATEv2 galaxies as a function of their numerical Hubble type (T). The color code indicates galaxies of different activity types.}
    \label{fig:MBH_vs_HT}
\end{figure}

In Fig. \ref{fig:MBH_vs_HT} we present the distribution of the black hole (BH) masses for HECATEv2 galaxies as a function of their numerical Hubble type (T) (see \cite{deVaucouleurs76}). The color coding indicates different galaxy activity types. As shown, the M$_{\rm BH}$ values in HECATEv2 span a broad range from $\sim$$3\times10^{2}$ to $10^{10}$ M$_{\odot}$, covering the full BH mass spectrum and extending well into the regime of intermediate-mass black holes \citep[IMBHs, M$_{\rm BH}$ = 100 - 10$^{5}$ M$_{\odot}$; ][]{green20}. This makes HECATEv2 a valuable parent catalogue for identifying candidate IMBH hosts in the local Universe. We also find a clear correlation between the M$_{\rm BH}$ of the galaxies and their morphological type. In particular, earlier-type galaxies (T $<<$ 0) host more massive BHs, while later-type (0 $<$ T $<$ 8) and irregular (T $>$ 8) galaxies tend to host BHs with lower masses. This trend supports the hierarchical model of galaxy evolution, where massive early-type galaxies represent the evolutionary end products of repeated mergers during which the BH also grows via intense accretion episodes. Finally, the distribution of activity types in the M$_{\rm BH}$–T plane reveals that early-type galaxies are predominantly passive, while later- and irregular-type galaxies are primarily star-forming, consistent with the observed link between morphology, star-formation activity, and BH-mass \citep{kormendy13,green20}. 

\subsubsection{Host Galaxies to multimessenger events}
All-sky galaxy catalogues play a vital role in the rapid identification of electromagnetic (EM) counterparts to gravitational wave (GW) sources (NS-NS, or NS-BH mergers) or candidate hosts to BH-BH mergers, which are not expected to produce EM emission. This is a crucial step toward constraining the physical nature of the events \citep[e.g][]{abbott17a}, shedding light on the formation and evolution of their progenitor systems \citep[e.g.][]{kalogera07,abbott17c}, and even enabling their use as standard "sirens" for measuring the Hubble constant \citep[e.g.][]{abbott17b,chen18}.

However, the search of the hosts of GW sources even when they are accompanied by EM emission, remains extremely challenging due to the poor sky localization of the current GW detectors. A recent analysis of the O3 run events by \cite{petrov22} showed that improvements in LIGO/Virgo/KAGRA (LVK) data analysis enabled the detection of weaker GW signals, but at the cost of significantly larger sky localization regions (median $\sim$2000 deg$^2$). Furthermore, based on these results, they predict that in the upcoming O4 and O5 observing runs, the median sky localization areas for such events will remain above 1000 deg$^{2}$, with the corresponding median detection distances reaching $\sim$ 350 Mpc for Binary Neutron Star (BNS) mergers and $\sim$ 620 Mpc for Neutron Star - Black Hole (NSB) mergers \citep{petrov22}. 

Given the large sky localization uncertainties, the identification of a galaxy hosting the EM counterpart of a GW event typically relies on targeted multiwavelength follow-up observations of a prioritized list of candidate host galaxies. This prioritization is based not only on sky position and distance, but also on observed properties (e.g. optical magnitudes or colors) and derived stellar population parameters. This is motivated by theoretical studies which, through population synthesis combined with cosmological simulations, suggest that GW events are correlated with the host galaxy properties such as optical-colors, \stellarmass{}, \sfr{}, and metallicity \citep[e.g.][]{mapelli17,toffano19,artale19,artale20}. Furthemore, recent studies have shown that the BNS merger rate is linked to the age of the stellar population and the merger delay time, with more massive galaxies, typically hosting older stellar populations, being associated with longer merger delay times \citep{adhikari20}.

Motivated by this need, several studies have proposed optimized follow-up strategies and developed galaxy catalogues that include not only positions and distances, but also photometric data \cite[e.g.][]{kopparapu08,gehrels16} and physical parameters such as \sfr{} and \stellarmass{} \cite[e.g.][]{dalya18,ducoin20,cook23}. However, all these catalogues lack detailed optical photometry across multiple bands and gas-phase metallicities, which can be an important factor to the calculation of GW rates of a galaxy \citep{artale19}.

HECATEv2 is currently the most complete all-sky galaxy catalogue in comparison to other similar resources, particularly in terms of its coverage of optical photometry and gas-phase metallicity (see Fig.~\ref{fig:radar_plot}). The combination of homogenized \sfr{}, \stellarmass{}, optical photometry from SDSS-DR17 and PS1-DR2 surveys, spectroscopic metallicity measurements, metallicities derived from the mass–metallicity relation, make HECATEv2 a well-suited reference catalogue for identifying electromagnetic counterparts to GW events from BNS mergers within a distance of 200 Mpc.  In addition, the host galaxy properties such as morphological type, \stellarmass{}, apparent magnitudes, activity type, and SMBH mass can be used for calculations of the SMBH merger rates and potential events for experiments like LISA and PTA \citep[e.g.;][]{charisi22,veronesi25}. 

\subsubsection{Statistical studies and characterisation of galaxy populations in multi-wavelength surveys}

HECATEv2 offers a robust reference galaxy catalogue for statistical studies of galaxy populations and their physical properties in the context of large multi-wavelength photometric and spectroscopic surveys \citep[Euclid, DESI, eROSITA;][]{euclid25, desi25, merloni24}. The combination of the basic and derived physical properties of a highly complete galaxy sample such as HECATEv2 with multi-wavelength information from other surveys, enables comprehensive statistical analyses as well as the identification of interesting rare galaxy subpopulations in the local Universe. 

For instance, \cite{kyritsis25} utilised HECATEv2 to study the integrated X-ray emission from star-forming galaxies as a function of their stellar population properties. By using the homogenised value-added information provided by our catalogue (i.e. photometric data, activity classification, \sfr{}, \stellarmass{}, metallicity) in combination with the X-ray data from the first complete all-sky scan of the eROSITA X-ray telescope \citep[eRASS1;][]{predehl21,merloni24}. They identified a sub-population of highly X-ray luminous starburst galaxies with higher specific SFRs, lower metallicities, and younger stellar populations. This result underscores the importance of a large and complete galaxy catalogue for uncovering rare galaxy populations. Furthemore, it suggests a complex picture where X-ray emission produced by X-ray binaries populations (XRBs) in metal poor dwarf star-forming galaxies can rival the X-ray or optical emission emitted by SMBHs of low-luminosity AGN (LLAGN).

\begin{figure}
    \includegraphics[width=\columnwidth]{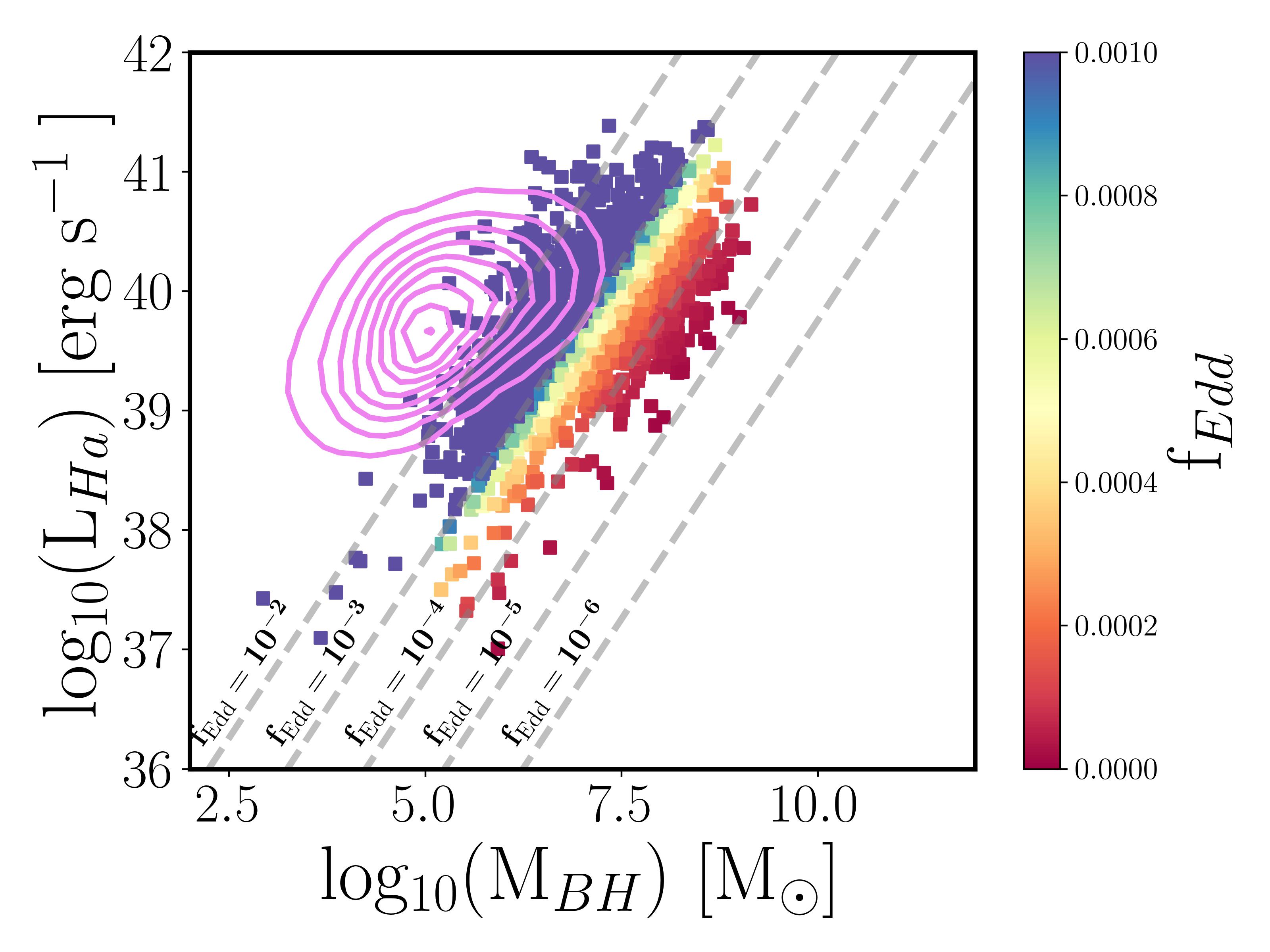}
    \caption{The H$_{\rm \alpha}$ luminosity (L$_{\rm H_{\rm \alpha}}$) as a function of the M$_{\rm BH}$ for all the HECATEv2 galaxies classified as AGN (filled squares) and star-forming (magenta contours). The color-code indicates the Eddington fraction (f$_{\rm Edd}$) of the AGN sample. The gray diagonal dashed lines represent different f$_{\rm Edd}$ values.}
    \label{fig:Ha_vs_MBH}
\end{figure}

The homogenised value-added information provided in the HECATEv2 catalogue is ideal for exploring this overlapping regime between XRBs and LLAGNs in large galaxy samples. This is illustrated in Fig. \ref{fig:Ha_vs_MBH} where we present the \ion{H}{$\alpha$} luminosity (L$_{\ion{H}{$\alpha$}}$) as a function of the M$_{\rm BH}$ for all the HECATEv2 galaxies classified as AGN (color-coded filled squares) and star-forming (magenta contours). The colour-code indicates the Eddington fraction (f$_{\rm Edd}$) of the AGN sample. The gray diagonal dashed lines represent different f$_{\rm Edd}$ values. 
For the calculation of the f$_{\rm Edd}$ we used the BH masses provided in our catalogue (see Sect. \ref{sec:SMBH_host_properties}) adopting the bolometric correction factor L$_{\rm bol} \sim 220$ L$_{H_{\rm \alpha}}$ from the LLAGN SED of \citet{ho08}. 

As shown, there is a region where AGN with relatively high f$_{\rm Edd}$ and low M$_{\rm BH}$ exhibit H$_{\rm \alpha}$ luminosities comparable to those of star-forming galaxies. Leveraging the extensive multi-wavelength and physical parameters provided by HECATEv2 we are able to better disentangle the two populations of star-forming galaxies and LLAGNs by using SED fitting to decompose the emission due to star-forming and AGN. That has important implications for understanding the population of XRBs contributing in the most X-ray luminous galaxies while it is also crucial because the population of LLAGN can be the result of lower accretion rate or accretion onto a lower-mass black hole. Forthcoming large surveys \citep[SDSS-V, Large Survey of Space and Time - LSST ; ][]{sdss5,lsst19} will supply deeper photometric, spectroscopic, and time‑domain data, that when cross‑matched to HECATEv2, will contribute on setting constrains on the low‑mass, high‑f$_{\rm Edd}$ regime and refine the census of faint accreting black holes in the local Universe.

\subsection{Current limitations and future work}
Although HECATEv2 represents a major update of the first release of the catalogue, there are still some limitations that potential users should take into account. 

First of all, since we did not change the parent sample of HECATE, the updated version still inherits the uncertainties associated with the unknown selection function of the HyperLEDA database. As a result, any generalization based on this galaxy compilation should be based on the completeness based on the $\rm L_{B}$, $\rm L_{Ks}$, \sfr{}, or \stellarmass{} depending on the relevant application. Furthermore, the current volume coverage remains unchanged with respect to HECATEv1, as in this work we prioritized a qualitative upgrade of the information provided rather than expanding the catalogue to include more distant objects. In a future release, however, we plan to extend the redshift coverage up to $\rm z \sim 0.5$. Such an expansion will be particularly valuable for forthcoming GW surveys (e.g. LIGO, Virgo), as well as for the preparation of simulations and the identification of SMBH mergers for PTA experiments or the LISA mission.

As discussed in Sect.~\ref{sec:sizes}, the majority of size estimates in HECATEv2 are drawn from the HyperLEDA database and correspond to the $\rm D_{25}$ isophote. While this provides a robust measure, in future releases we plan to update both the size information and the morphological classifications as new catalogues from deeper and wider area surveys become available. Despite efforts to minimise duplications, blends, or incorrect photometric matches (e.g. by selecting only G-type galaxies from HyperLEDA, performing internal cross-matching, and visual inspection), a certain level of contamination inevitably remains. In future releases, we aim to mitigate this by incorporating information from higher-quality, higher-resolution imaging surveys (e.g. DESI), which will allow us to identify additional problematic cases. 

While HECATEv2 provides coverage from the optical to the far-IR (i.e. SDSS-DR17/NSA, PS1-DR2, and AllWISE), future versions will also include UV measurements from GALEX, explicitly accounting for aperture effects in nearby extended galaxies. Similarly, while spectroscopic information in HECATEv2 is primarily drawn from SDSS, future versions will incorporate data from the DESI and LAMOST surveys, as well as future releases of SDSS, complemented by targeted observing campaigns.

In HECATEv2 we provide \sfr{} and \stellarmass{} measurements based on SED fits for galaxies in the SDSS footprint and monochromatic or hybrid calibrations for the remaining galaxies. In future releases, with the inclusion of UV photometry, we plan to perform a broadband SED fitting from the UV to the IR, which is the most robust approach for deriving \sfr{} and \stellarmass{}, as discussed in Sect.~\ref{subsec:SFR_Mstar_S18}, and use monochromatic indicators only when broadband SED data are not available.

Finally, HECATEv2 provides activity classifications based on spectroscopic diagnostics where available, while for the remaining galaxies we relied on photometric methods. Although the latter are based on state-of-the-art techniques (see Sect.~\ref{subsec:phot_activity_class}), spectroscopic classifications remain more robust. Future releases will take advantage of the additional spectroscopic coverage provided by DESI and LAMOST \citep{zhao12}, and future SDSS releases, combined with new machine learning-based spectroscopic activity diagnostics (Daoutis et al., 2025, submitted), to update the classifications. The same holds for the metallicity estimates provided in the catalogue, which will also be refined as new spectroscopic data become available.

\section{Conclusions}\label{sec:conclusion}
In this work we presented HECATEv2, the second release of the Heraklion Extragalactic Catalogue (HECATE) all-sky value-added galaxy catalogue \citep{kovlakas21}, which includes 204\,733 galaxies from the HyperLEDA database with
$\upsilon_{vir} < 14,000 \, \rm{km\,s^{-1}}$. In this release, we have implemented major updates focusing primarily on qualitative improvements to the information provided rather than expanding the catalogue with additional galaxies. These updates include:
\begin{enumerate}
\item a new distance estimation framework based on a cosmological model for galaxies without redshift-based distances, with special treatment for Virgo Cluster members. This approach ensures consistency and enables future extensions of the catalogue to higher redshifts,
\item significant improvement of the photometric coverage by incorporating and homogenising optical and mid-IR data from SDSS-DR17/NSA, PS1-DR2, and AllWISE, accounting for aperture effects in nearby galaxies;
\item additional quality checks, including flags for star contamination, incorrect photometry matches, and coordinate inconsistencies, which enhance the reliability of the photometry,
\item completion of the coverage for galaxy sizes, as well as validation against independent measurements from more detailed analyses, ensuring consistency and completeness of the size information provided,
\item derivation of stellar population parameters for a larger fraction of the sample compared to the first release. In particular, using the new homogenised photometric framework, updated mid-IR calibrations accounting for stellar population age and dust attenuation were applied to estimate \sfr{} and \stellarmass{} for over 70\% of galaxies, while gas-phase metallicities were computed for $\sim$90\% of the catalogue using spectroscopic data and the mass–metallicity relation,
\item improved activity classifications for more than 50\% of the sample, based on combined photometric and spectroscopic diagnostics.
\item M$_{\rm BH}$ measurements based on the empirical \stellarmass{} - M$_{\rm BH}$ relation provided by \cite{green20}.  
\end{enumerate}

In terms of $\rm L_{B}$, $\rm L_{Ks}$, \sfr{}, and \stellarmass{} completeness, HECATEv2 is among the most complete local-Universe galaxy samples with spectroscopic redshifts. Spatial completeness maps reveal significant variation across the sky due to the inhomogeneous depth and coverage of the contributing surveys. The full results of the completeness estimates, per distance bin and per sky region, are provided as online supplementary tables and are available through Vizier.
Comparison of HECATEv2 with two major catalogues (GLADE+ and NED-LVS) shows that our catalogue has, greater coverage in optical (SDSS/PS1-DR2), and far and near-IR (IRAS,2MASS) photometry, as well as, in metallicity, and activity classification. In addition, our coverage is similar to the other catalogues in mid-IR photometry and in terms of \stellarmass{}, and \sfr{}. 

These updates make HECATEv2 a robust all-sky galaxy catalogue within a volume of $\rm D \simeq 200~Mpc$, ideally suited for a wide range of scientific applications. These include: i) studies of the connection between SMBHs and their host galaxies, ii) identification and characterisation of host galaxies of GW and high-energy transients, iii) statistical studies and characterisation of extragalactic source populations in multi-wavelength surveys, and iv) discovery of rare populations such as low-metallicity dwarfs or IMBHs. With future planned extensions (i.e. HECATEv3) in redshift coverage, photometric and spectroscopic coverage, and refined activity diagnostics, HECATE will continue to provide a cornerstone reference for multi-wavelength and multi-messenger astrophysics.

\section*{Acknowledgements}
This work was supported by the \emph{Found\-ation for Re\-search \& Tech\-nology – He\-llas} (FORTH). EK  acknowledges support from the \emph{FORTH Synergy Grant PARSEC}, and from the Public Investments Program through a Matching Funds grant to the IA-FORTH. The research leading to these results has received funding from the European Union’s Horizon 2020 research and innovation programme under the Marie Skłodowska-Curie RISE action, Grant Agreement n. 873089 (ASTROSTAT-II). CD acknowledges support from the Public Investments Program through a Matching Funds grant to the IA-FORTH. The research leading to these results has received funding from the European Research Council under the European Union’s Seventh Framework Programme (FP/2007-2013)/ERC Grant Agreement n. 617001, the European Union’s Horizon 2020 research and innovation programme under the Marie Skłodowska-Curie RISE action, Grant Agreement n.873089 (ASTROSTAT-II). KK was supported an ICE Fellowship under the program Unidad de Excelencia Maria de Maeztu (CEX2020-001058-M). KK was supported by the institutional project RVO:67985815 by the INTER-COST LUC24023 project of the INTER-EXCELLENCE II programme, funded by the Czech Ministry of Education, Youth and Sports. AB-Z acknowledges support by NASA under award number 80GSFC24M0006.

%%%%%%%%%%%%%%%%%%%%%%%%%%%%%%%%%%%%%%%%%%%%%%%%%%
\section*{Data Availability}
The data underlying this article are provided as supplementary online material. They are also available at the HECATE Portal at \href{https://hecate.ia.forth.gr}{https://hecate.ia.forth.gr}. For legacy users, the two versions of HECATE will be provides as separate files. In addition, the HECATEv2 is also available at CDS via anonymous ftp to cdsarc.u-strasbg.fr (130.79.128.5) or via https://cdsarc.cds.unistra.fr/viz-bin/cat/J/MNRAS .


\bibliographystyle{mnras}
\bibliography{references} % if your bibtex file is called example.bib

%\begin{thebibliography}{99}
%\end{thebibliography}

%%%%%%%%%%%%%%%%%%%%%%%%%%%%%%%%%%%%%%%%%%%%%%%%%%

%%%%%%%%%%%%%%%%% APPENDICES %%%%%%%%%%%%%%%%%%%%%

\appendix
\section{Homogenisation of the optical photometries in the HECATEv2}\label{append:opt_phot_rescal}
To account for the systematic differences in the photometries provided by the various catalogues and surveys (i.e. SDSS-DR17, NSA, PS1-DR2) used in the HECATEv2, we rescaled all of them using as a reference the SDSS-DR17 photometry. 

To that end, we first compared the cModelMag with the PETRO\_MAG (adopted NSA photometry) and we found very good agreement between the two photometries. Specifically the  standard deviation of the cModelMag - PETRO\_MAG difference is $\sim 0.01$ mag for the g, r, i bands and $\sim 0.04$, and $\sim 0.08$ mag for the u, and z bands, respectively. Given this small difference, we use the NSA magnitudes without any further rescaling. A comparison between the fiberMag and the FIBER\_MAG (provided by the NSA) showed that the latter is systematically underestimated by $\sim 0.1$ mag. To correct for this, we calculated the median value of the fiberMag - FIBER\_MAG offset per band, and we rescaled the FIBER\_MAG accordingly. 

\begin{figure*}
    \centering 
        \includegraphics[width=18cm]{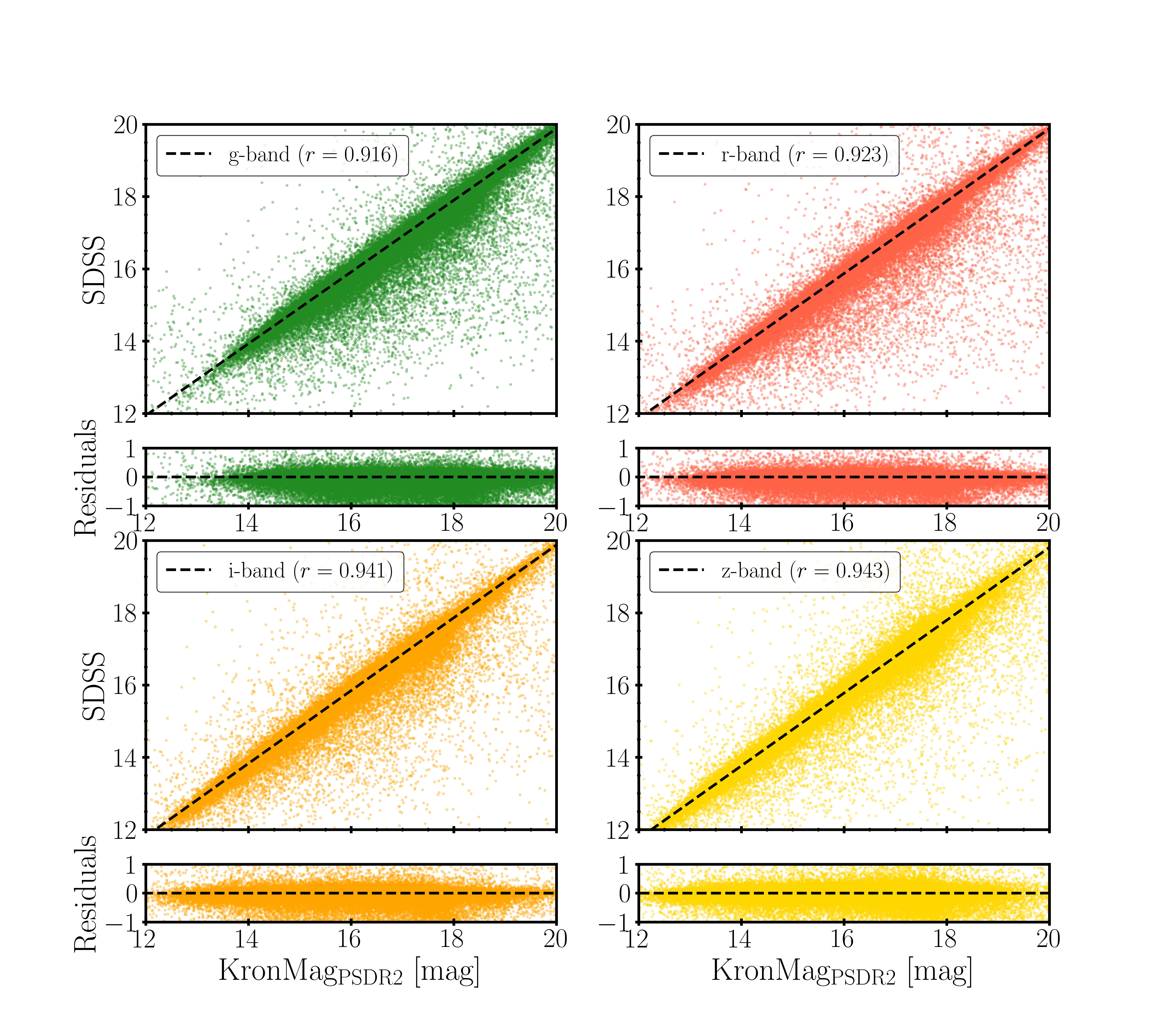}
    \caption{Comparison between SDSS/NSA and PS1-DR2 photometry across all bands. The black dashed line shows the best-fit relation per band along with their residuals. The r values indicate the Pearson Correlation Coefficient for each band.}
    \label{fig:SDSSDR17_NSA_PS1DR2_best_fit}
\end{figure*}

However, a similar comparison between the SDSS-DR17/NSA photometries with the adopted PS1-DR2 Kron photometry (i.e. cModelMag or PETRO\_MAG - KronMag) showed that the latter is systematically fainter by a factor of $\sim 0.1-0.2$ mag, depending on the band. To correct this, we rescaled the PS1-DR2 KronMag using a linear fit to the SDSS or NSA photometries as a function of the PS1-DR2 photometries for the common objects of the form:
\begin{equation}
    \rm{cModelMag\_? \, \text{or} \, PETRO\_MAG\_?} = \alpha \times \rm{?KronMag} + \beta
\end{equation}
where the question mark corresponds to the four common bands (g, r, i, z). The choice of this linear model is justified by the corresponding Pearson Correlation Coefficients (PCCs) values which indicate a strong linear correlation between the SDSS and PS1-DR2 magnitudes (see Fig. \ref{fig:SDSSDR17_NSA_PS1DR2_best_fit}).
The fit was performed using the Theil-Sen regression algorithm for robust fitting (\texttt{TheiSenRegressor}) provided by \texttt{scikit-learn} library \citep{scikit-learn}. Unlike ordinary least squares (OLS) regression, the Theil-Sen estimator determines the median of all possible pairwise slopes, making the fit more stable and less sensitive to strong outliers from the main data distribution. In Fig. \ref{fig:SDSSDR17_NSA_PS1DR2_best_fit} we present the results of the fit for each band together with their PCCs. The best-fit parameters and their Root-Mean Square Errors (RMSE) are listed in Table \ref{tab:PS1DR2-rescaling_factors}. By using these parameters we can calculate the rescaled Kron magnitude as:
\begin{equation}
    \rm{?KronMag_{\textrm{rescaled}}} = \alpha \times \rm{?KronMag} + \beta \,\,\, .
\end{equation}
We note that the rescaling of the Kron magnitudes was applied only to the common bands between SDSS-DR17/NSA and PS1-DR2 (i.e. g,r,i,z). For the y band we adopt the original Kron magnitudes as there is no corresponding reference standard for rescaling. Following the same methodology we also rescaled the 3\arcsec{}-aperture from PS1-DR2 (c6flxR5\_MAG) with respect to SDSS-DR17 fiberMag.

\begin{table}
% \centering
\caption{Best-fit parameters for the rescaling of PS1-DR2 magnitudes with respect to SDSS magnitudes. }
\label{tab:PS1DR2-rescaling_factors}
\begin{tabular}{cccc}
\hline\hline
Rescaled PS1-DR2 magnitudes & $\alpha$ & $\beta$ & RMSE \\\hline
gKronMag$_{\textrm{rescaled}}$ & 0.989 &  0.089  & 0.550 \\
rKronMag$_{\textrm{rescaled}}$ & 1.005 & -0.217  & 0.554\\
iKronMag$_{\textrm{rescaled}}$ & 1.012 & -0.349  & 0.500 \\
zKronMag$_{\textrm{rescaled}}$ & 1.011 & -0.394  & 0.509 \\
\hline
\end{tabular}
\medskip
\parbox{\columnwidth}{\footnotesize
Note: RMSE represents the Root Mean Square Error of our fits per photometric band.} 
\end{table}

\section{Homogenisation of the mid-IR photometries in the HECATEv2}\label{append:WISE_hyb_rescal}
\begin{figure*}
    \centering 
        \includegraphics[width=18cm]{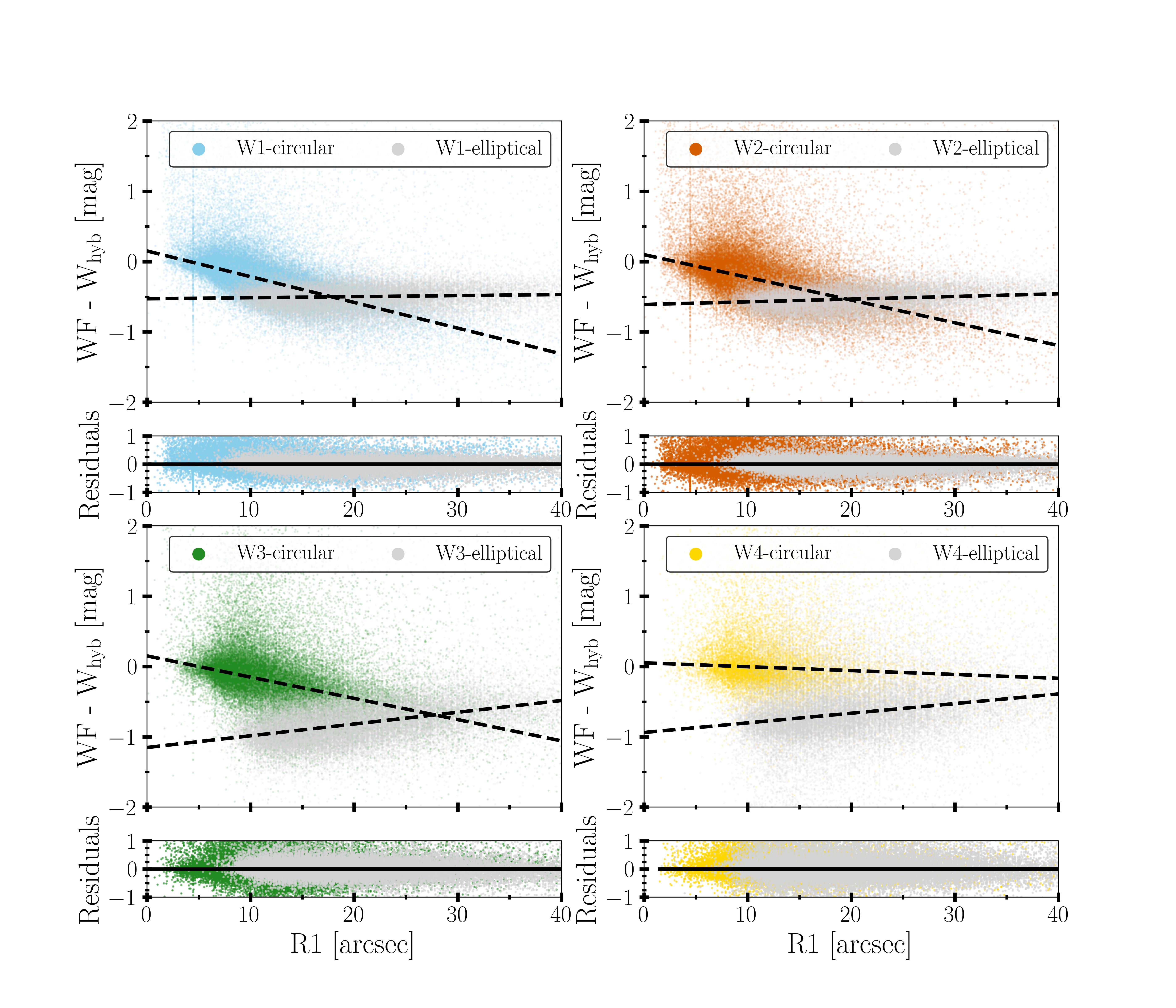}
    \caption{Difference between the forced photometry magnitudes (WF) and the hybrid photometric scheme magnitudes (W$_{\rm{hyb}}$) as a function of the galaxy semi-major axis (R1). The comparison is shown separately for galaxies with fixed-aperture photometry (w?mag\_2; different colors per band) and those with elliptical-aperture photometry (w?gmag; light grey points). The black dashed lines indicate the best-fit relations used to rescale the magnitudes in each band and each type of WISE photometry, along with their corresponding residuals.}
    \label{fig:WF_vs_Whyb_best_fit}
\end{figure*}

As we discussed in Sect. \ref{subsec:midir_phot_surveys_AllWISE} our mid-IR hybrid photometric scheme consists of two different types of photometry.
That is, elliptical apertures based on the 2MASS survey  when available, and fixed aperture (with the corresponding aperture corrections) for the remaining galaxies. For that reason it is of high importance to homogenise them by rescaling them using the reliable forced photometry \citep[photometry from the unWISE coadds based on the SDSS-D10 photometry; WF; see Sect.~\ref{subsec:midir_phot_surveys_AllWISE};][]{lang16} from HECATEv1 as a reference. Additionally, since our hybrid scheme includes photometries with different aperture sizes, we expect a correlation between the measured fluxes and galaxy size. Indeed, in Fig. \ref{fig:WF_vs_Whyb_best_fit}, we present the difference between the forced photometry (WF) and the hybrid photometric scheme (W$_{\rm{hyb}}$) as a function of the galaxy semi-major axis (R1), separately for galaxies with fixed-aperture photometry (w?mag\_2) and those with elliptical-aperture photometry (w?gmag). Galaxies measured with the circular aperture show an anti-correlation with size, as larger galaxies appear fainter compared to their forced photometry values. This is expected, as the fixed circular aperture does not fully encompass the flux from extended sources, resulting to an underestimation of their brightness. On the other hand, elliptical apertures systematically yield fainter magnitudes than forced photometry, though this difference does not exhibit a strong correlation with galaxy size. Based on this results, we rescaled separately the W$_{\rm{hyb}}$ for each of the elliptical and fixed-aperture photometries using a linear model:
\begin{equation}
    \rm{WF? - W?_{hyb}} = \alpha \times \rm{R1} + \beta
\end{equation}
where the question mark corresponds to the four AllWISE bands (1,2,3, and 4) and R1 is the galaxy's semi-major axis in arcseconds. We performed the fit independently in two different galaxy groups based on the origin of their aperture in the hybrid photometry scheme. For the fits we used the same robust fitting algorithm as in Appendix \ref{append:opt_phot_rescal}. The best-fit parameters and their Root-Mean Square Errors (RMSE) are listed in Table \ref{tab:WISE-rescaling_factors}. In addition, Fig. \ref{fig:WF_vs_Whyb_best_fit} shows along with the data and the fits the residuals, demonstrating excellent homogenization between the two types of photometry.
By using these parameters we can calculate the rescaled hybrid photometric scheme  magnitude as follows:
\begin{equation}
    \rm{W?_{hyb, rescaled}} = \alpha \times \rm{R1} + \beta + \rm{W?_{hyb}} \,\,\, .
\end{equation}

\begin{table}[h]
% \centering
\caption{Best-fit parameters for the rescaling of our hybrid mid-IR photometric scheme with respect to forced photometry.}
\label{tab:WISE-rescaling_factors}
\begin{tabular}{cccc}
\hline\hline
Rescaled W$_{\rm{hyb}}$ magnitudes& $\alpha$ & $\beta$ & RMSE  \\ 
\hline
\multicolumn{4}{c}{Circular aperture 8.25\arcsec{}} \\
W1$_{\rm{hyb, rescaled}}$ & -0.036 &    0.151  & 0.615 \\
W2$_{\rm{hyb, rescaled}}$ & -0.032 &    0.098  & 0.582 \\
W3$_{\rm{hyb, rescaled}}$ & -0.030 &    0.162  & 0.596 \\
W4$_{\rm{hyb, rescaled}}$ & -0.005 &    0.052 & 0.727 \\
\multicolumn{4}{c}{Elliptical aperture} \\
W1$_{\rm{hyb, rescaled}}$ & 0.001 &  -0.529  & 0.457 \\
W2$_{\rm{hyb, rescaled}}$ & 0.003 &  -0.610  & 0.411 \\
W3$_{\rm{hyb, rescaled}}$ & 0.016 &  -1.152  & 0.486 \\
W4$_{\rm{hyb, rescaled}}$ & 0.013 &  -0.938  & 0.784 \\
\hline
\end{tabular}
\medskip
\parbox{\columnwidth}{\footnotesize
Note: RMSE represents the Root Mean Square Error of our fits per photometric band.
}
\end{table}

\section{Homogenisation of the \sfr{} and \stellarmass{} in the HECATEv2}\label{append:SFR_Mstar_rescales}
The \sfr{} and \stellarmass{} provided in HECATEv2 originate from three main sources. The first one is the z0MOGS catalogue by \citet{leroy19}, which offers accurate measurements for $\sim 15\,000$ nearby galaxies, carefully accounting for their extended emission. The second source is the SED-based estimates from the GSWLC-2 catalogue by \cite{salim16,salim18}, and the third one is based on the mid-IR/optical calibrations from \cite{kouroumpatzakis23}. For more details on the methods used, see Sects. \ref{subsec:SFR_Mstar_S18_K23}, and \ref{subsec:SFR_Mstar_L19}. However, these different calibrations can introduce systematic offsets that must be addressed to ensure consistency of the SFR and M for the overall sample. To this end, we opted to homogenise all available \sfr{} and \stellarmass{} estimates using the K23 calibrations as our reference standard, enabling future extensions of our catalogue to higher redshifts. 

As a first step, and to mitigate potential biases introduced by the different distance methodologies used in S18 and L19, we rescaled their \sfr{} and \stellarmass{} measurements, using as a reference the HECATEv2 distances, as follows:
\begin{equation}
    \begin{split}
    \log_{10} M_{\star , \text{D-rescaled}}^{S18 \, \rm{or} \, L19} &= \log_{10} M_{\star}^{S18 \, \rm{or} \, L19} + 2 \log_{10} \left( \frac{D^{\text{HECv2}}}{D^{S18 \, \rm{or} \, L19}} \right)\\
    \log_{10} SFR_{\text{D-rescaled}}^{S18 \, \rm{or} \, L19} &= \log_{10} SFR^{S18 \, \rm{or} \, L19} + 2 \log_{10} \left( \frac{D^{\text{HECv2}}}{D^{S18 \, \rm{or} \, L19}} \right)
    \end{split}
    \label{eq:SFR_Mstar_D_resc}
\end{equation}

In case of L19 the galaxy distances are provided in their Table 4, allowing a direct application of the above equation. In the case of S18, however, distance are not directly provided, only spectroscopic redshifts instead. Therefore, we calculated the corresponding luminosity distances using these redshifts and the same cosmological parameters adopted by S18\footnote{\cite{salim16,salim18} assumed the WMAP7 cosmology with the following parameters: H$_{0} = 70.4 \, \rm{km/s/Mpc}$, $\Omega_{m} = 0.272$, and $\Omega_{\Lambda} = 0.728$. }.
While in our analysis we only included galaxies with good redshift measurements, for 23 galaxies our calculation of the luminosity distance resulted in extremely high distances ($>400$ Mpc). These outliers are likely due to either incorrect spectroscopic redshifts assigned to GSWLC-2 galaxies, affecting the calculation of $D_{\text{S18}}$, or wrong heliocentric radial velocities in HECATEv1, which impact the derivation of the D$_{\text{HECv2}}$. In both cases, these outliers result in erroneous \sfr{}, \stellarmass{} values and were therefore flagged accordingly in our catalogue.

\begin{figure*}
 \centering
        \includegraphics[width=0.47\textwidth]{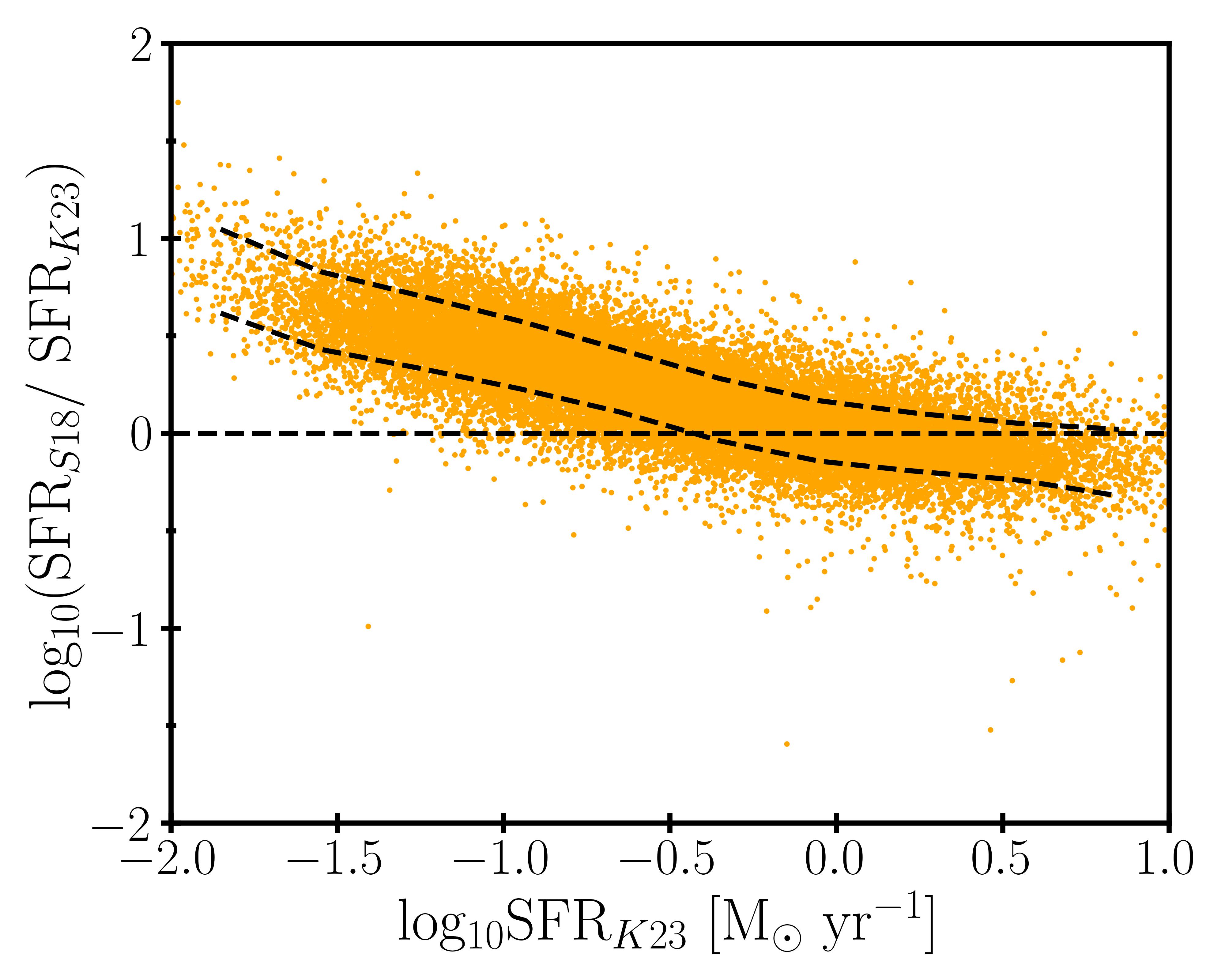}
        \includegraphics[width=0.47\textwidth]{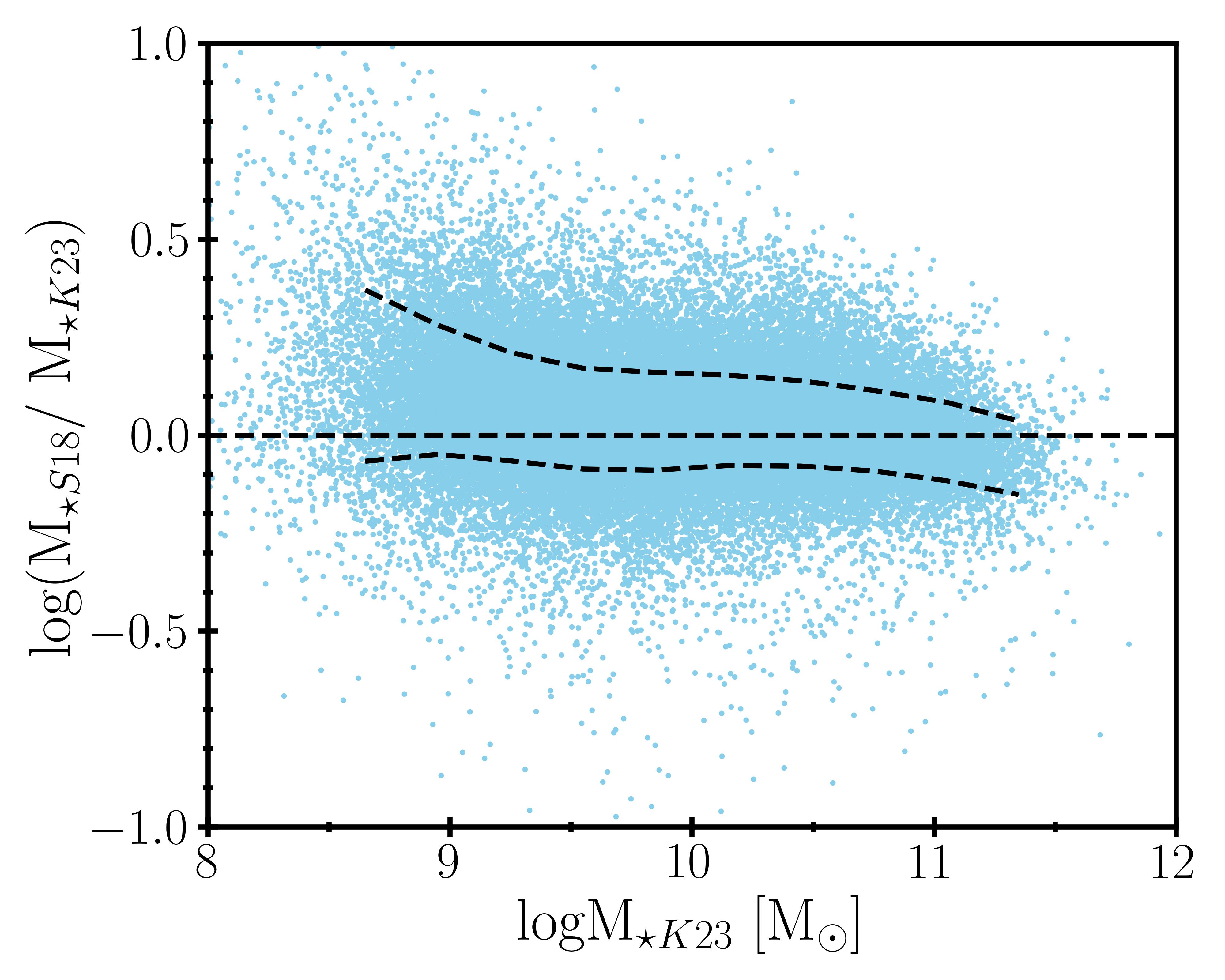}
        \includegraphics[width=0.47\textwidth]{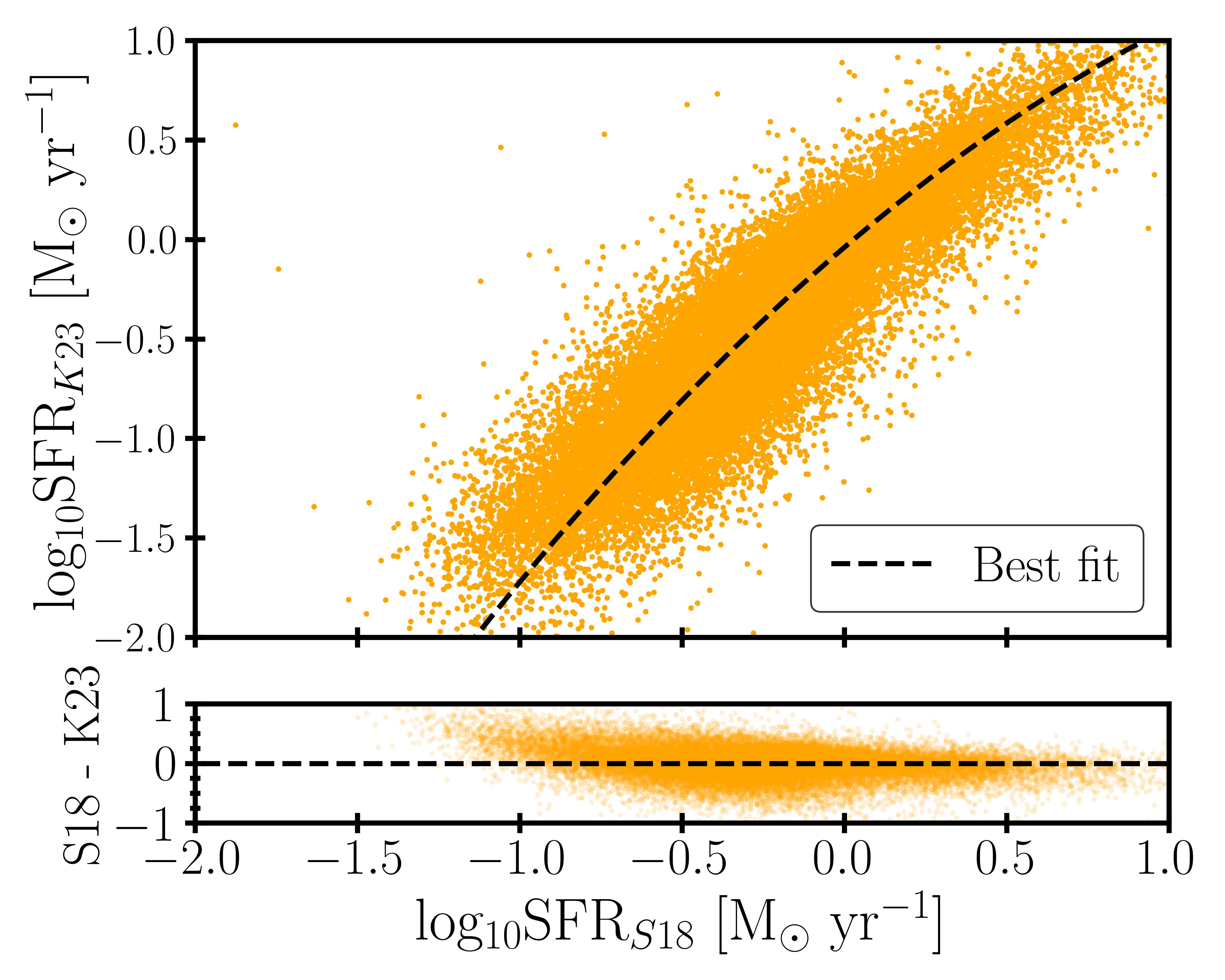}
        \includegraphics[width=0.47\textwidth]{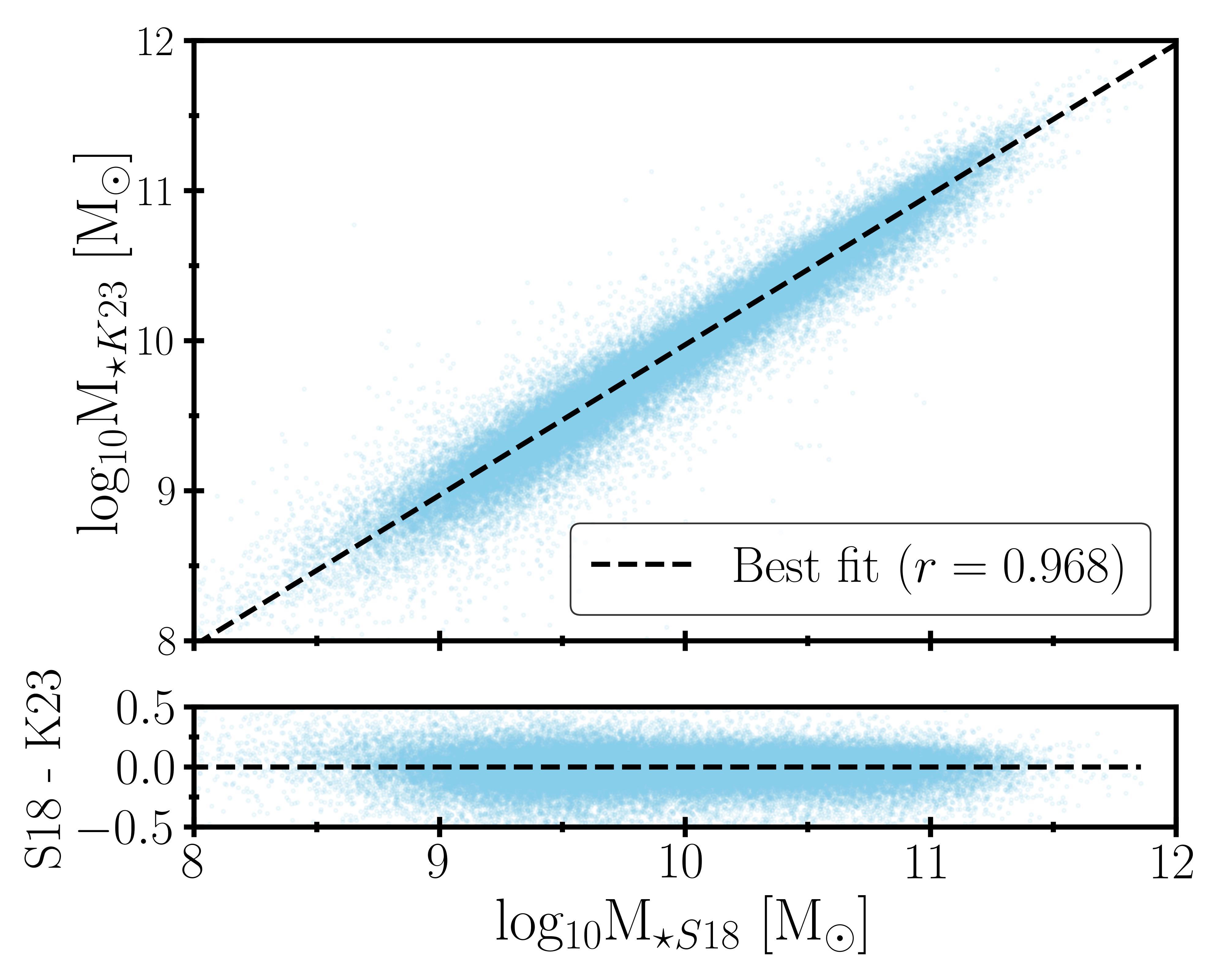}
    \caption{Top panel: Comparison between the \sfr{} (left panel) and \stellarmass{} (right panel) estimates derived from S18 and K23 calibrations. The y-axis shows the logarithmic ratio of the values (i.e. $\log_{10}$(SFR${\rm S18}$/SFR${\rm K23}$)), with the black dashed horizontal line at zero indicating equality between the two methods. The additional dashed lines represent the 16\% and 84\% percentiles, respectively.\\
    Bottom panel: Best-fit (black dashed line) for the rescaling of the \sfr{} (left panel) and \stellarmass{} (right panel) from S18 with respect to K23. We also show their corresponding residuals. To ensure that we include high-quality data in the fit, we selected only star-forming galaxies (probability $> 75\%$, see Sect. \ref{subsec:activity_class}) with S/N $\geq 5$ and S/N $\geq3$ in the W1, and W3 bands respectively. In addition, we only included galaxies from S18  with no problems in their UV/Optical/IR SED analysis (i.e. FLAG\_SED\_GSW = 0, FLAG\_MIDIR\_GSW $\neq$ 0). 
    }
    \label{fig:SFR_Mstar_K23_best_fit}
\end{figure*}

Using the distance-rescaled \sfr{} and \stellarmass{} values from S18, we compare them with the corresponding K23 measurements in the top panel of Fig.~\ref{fig:SFR_Mstar_K23_best_fit}. As shown in the top left panel, a systematic offset is observed between the S18 and K23 \sfr{} estimates. The S18 calibration tends to underestimate \sfr{} by $\sim$ 0.5 dex in the high-\sfr{} regime, while overestimating it by a similar factor in the low-\sfr{} regime. In contrast, as shown in the top right panel, the \stellarmass{} estimates from S18 are systematically lower than those from K23 by $\sim$ 0.3 dex across the entire stellar mass range. 

To correct for these offsets between the different methods we first rescaled the \sfr{} from S18 by using a second order polynomial of the form:

\begin{align}
\log_{10} \mathrm{SFR}_{\textrm{K23}}
   &= \alpha \times \bigl(\log_{10}\mathrm{SFR}_{\text{D-rescaled}}^{S18}\bigr)^{2} \notag \\
   &\quad + \beta \times \log_{10}\mathrm{SFR}_{\text{D-rescaled}}^{S18} + \gamma \,.
\label{eq:SFR_S18_resc_a}
\end{align}

We also tested a simple linear rescaling, but we found that it was not adequate to reproduce the full data distribution, in particular at high \sfr{}, motivating the adoption of a second-order polynomial.
The fit was performed using the Orthogonal Distance Regressor (ODR) provided by the \texttt{scipy} library \citep{virtanen20}, considering all galaxies with both K23 and S18 measurements. We included both the $E(B–V)$-dependent and $E(B–V)$-independent cases, to avoid introducing biases as the $E(B–V)$-dependent sample is skewed toward galaxies with higher \sfr{}s. To ensure that we include high-quality data in the fit, we selected only star-forming galaxies (probability $> 75\%$, see Sect. \ref{subsec:activity_class}) with S/N $\geq 5$ and S/N $\geq3$ in the W1, and W3 bands respectively. In addition, we only included galaxies from S18  with no problems in their UV/Optical/IR SED analysis (i.e. FLAG\_SED\_GSW = 0, FLAG\_MIDIR\_GSW $\neq$ 0). 
Similarly, for the rescaling of the \stellarmass{} from S18 we fitted the linear model of the form: 
\begin{equation}
\log_{10}M_{\star K23} = \alpha \times \log_{10}M_{\star, \textrm{D-rescaled}}^{S18} + \beta    
    \label{eq:Mstar_S18_resc_b}
\end{equation}
in all galaxies (not only star-forming) with with S/N $\geq 5$ in the W1 band and good quality S18 measurements. In the bottom pannel of Fig. \ref{fig:SFR_Mstar_K23_best_fit} we present the best-fit for the \sfr{} and \stellarmass{} as a function of the corresponding distance-rescaled \sfr{} and \stellarmass{} from S18. The best-fit parameters and their RMSE are listed in Table \ref{tab:SFR_Mstar_rescaling_best_fit}. By using these values we rescaled the \sfr{} and \stellarmass{}, accordingly.  

\begin{table}
\centering
\caption{Best-fit parameters for the rescaling of the \sfr{} and \stellarmass{} estimates from S18 with respect to K23.}
\label{tab:SFR_Mstar_rescaling_best_fit}
\begin{tabular}{ccccc}
\hline\hline
Parameter & $\alpha$ & $\beta$ & $\gamma$ & RMSE \\\hline
$\log$\sfr{}$^{\textrm{K23}}$& -0.290 & 1.392 & -0.039&0.276 \\
$\log$\stellarmass{}$^{\textrm{K23}}$ & 1.005 &-0.078  & - &0.166 \\
\hline
\end{tabular}
\medskip
\parbox{\columnwidth}{\footnotesize
Note: RMSE represents the Root Mean Square Error of our fits per photometric band.
}
\end{table}

As discussed in Sect.~\ref{subsec:SFR_Mstar_S18_K23}, no additional rescaling was applied to the adopted \sfr{} and \stellarmass{} values from L19 to match the K23 calibrations, apart from the adjustment for differences in distance estimates. This is because differences in the photometric apertures for these nearby galaxies dominate over any inconsistencies between the calibration methods.

\section{Study the completeness of HECATEv2 as a function of distance, luminosity, and sky area}\label{append:completeness_per_lum_bin_sky_patch}
To study the completeness of HECATEv2 by number we also compared the observed number of galaxies in different distance and luminosity bins with expectations from the corresponding LFs in the B and Ks bands. The expected number of objects ($\rm{N}_{\rm{bin}}$) under a Shechter LF in a given luminosity bin and in a given distance bin can be computed as 

\begin{align}
N_{\rm bin} = \Delta N
  &= \int_{L_{1}}^{L_{2}} \phi(L)\, dL \times \Delta V \notag \\
  &= \phi^{\ast}\,\biggl[\gamma\!\left(\alpha + 1, \tfrac{L_{2}}{L^{\ast}}\right)
     - \gamma\!\left(\alpha + 1, \tfrac{L_{1}}{L^{\ast}}\right)\biggr] \times \Delta V
\end{align}
where L$_{1}$, L$_{2}$ are the lower and upper bounds  of the luminosity bin and $\Delta V$ is the volume element
$\Delta V = \frac{4}{3}\pi\big[D_{2}^{3} - D_{1}^{3}\big]$. The $\phi^{\ast}$, $\gamma$, and L$^{\ast}$ are the best-fit parameters of the corresponding Shechter Functions in the B and Ks bands (see Sect. \ref{sec:completeness}). After dividing our sample into 20 equally spaced distance bins over the distance range of HECATEv2 (D = 0-200 \distanceunits{}) and luminosity bins spanning the range $10^{7}$–$10^{12}$ $L_{\odot}$ with a step of 0.2 dex, we performed the numerical integration using the \texttt{quad} function from the \texttt{scipy} python library \citep{virtanen20}. 

\begin{figure*}
 \centering
        \includegraphics[width=0.97\textwidth]{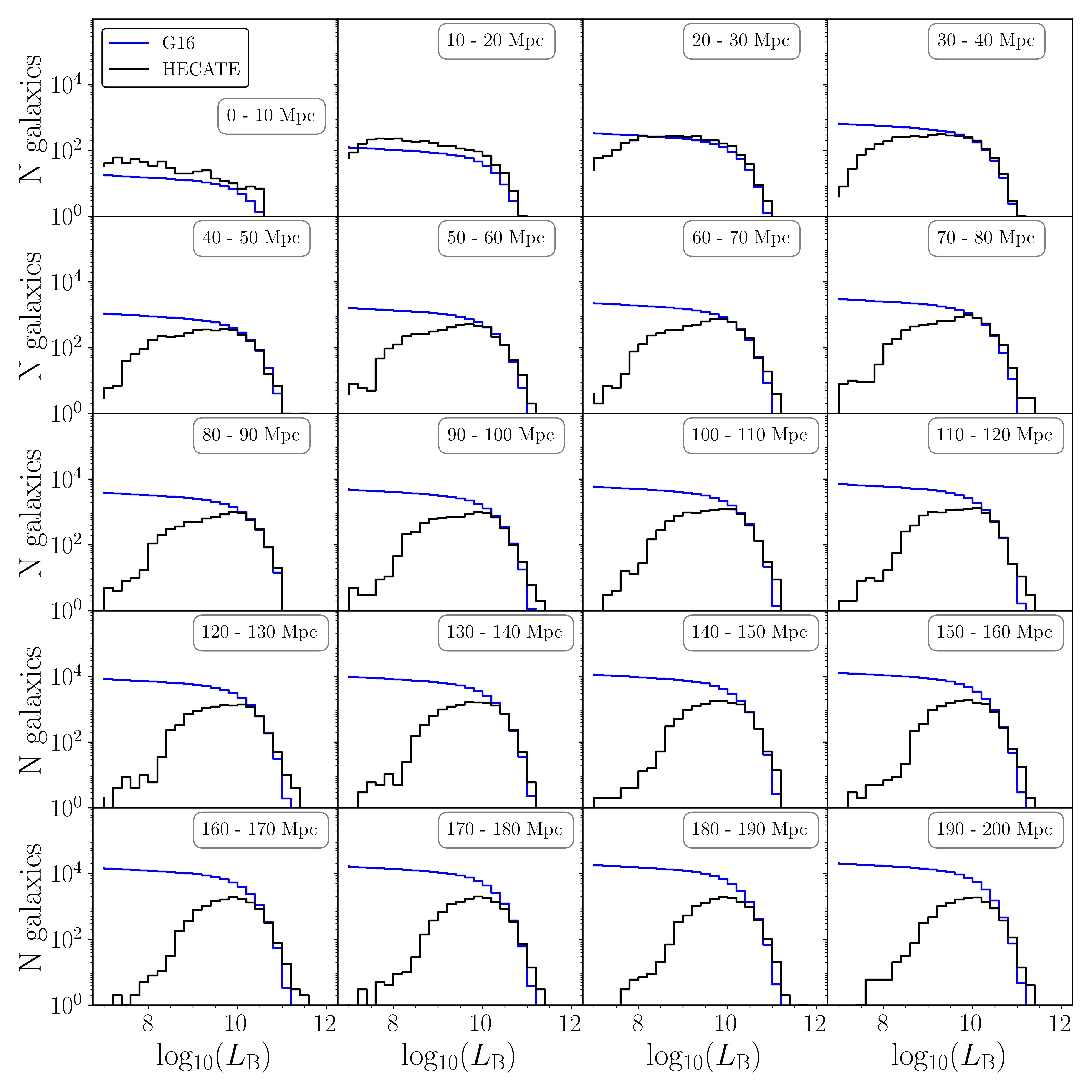}
    \caption{The distribution of B-band luminosities for objects in HECATEv2 (black histogram) in distance bins from 0 to 200 Mpc. The blue histogram represents the expected distribution derived from the B-band LF of \protect\cite{gehrels16} (G16).}
    \label{fig:Completeness_hist_B}
\end{figure*}

\begin{figure*}
 \centering
        \includegraphics[width=0.97\textwidth]{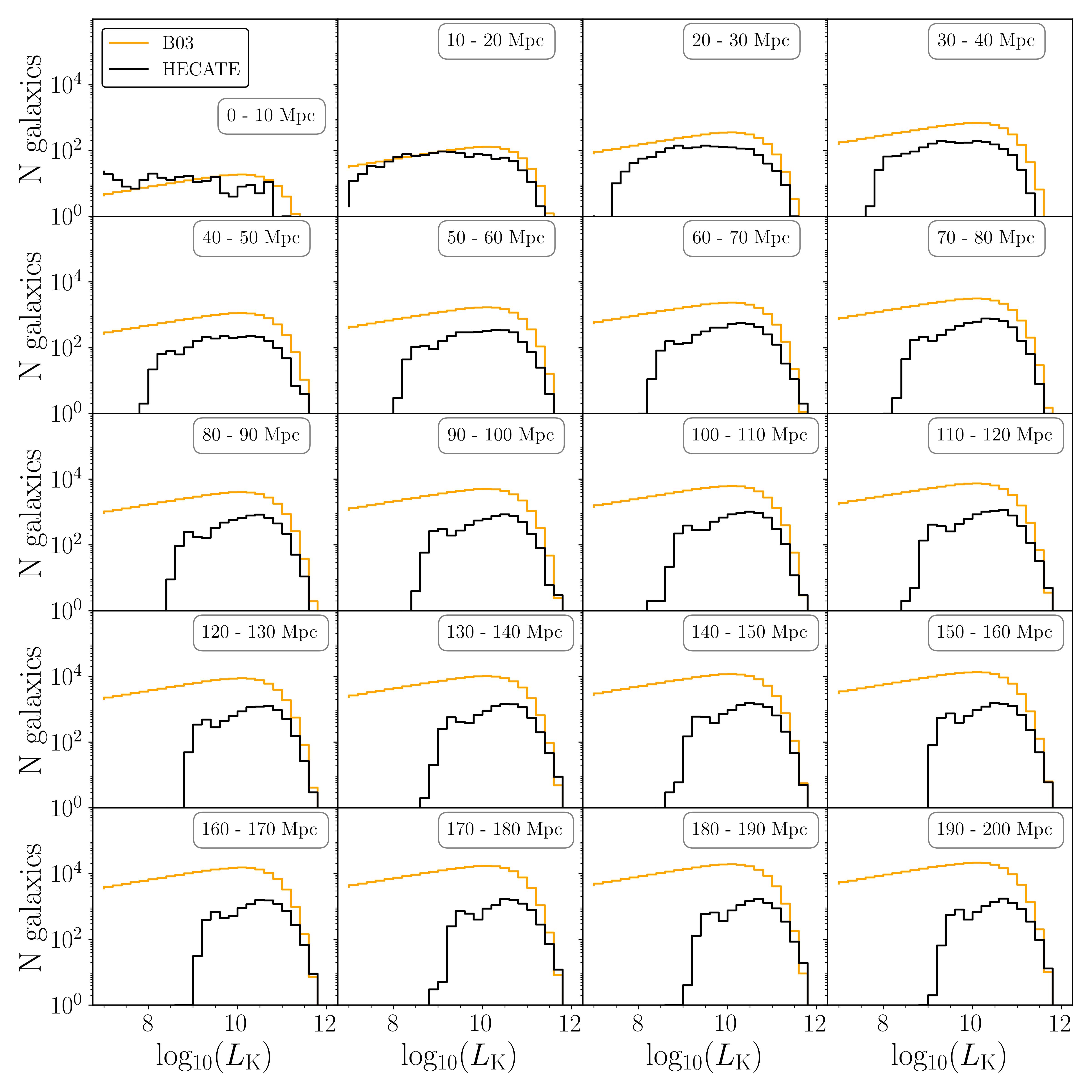}
    \caption{The distribution of Ks-band luminosities for objects in HECATEv2 (black histogram) in distance bins from 0 to 200 Mpc. The orange histogram represets the expected distribution derived from the Ks-band LF of \protect\cite{bell03} (B03)}
    \label{fig:Completeness_hist_Ks}
\end{figure*}

In Fig. \ref{fig:Completeness_hist_B} and Fig. \ref{fig:Completeness_hist_Ks}
we present the corresponding luminosity distribution of the galaxies in the HECATEv2 with B and the Ks-band luminosities in distance bins from 0 to 200 Mpc. We find that HECATEv2 is complete down to L$_{B} \sim 10^{7.1} \, \rm L_{B,\odot}$ in the nearest bin (i.e. 0$<$D$<$10 Mpc) and down to L$_{B} \sim 10^{10.7} \, \rm L_{B,\odot}$ in the distance bin 40$<$D$<$50 Mpc. At distances beyond 100 Mpc, our catalogue is 100\% complete for galaxies more luminous than L$_{B} \sim 10^{11.1} \, \rm L_{B,\odot}$. Similarly, HECATEv2 is 100\% complete down to L$_{Ks} \sim 10^{8.3} \, \rm L_{Ks,\odot}$ within 10$<$D$<$20 Mpc, and remains $>$50\% complete down to L$_{Ks} \sim 10^{9} \, \rm L_{Ks,\odot}$ for D $>70$ Mpc. 

\begin{figure*}
    \centering
        \includegraphics[width=0.49\textwidth]{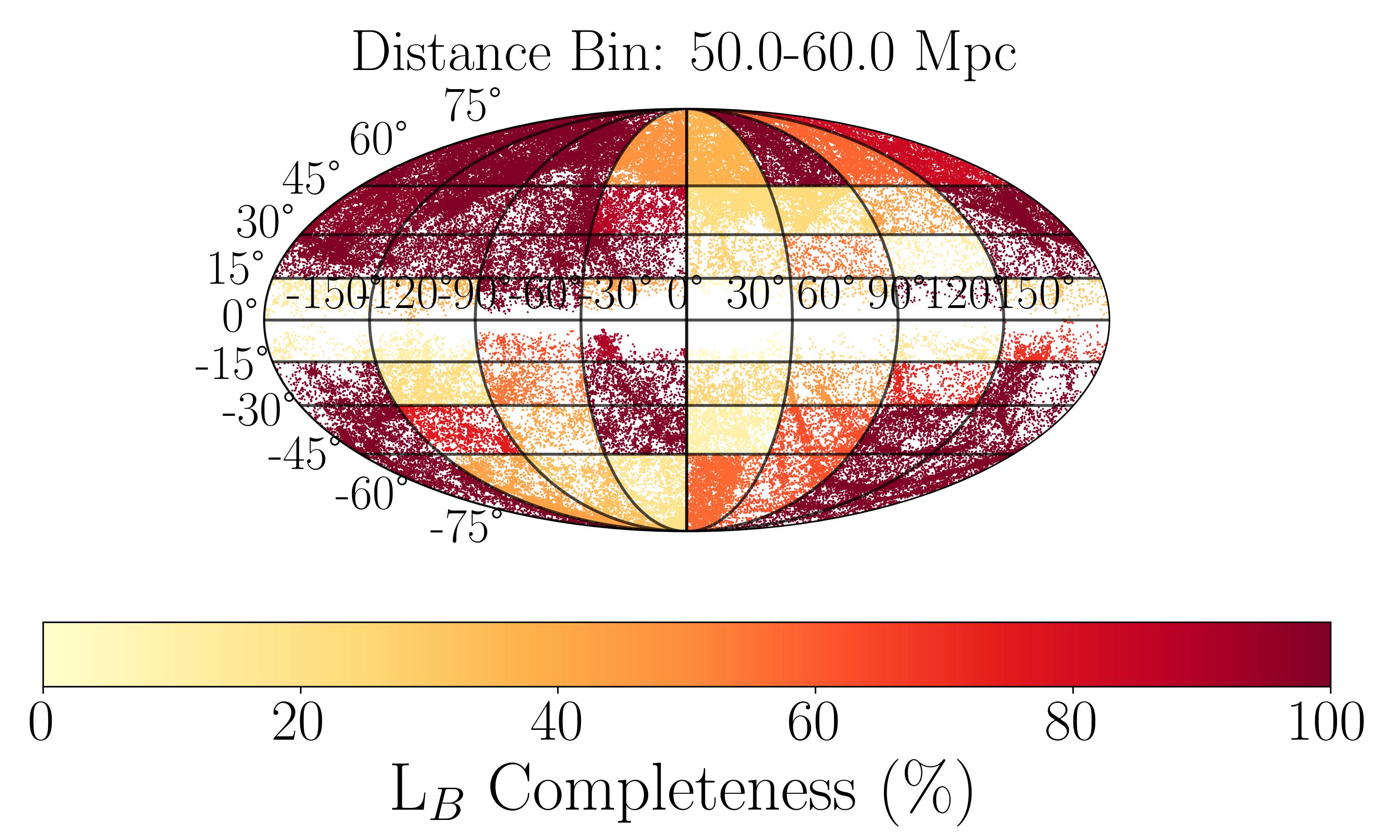}
        \includegraphics[width=0.49\textwidth]{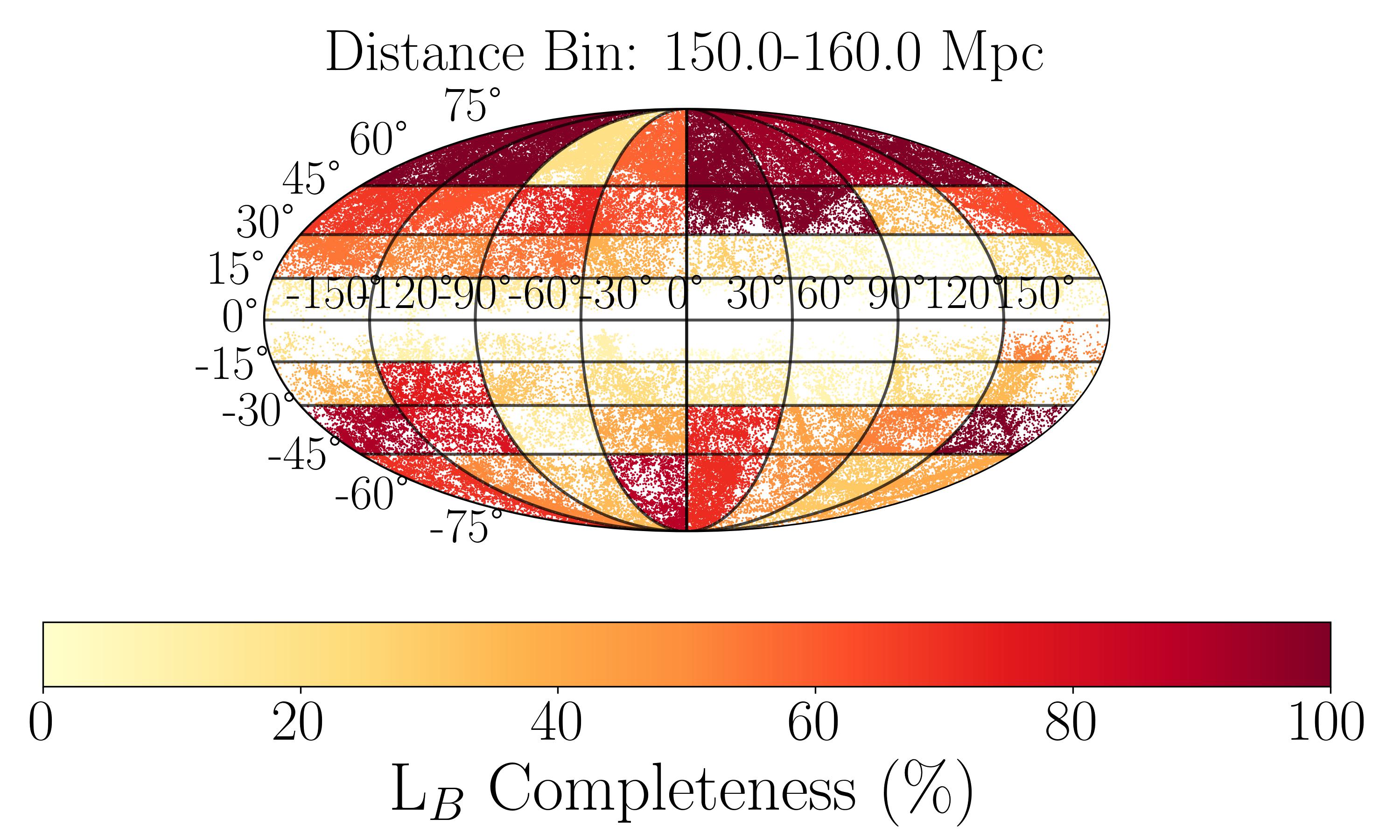}
        \includegraphics[width=0.49\textwidth]{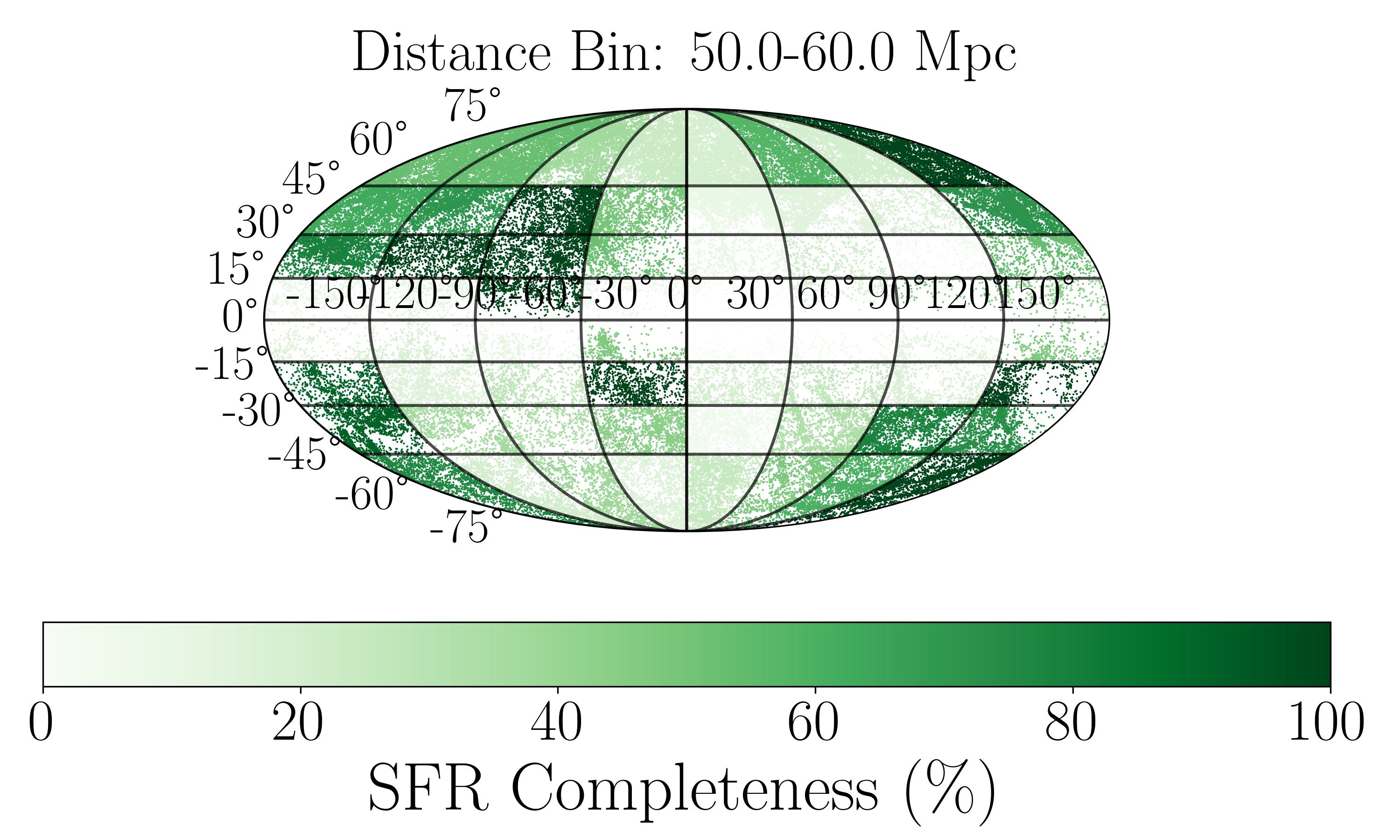}     
       \includegraphics[width=0.49\textwidth]{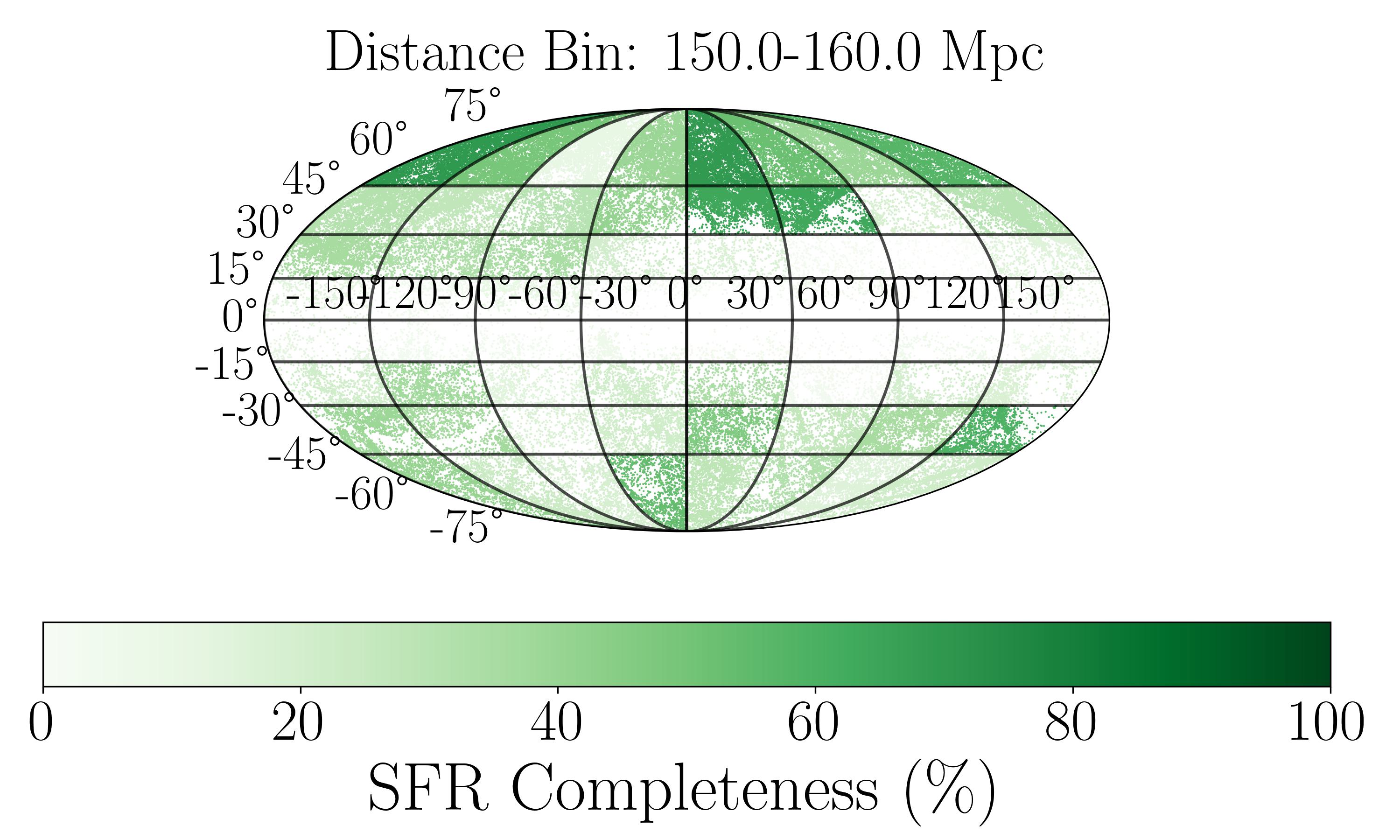}
        \includegraphics[width=0.49\textwidth]{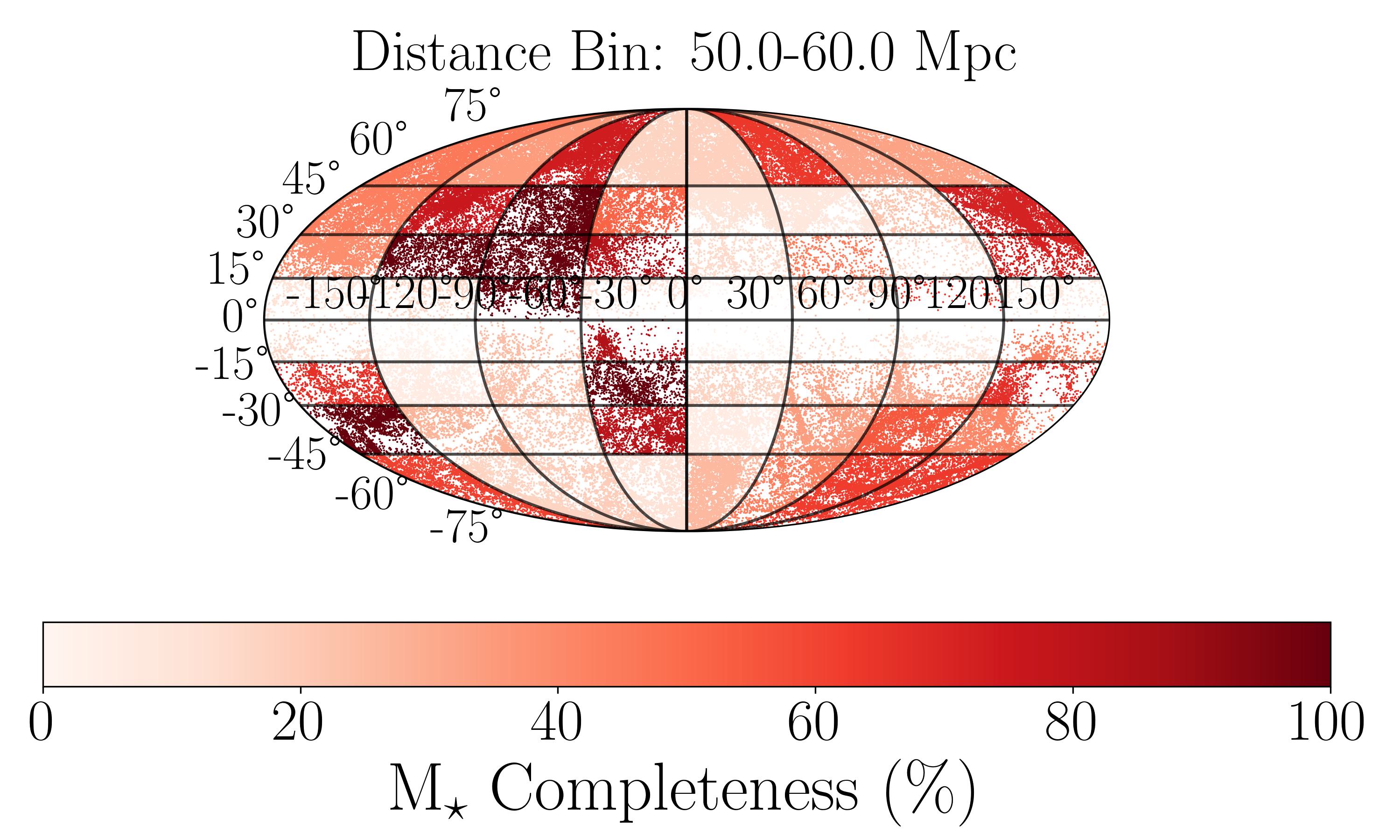}   
       \includegraphics[width=0.49\textwidth]{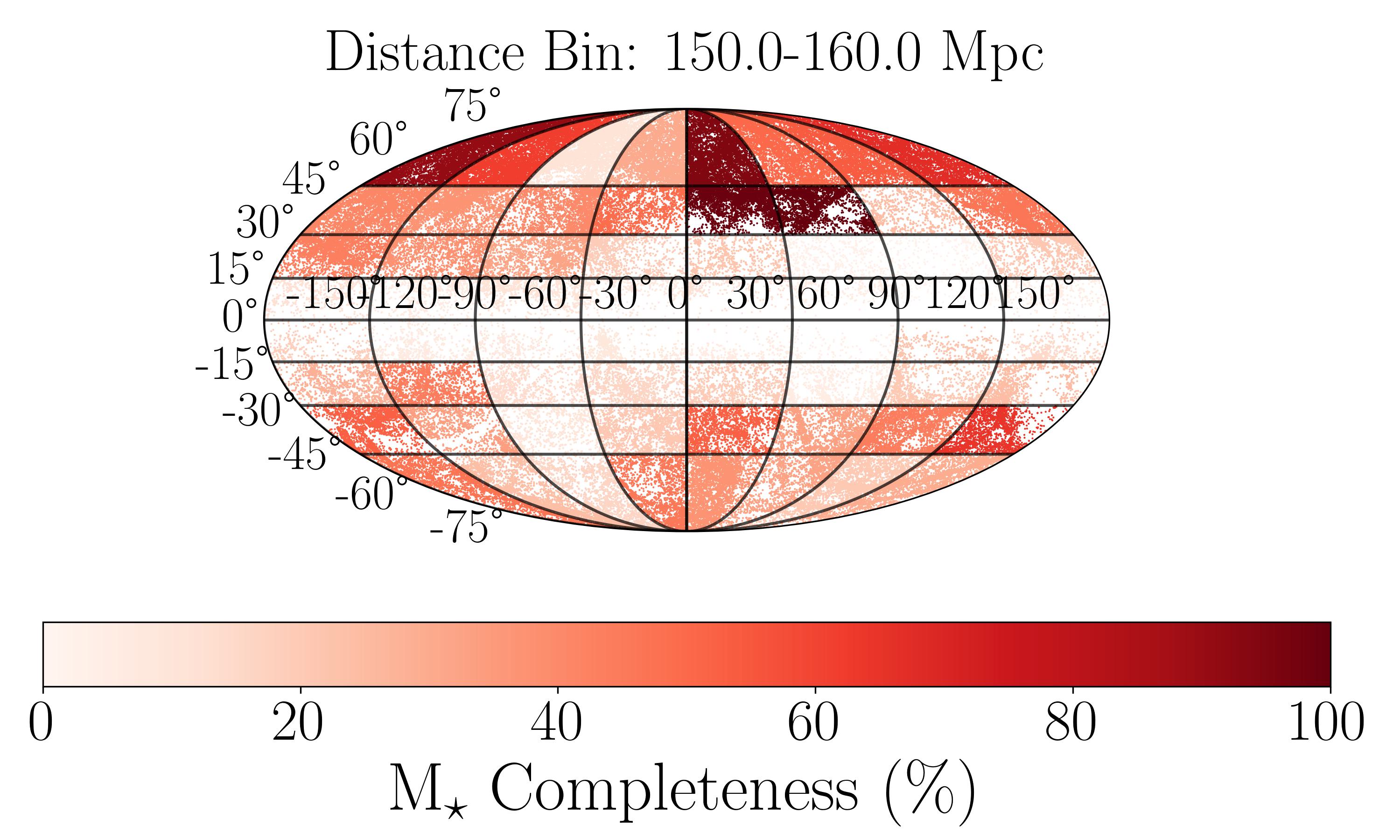}
        \includegraphics[width=0.49\textwidth]{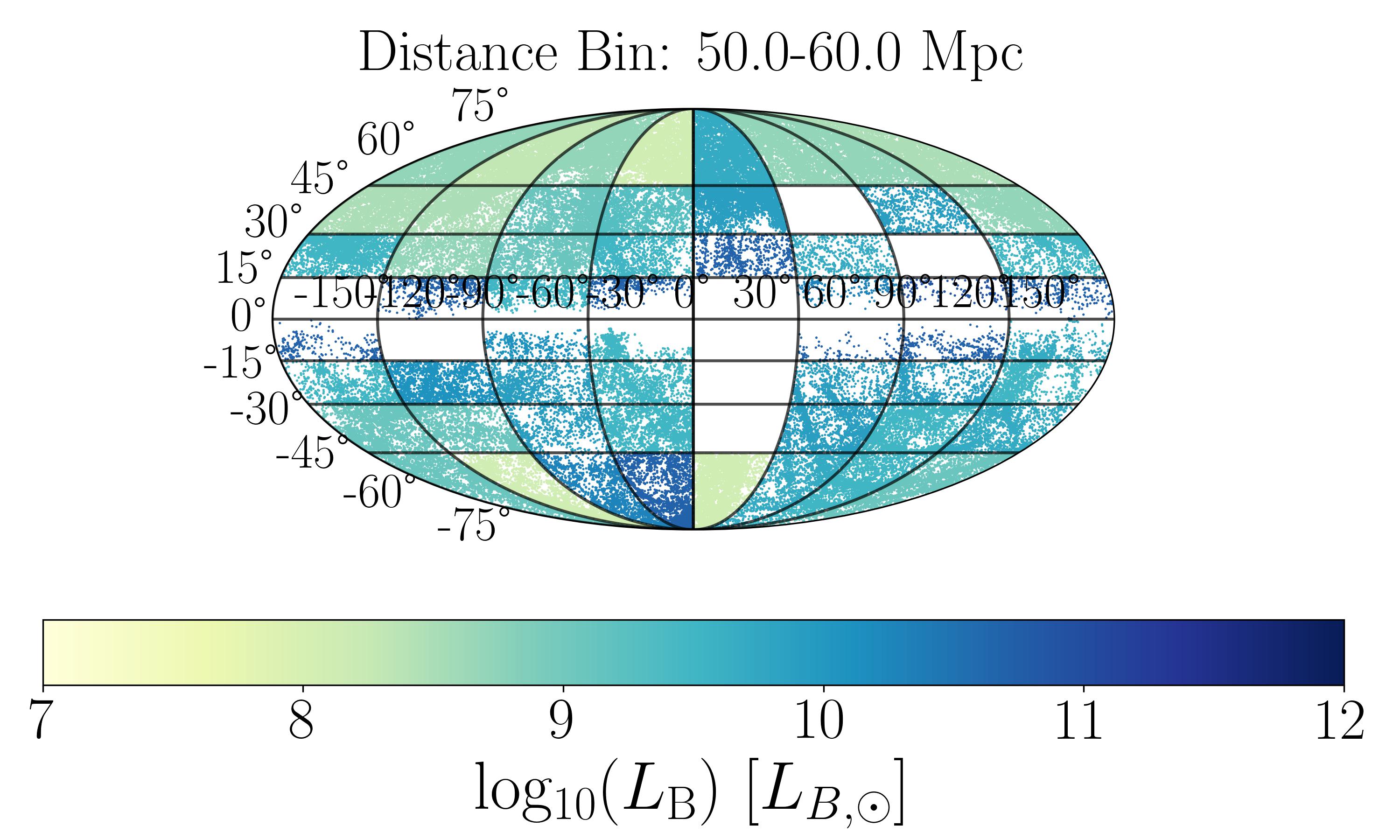}
       \includegraphics[width=0.49\textwidth]{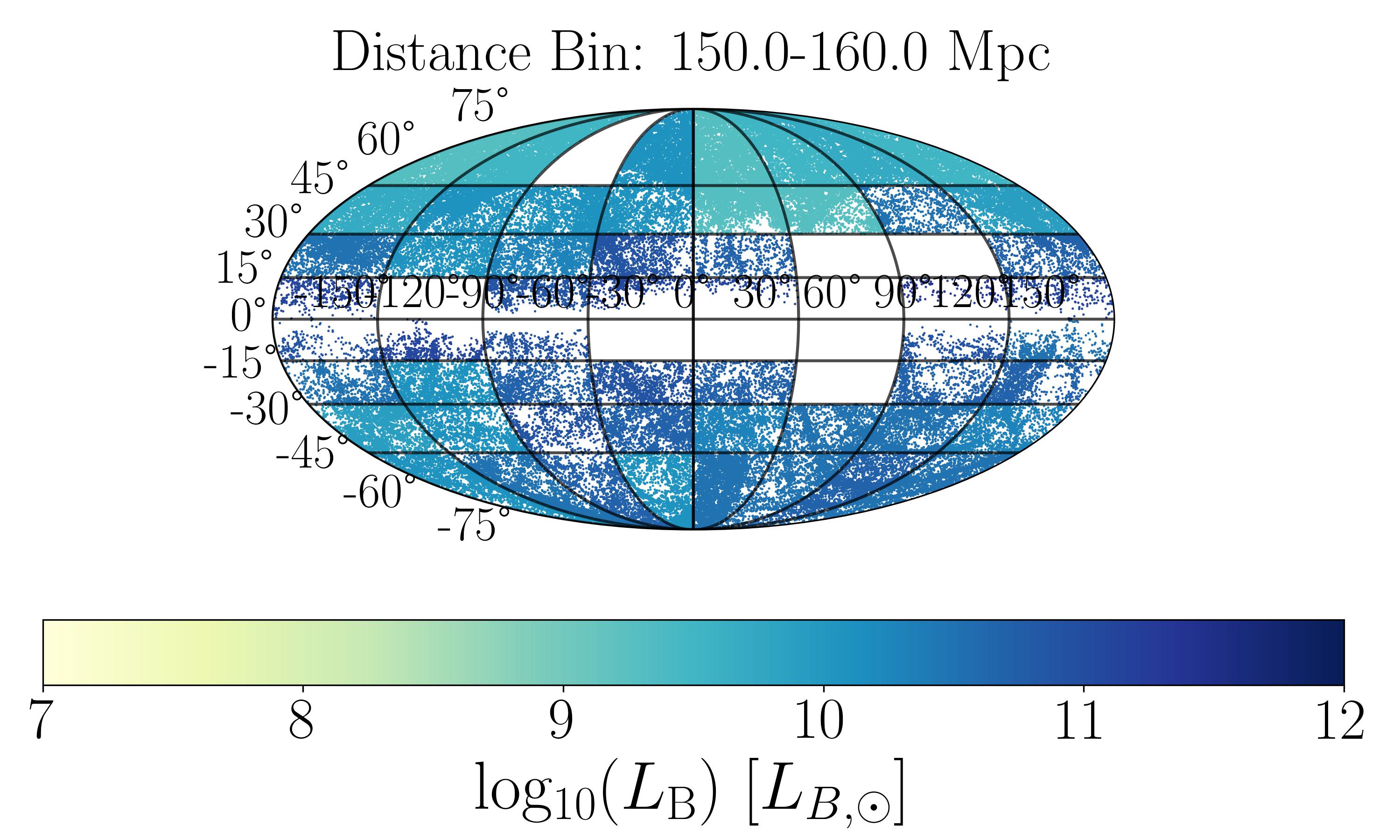}      
    \caption{Completeness of HECATEv2 per sky-patches and per distance bin in terms of B-band luminosity, \sfr{} and \stellarmass{}. The completeness maps are not smoothed to reflect the actual completeness in each patch, and for this reason there may be sharp transitions between neighboring patches. We only present representative examples for two distance bins, 50$<$D$<$60 Mpc and 150$<$D$<$160 Mpc, illustrating how the completeness varies across different sky regions and different distances. The bottom panel presents the sky distribution of the L$_{B}$ threshold above which the completeness exceeds 50\%. The full set of the L$_{B}$-, \sfr{}, \stellarmass{} completeness sky maps per Distance bin is available online.}
    \label{fig:completeness_per_sky}
\end{figure*}

HECATEv2 incorporates photometric data from a wide range of multi-wavelength surveys (e.g. SDSS-DR17, PS1-DR2, AllWISE), each with differing sky coverage and depth. Consequently, the catalogue’s completeness is not uniform across the sky. To quantify this spatial variation, we divided the celestial sphere into 64 equal-area sky patches. For each patch, we estimated the completeness using again the Equation \ref{eq:completeness_per_Dbin} scaled by the fractional sky area covered in each patch as:
\begin{equation}
\text{Completeness per sky-patch} = \frac{\sum_{i=1}^{N}L_{\rm{band}}^{i}}{p_{L}\times V_{\rm{D,bin}} \times S^{sky-patch}}
\end{equation}\label{eq:completeness_per_Dbin_per_sky}
where $S^{sky-patch} = \frac{\Delta \Omega^{\rm{sky-patch}}}{\Omega^{\text{total sky}}} = \frac{1}{64}$
is the fractional sky area of each patch. By using again the same 20 distance bins we calculated the completeness per sky region in the B-band, as well as in terms of \sfr{} and in terms of \stellarmass{}. In addition, for each sky patch we determined the L$_{B}$ threshold above which the completeness exceeds 50\%. Figure \ref{fig:completeness_per_sky} shows representative examples of this analysis for two distance bins, 50$<$D$<$60 Mpc and 150$<$D$<$160 Mpc, illustrating how the completeness varies across different sky regions and different distances. These variations reflect the non-uniform depth and sky coverage of the different surveys used in HECATEv2. Table~\ref{tab:HECv2_Lb_complet_per_D_bin} and Table~\ref{tab:HECv2_SFR_Mstar_complet_per_D_bin} present a subset of the L$_{B}$, \sfr{}, and \stellarmass{} completeness per distance bin. Table~\ref{tab:HECv2_Lb_complet_per_D_bin_per_sky}, and Table~\ref{tab:HECv2_SFR_Mstar_complet_per_D_bin_per_sky} present a subset of the L$_{B}$, \sfr{}, and \stellarmass{} luminosity per distance bin and per sky patch. The full tables of the completeness estimates  per distance bin and per sky region in terms of B-band luminosity, \sfr{}, \stellarmass{} are provided as online supplementary tables and are available through Vizier. 

\begin{table}
\centering
\caption{HECATEv2 L$_{B}$ completeness per Distance bin. The full table is available as supplementary online material.}
\label{tab:HECv2_Lb_complet_per_D_bin}
\begin{tabular}{cc}
\hline\hline
D-bin  &  L$_{B}$\\
 & Completeness\\
Mpc &   (\%) \\\hline
0.0-10.0  & 218.4 \\
10.0-20.0 & 200.3\\
20.0-30.0 & 141.1\\
30.0-40.0 & 103.1\\
40.0-50.0 & 82.20\\
...&...\\
190.0-200.0 & 34.1\\
\hline
\end{tabular}
\end{table}

\begin{table}
\centering
\caption{HECATEv2 \sfr{} and \stellarmass{} completeness per Distance bin. The full table is available as supplementary online material.}
\label{tab:HECv2_SFR_Mstar_complet_per_D_bin}
\begin{tabular}{ccc}
\hline\hline
D-bin  & \sfr{} & \stellarmass{}\\
& Completeness& Completeness\\
Mpc & (\%)  &(\%)  \\\hline
0.0-10.0  & 147.6 & 72.40 \\
10.0-20.0 & 124.1 & 96.7\\
20.0-30.0 & 78.0 & 71.1\\
30.0-40.0 & 63.3 & 49.2\\
40.0-50.0 & 84.2 & 44.3\\
...&...&...\\
190.0-200.0 & 24.1 & 22.8\\
\hline
\end{tabular}
\end{table}

\begin{table}
\centering
\caption{HECATEv2 L$_{B}$ completeness per Distance bin and per sky patch. The full table is available as supplementary online material.}
\label{tab:HECv2_Lb_complet_per_D_bin_per_sky}
\begin{tabular}{ccccc}
\hline\hline
D-bin & (l$_{1}$,l$_{2}$) & (b$_{1}$,b$_{2}$) & L$_{B}$ \\
 & & & Completeness\\
Mpc & deg & deg & (\%)\\\hline
0.0-10.0 & (0.0,45.0) & (-90.0,-48.59) & 75.4\\
0.0-10.0 & (0.0,45.0) &(-48.59,-30.0) &0.1\\
0.0-10.0 & (0.0,45.0) &(-30.0,-14.478)&0.5\\
0.0-10.0 & (0.0,45.0) &(-14.478,0.0) &4.2\\
0.0-10.0 & (0.0,45.0) &(0.0,14.478) &4.3\\
...&...&...&...\\
190.0-200.0 & (315.0,360.0) & (-14.478,0.0) & 5.8\\
190.0-200.0 & (315.0,360.0) & (0.0,14.478) & 4.0\\
190.0-200.0 & (315.0,360.0) & (14.478,30.0) & 13.9\\
190.0-200.0 & (315.0,360.0) & (30.0,48.59) & 54.7\\
190.0-200.0 & (315.0,360.0) & (48.59,90.0) & 40.9\\

\hline
\end{tabular}
\medskip
\parbox{\columnwidth}{\footnotesize
Note: l,b are in galactic coordinate system.
}
\end{table}

\begin{table}
\centering
\caption{HECATEv2 \sfr{} and \stellarmass{} completeness per Distance bin and per sky patch.The full table is available as supplementary online material.}
\label{tab:HECv2_SFR_Mstar_complet_per_D_bin_per_sky}
\setlength{\tabcolsep}{4pt}
\begin{tabular}{cccccc}
\hline\hline
D-bin & (l$_{1}$,l$_{2}$) & (b$_{1}$,b$_{2}$) & \sfr{}& \stellarmass{} \\
 &  &  & Completeness  & Completeness \\
Mpc & deg & deg & (\%) & (\%)\\\hline
0.0-10.0 & (0.0,45.0) & (-90.0,-48.59) & 33.7 & 14.9\\
0.0-10.0 & (0.0,45.0) &(-48.59,-30.0) &0 & 0\\
0.0-10.0 & (0.0,45.0) &(-30.0,-14.478)&0 & 0\\
0.0-10.0 & (0.0,45.0) &(-14.478,0.0) &6.2 & 0.6\\
0.0-10.0 & (0.0,45.0) &(0.0,14.478) &0 & 0\\
...&...&...&...&...\\
190.0-200.0 & (315.0,360.0) & (-14.478,0.0) & 5.2 & 4.5 \\
190.0-200.0 & (315.0,360.0) & (0.0,14.478) & 2.9 & 1.8 \\
190.0-200.0 & (315.0,360.0) & (14.478,30.0) & 8.6 & 8.1 \\
190.0-200.0 & (315.0,360.0) & (30.0,48.59) & 33.3 & 42.2 \\
190.0-200.0 & (315.0,360.0) & (48.59,90.0) & 23.6 & 23.7 \\ 
\hline
\end{tabular}
\medskip
\parbox{\columnwidth}{\footnotesize
Note: l,b are in galactic coordinate system.
}
\end{table}

\section{Column description of HECATEv2}\label{append:HECv2_columns}

In this section, we describe the columns of HECATEv2. Table~\ref{tab:columns} provides an overview of all columns, while the following paragraphs give additional details for specific entries.
\begin{itemize}
    \item RFLAG: Provides information for the availability of galaxy sizes. A value of 0 indicates that both R1 and R2 were not available; in such cases, a fiducial median circular aperture was adopted, based on photometric data from PS1‑DR2. A value of 1 means that all size parameters are well-defined. A value of 2 indicates that either R2 or the position angle (PA) was missing, and a circular aperture with radius R1 was adopted instead. A value of 3 or 4 signifies that our visual inspection identified a \textit{large} or \textit{wrong} aperture, respectively. In some cases, more than one flag value may coexist. For instance, an entry with RFLAG=
    "14" means that the size information is defined, but visual inspection revealed that it is Wrong. Such cases should therefore be treated with caution. For more details, see Sect.~\ref{subsec:sizes}.
    \item R\_ORIGIN: Provides information for the origin of the size information. For the galaxy sizes retained from the HECATEv1 we adopted their origin flags as described in Table B1 in \cite{kovlakas21}. For the additional galaxy sizes that we provide in HECATEv2 the origin flags are the following:  "PS1 - circular/median", "PS1 - elliptical", "PS1 - circular" refer to the sizes derived using photometric information from  PS1-DR2 as it is described in Sect.~\ref{subsec:sizes}. The flag "SGA-20" means that the size is adopted from SGA catalogue (see Sect.~\ref{subsect:size_comparison}). 
    \item DIST\_METH: Methodology for the distance derivation. A value of 0 indicates a redshift-independent distance adopted from HECATEv1. A value of 1 means that the galaxy is not a Virgo Cluster member and the distance is calculated from the cosmological model. A value of 2 indicates that the galaxy is a Virgo Cluster member and the distance is calculated based on the Virgo Cluster model. Finally a value of 3 signifies a fiducial distance adopted for the blushifted galaxies which are not Virgo Cluster members. We recommend the users to treat distances flagged as "3", with caution. For more details see Sect.~\ref{sec:distances}.
    \item ALLWISE\_QF: Eight-character AllWISE quality flag. A value of 1 means that the photometry is reliable and a value of 0 means unreliable photometry. The first four characters refer to detector artifacts in W1\_m–W4\_m and last four to image artifacts in W1\_m–W4\_m. For example a flag of "10101010" indicates that the galaxy has reliably photometry in W1\_m and W3\_m bands and unreliable photometry due to detector and image artifacts in the W2 and W4 bands. For more details, see Sect.~\ref{subsec:midir_phot_surveys_AllWISE}.
    \item OPTPHOT\_ORIGIN: Six-character origin flag of the homogenised optical photometry with each character corresponding to each of the ugrizy bands. A value of "S" means that the photometry was adopted from SDSS-DR17 and a value of "N" means that the photometry originates from NSA. A value of "P" means that the photometry was adopted from PS1-DR2. Finally, a value of "M" means that the photometry is missing. For example a flag of "SSSSSP" means that SDSS-DR17 photometry was adopted for the u,g,r,i,z - bands and PS1-DR2 photometry for the Y-band. For more details, see Sect.~\ref{subsec:optical_phot_homogenization}
    \item OPTPHOT\_QF: Six-character quality flag of the homogenised optical photometry with each character corresponding to  each of the ugrizy bands. A value of 1 means that the photometry is reliable and a value of 0 means unreliable photometry. A value of 2 indicates that the photometry is missing. For example a quality flag of the form "211101" means that the U-band is missing, the g,r,i,y-bands are reliable, and z-band is unreliable. For more details, see Sect.~\ref{subsec:optical_phot_homogenization}.
    \item <LINE>\_FLUX, <LINE>\_FLUX\_ERR: Spectral line flux and its uncertainty. In the current version of the HECATE catalogue we only provide line fluxes for the lines adopted from the MPA-JHU DR8 catalogue. In future releases we will also provide line fluxes from our own customise analysis of interesting objects and othe spectroscopic surveys. The available lines are the following: <LINE> = H\_BETA, OIII\_5007, OI\_6300, H\_ALPHA, NII\_6584, SII\_6717, SII\_6731, respectively. For more details see Sect.~\ref{subsec:optical_spec_data}.
    \item SPEC\_ORIGIN: Indicates the origin of the spectroscopic information included in HECATEv2. A value of "SDSSDR8" means that the spectroscopic data adopted from the MPA-JHU DR8 catalogue. A value of "SDSSDR17", or "SKI" indicate additional spectroscopic data obtained for other projects from the SDSS-DR17 database or observing runs at Skinakas Observatory, respectively. For more details see Sect.~\ref{subsec:optical_spec_data}.
    \item  ACTIVITY\_CLASS\_METH: Indicates the diagnostic that was used for the nuclear activity classification. A value of "SoDDA" means that the classification is based on the four-dimensional spectroscopic diagnostic from \cite{stampoulis19}, while a value of "SVM3d" means that the three-dimensional SVM-based spectroscopic diagnostic from \cite{stampoulis19} was used. Finally, a value of "PH" indicates that the classification is provided by the optical/mid-IR photometric diagnostic of \cite{daoutis23}.\\
    \item LOG\_SFR\_HEC\_QF: This column provides a quality flag for the adopted value of the \sfr{}. Specifically, for galaxies whose \sfr{} is adopted from L19 \citep{leroy19}, the flag values are taken from the "Flags" column in Table 4 of L19. For galaxies whose \sfr{} values are adopted from S18 \citep{salim16,salim18} we retained the flags from the "flag\_sed" column in Table 1 of S18. In general, users are advised to treat with caution all entries where LOG\_SFR\_HEC\_QF $\neq$ 0 or LOG\_SFR\_HEC\_QF $\neq$ NULL, as these values may be uncertain or unreliable.
    \item LOG\_SFR\_HEC\_METH: This column indicates the method used for the derivation or adoption of the \sfr{}. For galaxies whose \sfr{} values are adopted from the SED fits of S18, the flag is "UVOIR-SED". For galaxies where we adopt the hybrid \sfr{} estimates from L19, we retained the same flags as described in their Table 4 ("SFR Method" column), namely "FUV+WISE4", "WISE4", and "NUV+WISE4". Finally, for galaxies with \sfr{} values derived from the K23 calibrations, we use the flags "EBV" and "no-EBV", depending on whether the extinction-dependent or extinction-independent calibrations of K23 were applied (see Sect.~\ref{subsec:SFR_Mstar_S18_K23}).
    \item LOG\_MSTAR\_HEC\_QF: This column provides a quality flag for the adopted value of the \stellarmass{}. For galaxies whose \stellarmass{} values adopted from L19 we retained the flags from column "Flags" in L19, and the column "flag\_sed" from S18. In addition, galaxies flagged as "C" have suspiciously low \stellarmass{} values, and as it is discussed in Sect.~\ref{subsec:homo_SFR_Mstar} they should be used with caution. In general, users are advised to treat with caution all entries where LOG\_MSTAR\_HEC\_QF $\neq$ 0 or LOG\_MSTAR\_HEC\_QF $\neq$ NULL, as these values may be uncertain or unreliable.
    \item LOG\_MSTAR\_HEC\_METH: This column indicates the method used for the derivation or adoption of the \stellarmass{}. For galaxies whose \stellarmass{}{} values are adopted from the SED fits of S18, the flag is "UVOIR-SED". For galaxies where we adopt the hybrid \stellarmass{} estimates from L19, we retained the same flags as described in their Table 4 ("$\Upsilon^{3,4}_{\star}$ Method" column), namely "W4W1", "FIXEDUL", and "SSFRLIKE". Finally, for galaxies with \stellarmass{}{} values derived from the K23 calibrations, we use the flags "u\_r\_SDSSDR17\_EBV", "u\_r\_SDSSDR17\_noEBV", "g\_r\_SDSSDR17\_EBV","g\_r\_PSDR2\_EBV", "g\_r\_PSDR2\_noEBV", "u\_r\_SDSSDR17\_EBV\_median", and "u\_r\_SDSSDR17\_noEBV\_median" depending on the availability of the optical colors and the extinction measurements (see Sect.~\ref{subsec:SFR_Mstar_S18_K23}). 
    \item METAL\_METH: This column specifies the method used for deriving the gas-phase metallicity. Galaxies with metallicities obtained from the spectroscopic calibration of C20 \citep{curti20}, based on the O3N2 line ratio, are flagged as "O3N2/C20". Galaxies with metallicities estimated using the mass–metallicity relation from C20 are flagged as "MZ/C20" (see Sect.~\ref{subsec:metal_estimates}).
    \item LOG\_MBH\_ORIGIN: This column specifies the scaling relation from \cite{green20} used for the calculation of the M$_{\rm BH}$. More specifically, for the star-forming and composite galaxies we adopted the scaling relation corresponding to the late-type galaxies ('G20-Late'), while for passive galaxies we used the prescription for early-type galaxies ('G20-Early'). For AGN and LINERs we applied the global scaling relation ('G20-All').
    \item DUP\_FLAG: This column identifies whether a galaxy appears under two different PGC numbers within the catalogue and provides the PGC number of the corresponding duplicate. Users are advised to use the galaxy entry with the lowest PGC number (see Sect.~\ref{subsec:sizes_of_interacting}).
    \item BLEND\_FLAG: Indicates whether a galaxy is blended with one or more HECATE galaxies. The PGC numbers of the blended counterparts are provided as a list. Users are advised to treat galaxies flagged as blended with caution (see Sect.~\ref{subsec:sizes_of_interacting}).
    \item STAR: Flags objects that are either stars misclassified as galaxies ("S") or objects that show indications of star contamination ("S?"). Users are strongly advised not to use objects flagged as secure stars (STAR = "S") and to treat objects flagged as candidate stars (STAR = "S?") with caution. See Sect.~\ref{sec:star_contamination} for more details.
\end{itemize}

\section{Columns comparison between HECATEv1 and HECATEv2}\label{append:HECv1_HECv2_columns}
For legacy users of the first version of the HECATE catalogue, in Table~\ref{tab:column_status} we summarise the status of each HECATEv1 column with respect to the updated version (HECATEv2) presented in this work. 
In particular, a column status of "Unchanged" indicates that the column name and in some cases the values in HECATEv2 are retained from HECATEv1. A column status of "Renamed" means that only the column name has been modified, while the values remain identical. A column status of "Updated" indicates that data for more objects are available in the HECATEv2. A column status of "New calculation" means that the column has been entirely re-derived using the updated methods presented in this work. Finally, a column status of "Deprecated" indicates that the column is no longer included in HECATEv2, while "New column" denotes a column introduced in HECATEv2 that has no counterpart in HECATEv1. We note that one column may have a double status. For example a column that has been updated and renamed will have a status of "Updated, Renamed".

\begin{onecolumn}
\footnotesize
\setlength{\tabcolsep}{2pt}

\begin{longtable}{p{0.22\textwidth}p{0.05\textwidth}p{0.12\textwidth}p{0.61\textwidth}}
\caption{Description of the HECATEv2 columns.}\label{tab:columns}\\
\toprule
Column & Data Type & Units & Description \\
\midrule
\endfirsthead

\multicolumn{4}{c}{{\bfseries \tablename\ \thetable{} -- continued from previous page}} \\
\toprule
Column & Data Type & Units & Description \\
\midrule
\endhead

\midrule \multicolumn{4}{r}{{Continued on next page}} \\ \bottomrule
\endfoot

\bottomrule
\endlastfoot

% ---- table content starts here ----
                 PGC &          I &                                                  - &             Principal Galaxies Catalogue number. \\
             OBJNAME &    S &                                                  - &                           Object name in HyperLEDA. \\
          ALLWISE\_ID &    S &                                                  - &          Object ID in the AllWISE Source Catalogue. \\
        SDSS17\_OBJID &    S &                                                  - &                  Photometric object ID in SDSSDR17. \\
    SDSS17\_SPECOBJID &    S &                                                  - &                Spectroscopic object ID in SDSSDR17. \\
              PS2\_ID &    S &                                                  - &                     Photometric object ID in PSDR2. \\
              RA\_HEC, DEC\_HEC &   F &                                        deg &             Decimal J2000.0 equatorial coordinates. \\
            F\_ASTROM &     I &                                                  - & Astrometric precision flag: -1 for ${\sim}0.1$\arcsec{}; 0 for ${\sim}1$\arcsec{}; 1 for ${\sim}10$\arcsec{}; and so on.\\
                  R1, R2 &   F &                                            arcmin &                            D$_{25}$ semimajor and semiminor axes \\
                  PA &   F &                                                deg &                  North-to-Northeast position angle \\
               RFLAG &    S &                                                  - & Flag of the size information: ["0", "1", "2", "03", "13", "23", "04", "14", "24"]. See Appendix \ref{append:HECv2_columns} for more details.\\
            R\_ORIGIN &    S &                                                  - & Origin of the size information. See Appendix \ref{append:HECv2_columns} for more details. \\
            T, T\_E  &   F &                                                  - & Numerical Hubble Type and its uncertainty. \\
            INCL &   F &                                                deg &                                        Inclination \\
            V, V\_E &   F &                      $\mathrm{km} \mathrm{s}^{-1}$ &  Heliocentric radial velocity and its uncertainty. \\
            V\_VIR, V\_VIR\_E &   F &            $\mathrm{km} \mathrm{s}^{-1}$ &  Virgo-infall corrected radial velocity and its uncertainty.\\
            DIST, DIST\_E &   F &                                                Mpc &  Distance and its uncertainty. \\
           DIST\_METH &     I &                                                  - & Methodology for the distance derivation in HECATEv2. [0, 1, 2, 3]. See Appendix \ref{append:HECv2_columns} for more details. \\
        AG, AI &   F &                                                mag &   Galactic and Intrinsic absorption. \\
        <UBVI>T\_m &   F &                                                mag &             Total U,B,V, and I bands apparent magnitudes. \\
        <UBVI>T\_m\_e  &   F &                                                mag &Uncertainties on U,B,V, and I bands apparent magnitudes. \\
        S?\_FLUX &   F &                                                 Jy &  IRAS flux. ? = 12, 25, 60, and 100 $\mathrm{\mu m}$, respectively. \\
        S?\_FLUX\_QF &   I & - & Quality flag, 0 = not in IRAS, 1 = upper limit , 2 = moderate quality, 3 = high quality in FSC, 4 = flux from RBGS. \\
        W?\_m, W?\_m\_e  &   F &                                         mag (Vega) & Homogenised hybrid AllWISE magnitudes and their uncertainties. ? = 3.4, 4.6, 12, and 22 $\mathrm{\mu m}$, respectively. \\
    ALLWISE\_APORIGIN &    S &                                                  - & AllWISE aperture adopted. ["Circular 8.25", "Elliptical"]. See Sect.~\ref{subsec:midir_phot_surveys_AllWISE} \\
          ALLWISE\_QF &    S &                                                  - & AllWISE 8-characters quality flag. See Appendix \ref{append:HECv2_columns} for more details.  \\
             WF?\_m, WF?\_m\_e  &   F &                                         mag (Vega) & WISE forced photometry magnitudes and their uncertainties. ? = 3.4, 4.6, 12, and 22 $\mathrm{\mu m}$, respectively. \\
             <JHK>\_m, <JHK>\_m\_e  &   F &                                                mag (Vega) &  2MASS magnitude in J, H, and K bands and their uncertainties.\\
        2MASS\_ORIGIN &     I &                                                  - & Origin of the 2MASS photometry: 0 = none, 1 = LGA, 2 = XSC, 3 = PSC. \\
        <ugrizy>\_m &   F &                                           mag (AB) &    Homogenised SDSS-DR17/NSA/PS1-DR2 u,g,r,i,z,and y band apparent magnitudes. \\
        <ugrizy>\_m\_e &   F &                                           mag (AB) & Uncertainties on homogenised SDSS-DR17/NSA/PS1-DR2 u,g,r,i,z,and y apparent magnitudes. \\
      OPTPHOT\_ORIGIN &    S &                                                  - & Six-character origin flag of the homogenised optical photometry. See Appendix \ref{append:HECv2_columns} for more details. \\
          OPTPHOT\_QF &    S &                                                  - & Six-character quality flag of the homogenised optical photometry. See Appendix \ref{append:HECv2_columns} for more details.\\
         <LINE>\_FLUX &   F & $10^{-17}\ \mathrm{erg}\mathrm{s}^{-1}\mathrm{cm}^{-2}$ & Line flux adopted from the MPA-JHU DR8 catalogue. <LINE>=["H\_BETA", "OIII\_5007", "OI\_6300", "H\_ALPHA", "NII\_6584", "SII\_6717", "SII\_6731"]. See Appendix \ref{append:HECv2_columns} for more details. \\
         <LINE>\_FLUX\_ERR & F& $10^{-17}\ \mathrm{erg}\mathrm{s}^{-1}\mathrm{cm}^{-2}$ & Line flux uncertainty adopted from the MPA-JHU DR8 catalogue. See Appendix \ref{append:HECv2_columns} for more details. \\
         SPEC\_ORIGIN &    S &                                                  - & Origin of the spectroscopic information: ["SDSSDR8", "SDSSDR17", "SKI"]. See Appendix \ref{append:HECv2_columns} for more details.  \\
    W1\_W2\_RF, W2\_W3\_RF &   F &                                                - & Mid-IR colors used for the Photometric activity classification. We note that these colors are derived from the unrescaled no aperture corrected fiber magnitude since the classifier was trained on these magnitudes. See Sect.~\ref{subsec:phot_activity_class}.\\
     G\_R\_FIB\_RF  &   F &                                                - & Optical colors used for the Photometric activity classification. We note that these colors are derived from the unrescaled no aperture corrected fiber magnitude since the classifier was trained on these magnitudes. See Sect.~\ref{subsec:phot_activity_class}. \\
      ACTIVITY\_CLASS &   F &                                                  - & Nuclear activity classification: 0=Star-forming, 1=AGN, 2=LINERS, 3=Composites, 4=Passive, 5=DO NOT USE.\\
 ACTIVITY\_CLASS\_METH &    S &                                                 - & Method used for the classification. ["SoDDA", "SVM3d", "PH"]. See Appendix \ref{append:HECv2_columns} for more details.\\
            <CLASS>\_PROB &   F &                                                  - &  Probability of a galaxy to belong in each class. <CLASS>= SFG, AGN, LINER, COMP, PASS, respectively. \\
             EBV, EBV\_e &   F &                                                  - &                                         Extinction and its uncertainty. See Sect.~\ref{subsec:SFR_Mstar_S18_K23}  \\
         LOG\_SFR\_GSW &   F &              $\mathrm{M}_{\odot} \mathrm{yr^{-1}}$ &            Distance-rescaled decimal logarithm of the SFR in GSWLC-2. \\
       LOG\_MSTAR\_GSW &   F &                               $\mathrm{M}_{\odot}$ & Distance-rescaled decimal logarithm of the $\mathrm{M}_{\odot}$ in GSWLC-2. \\
         LOG\_SFR\_HEC &   F &             $\mathrm{M}_{\odot}\ \mathrm{yr}^{-1}$ & Homogenised decimal logarithm of the SFR in HECATEv2. \\
      LOG\_SFR\_HEC\_QF &    S &                                                  - & Star-formation quality flag. See Appendix \ref{append:HECv2_columns} for more details. \\
  LOG\_SFR\_HEC\_ORIGIN &    S &                                                  - & Reference paper from which we adopted or derived the SFR: ["S18", "K23", "L19"]. See Sect.~\ref{subsec:SFR_Mstar_S18_K23} and Sect.~\ref{subsec:SFR_Mstar_L19}.\\
    LOG\_SFR\_HEC\_METH &    S &                                                  - & Method used for the derivation/adoption of the SFR. See Appendix \ref{append:HECv2_columns} for more details. \\
       LOG\_MSTAR\_HEC &   F &                               $\mathrm{M}_{\odot}$ & Homogenised decimal logarithm of the $\mathrm{M}_{\odot}$ in HECATEv2. \\
    LOG\_MSTAR\_HEC\_QF &    S &                                                  - & Stellar mass quality flag. See Appendix \ref{append:HECv2_columns} for more details.\\
LOG\_MSTAR\_HEC\_ORIGIN &    S &                                                  - & Reference paper from which we adopted or derived the \stellarmass{}: ["S18", "K23", "L19"]. See Sect.~\ref{subsec:SFR_Mstar_S18_K23} and Sect.~\ref{subsec:SFR_Mstar_L19}\\
  LOG\_MSTAR\_HEC\_METH &    S &                                                  - & Method used for the derivation/adoption of the \stellarmass{}. See Appendix \ref{append:HECv2_columns} for more details.\\
               METAL &   F &                                                  - &                 Gas phase metalliciy in HECATEv2 [12+log(O/H)]. \\
          METAL\_METH &    S &                                                  - & Method used for the derivation of the metallicity. ["O3N2/C20", "MZ/C20"]. See Appendix \ref{append:HECv2_columns} for more details.\\
          LOG\_MBH & F & $\mathrm{M}_{\odot}$ & SMBH mass as it was derived from the \stellarmass{} - M$_{\rm BH
          }$ relation. See Sect.~\ref{sec:SMBH_host_properties}\\   
          LOG\_MBH\_ORIGIN & S & - & Method used for the derivation of SMBH mass: ["G20-Late", "G20-Early", "G20-All"]. See Appendix \ref{append:HECv2_columns} for more details.\\   
            DUP\_FLAG &   F &                                                  - & Duplication flag. See Appendix \ref{append:HECv2_columns} for more details. \\
          BLEND\_FLAG &    S &                                                  - & Blending flag. See Appendix \ref{append:HECv2_columns} for more details. \\
                STAR &    S &                                                  - & Star contamination flag. ["S", "S?"] See Appendix \ref{append:HECv2_columns} for more details. \\
            WC\_NOTES &    S &                                                  - & Wrong coordinates flag: "WC?" The galaxy may has wrong coordinates. See Sect.~\ref{subsec:wrong_phot_wrong_coord} \\
               NOTES &    S &                                                  - &     General notes for a galaxy when is needed. Users are advised to use objects with "NOTES = NULL". \\
\end{longtable}
\medskip
\parbox{\columnwidth}{\footnotesize
Note: Columns 2 and 3 present the data types (I=Integer, S=String, F=Float) and the units for the HECATEv2 columns.}
\end{onecolumn}

%%%%%%%%%%%%%%%%%%%%%%%%%%%%%%%%%%%%%%%%%%%%%%%%%%
% Don't change these lines
\bsp	% typesetting comment
\label{lastpage}

\begin{onecolumn}
\footnotesize
\setlength{\tabcolsep}{2pt}

\begin{longtable}{p{0.31\textwidth}p{0.31\textwidth}p{0.20\textwidth}}
\caption{Column comparison between HECATEv1 and HECATEv2.}\label{tab:column_status}\\
\toprule
HECATEv1 & HECATEv2 & Column status \\
\midrule
\endfirsthead

\multicolumn{3}{c}{{\bfseries \tablename\ \thetable{} -- continued from previous page}} \\
\toprule
\midrule
\endhead

\midrule \multicolumn{3}{r}{{Continued on next page}} \\ \bottomrule
\endfoot

\bottomrule
\endlastfoot

% ---- table content starts here ----
                 PGC &          PGC &             Unchanged \\
             OBJNAME &    OBJNAME &    Unchanged             \\
            ID\_NED,ID\_NEDD,ID\_IRAS,ID\_2MASS & - & Deprecated \\
            SDSS\_PHOTID & SDSS17\_OBJID &  Updated, Renamed \\
            SDSS\_SPECID & SDSS17\_SPECOBJID &  Updated, Renamed\\ 
            - & PS2\_ID& New column\\ 
            RA, DEC &  RA\_HEC, DEC\_HEC & Renamed \\
            F\_ASTROM, R1, R2, PA &  F\_ASTROPM, R1, R2, PA & Unchanged \\
            RFLAG & RFLAG & Updated \\
            RSOURCE & R\_ORIGIN & Updated, Renamed\\
            T,E\_T,INCL,V, V\_VIR & T,E\_T,INCL,V,V\_VIR & Unchanged\\
            E\_V, E\_V\_VIR & V\_E, V\_VIR\_E & Renamed \\
            NDIST, EDIST & - & Deprecated \\
            D, E\_D & DIST, DIST\_E & New calculation, Renamed \\
            D\_LO68, D\_HI68, D\_LO95, D\_HI95 & - & Deprecated \\
            DMETHOD & DIST\_METHOD & New calculation, Renamed\\
            UT,BT,VT,IT & UT\_m, BT\_m, VT\_m, IT\_m & Renamed\\
            E\_UT,E\_BT,E\_VT,E\_IT & UT\_m\_e,BT\_m\_e,VT\_m\_e,IT\_m\_e & Renamed\\
            AG, AI & AG, AI & Unchanged \\
            S12, S25, S60, S100 & S12\_FLUX, S25\_FLUX,S60\_FLUX,S100\_FLUX & Renamed\\
            Q12, Q25, Q60, Q100 & S12\_FLUX\_QF,S25\_FLUX\_QF, S60\_FLUX\_QF, S100\_FLUX\_QF & Renamed\\
            - & ALLWISE\_APORIGIN & New column \\
            - & ALLWISE\_QF & New column \\
            - & W1\_m, W2\_m, W3\_m, W4\_m & New column \\
            - & W1\_m\_e,W2\_m\_e,W3\_m\_e,W4\_m\_e & New column \\
            WF1, WF2, WF3, WF4 & WF1\_m, WF2\_m, WF3\_m, WF4\_m & Renamed\\
            E\_WF1, E\_WF2, E\_WF3, E\_WF4 & WF1\_m\_e,WF2\_m\_e,WF3\_m\_e, WF4\_m\_e & Renamed\\
            WFPOINT, WFTREAT & - & Deprecated \\
            J,H,K & J\_m, H\_m, K\_m & Renamed \\
            E\_J, E\_H, E\_K, & J\_m\_e, H\_m\_e, K\_m\_e & Renamed \\
            FLAG\_2MASS & 2MASS\_ORIGIN & Renamed \\
            U,G,R,I,Z & u\_m, g\_m, r\_m, i\_m, z\_m & Updated, Renamed\\
            E\_U, E\_G, E\_R, E\_I, E\_Z & u\_m\_e, g\_m\_e, r\_m\_e, i\_m\_e, z\_m\_e & Updated, Renamed\\
            - & y\_m, y\_m\_e & New column \\
            - & OPTPHOT\_ORIGIN & New column \\
            - & OPTPHOT\_QF & New column \\
            - & <LINE>\_FLUX$^{\ast}$ & New column\\
            - & <LINE>\_FLUX\_ERR$^{\ast}$ & New column\\
            - & SPEC\_ORIGIN & New column \\
            - & W1\_W2\_RF, W2\_W3\_RF & New column\\
            - & G\_R\_FIB\_RF & New column \\ 
            logL\_TIR, logL\_FIR & - & Deprecated \\
            logL\_60u, logL\_12u, logL\_22u, logL\_K & - & Deprecated \\
            ML\_RATIO& - & Depricated \\
            logSFR\_TIR, logSFR\_FIR & - & Deprecated \\
            logSFR\_60u, logSFR\_12u, logSFR\_22u & - & Deprecated \\
            logSFR\_HEC & LOG\_SFR\_HEC & New calculation, Renamed \\
            FLAG\_SFR\_HEC & - & Deprecated\\
            -& LOG\_SFR\_HEC\_QF & New column\\
            -& LOG\_SFR\_HEC\_ORIGIN & New column\\
            -& LOG\_SFR\_HEC\_METH & New column\\
            logM\_HEC & LOG\_MSTAR\_HEC & New calculation, Renamed \\
            -& LOG\_MSTAR\_HEC\_QF & New column\\
            -& LOG\_MSTAR\_HEC\_ORIGIN & New column\\
            -& LOG\_MSTAR\_HEC\_METH & New column\\
            logSFR\_GSW, logM\_GSW & LOG\_SFR\_GSW, LOG\_MSTAR\_GSW & New calculation, Renamed\\
            MIN\_SNR & - & Deprecated \\
            METAL& METAL & Unchanged, New calculation\\
            FLAG\_METAL & - & Deprecated \\
            - & METAL\_METHOD & New column\\
            CLASS\_SP & ACTIVITY\_CLASS & New calculation, Renamed \\
            - & ACTIVITY\_CLASS\_METH & New column \\
            - & <CLASS>$^{\ast \ast}$\_PROB & New column \\
            - & EBV, EBV\_e & New column\\
            AGN\_S17, AGN\_HEC & - & Deprecated \\

\end{longtable}
\medskip
\parbox{\columnwidth}{\footnotesize
$^{\ast}$ See Table~\ref{tab:columns} for the available spectral lines in <LINE>.
$^{\ast \ast}$ See Table~\ref{tab:columns} for the available activity classes in <CLASS>.}
\end{onecolumn}
%%%%%%%%%%%%%%%%%%%%%%%%%%%%%%%%%%%%%%%%%%%%%%%%%%
% Don't change these lines
\bsp	% typesetting comment
\label{lastpage}

\end{document}